\documentclass[11pt]{article}\usepackage{geometry}
\usepackage{graphicx,amssymb,epstopdf,amsmath,cite,subcaption,breqn,multirow,array,tikz,verbatim,amsthm,citesort,wasysym}
\usetikzlibrary{arrows,chains,matrix,positioning,scopes}
\usepackage{color}

\newtheorem{lemma}{Lemma}[section] 

\newcommand{\btau}{\boldsymbol{\tau}}

\newcommand{\real}{\mathbb{R}}
\newcommand{\ealpha}{\hat{E}}

\title{Technical Report: \\ Compressive Temporal Higher Order Cyclostationary Statistics}
\author{Chia Wei Lim and Michael B. Wakin%
\thanks{Department of Electrical Engineering and Computer Science, Colorado School of Mines. E-mail: clim,mwakin@mines.edu. This work was partially supported by DSO National Laboratories of Singapore. A preliminary version of this work was first presented at the 2012 SPIE Defense, Security, and Sensing Conference \cite{cwlim_dss_2012}.}
}

\begin{document}
\maketitle{}

\begin{abstract}
The application of nonlinear transformations to a cyclostationary signal for the purpose of revealing hidden periodicities has proven to be useful for applications requiring signal selectivity and noise tolerance. The fact that the hidden periodicities, referred to as cyclic moments, are often compressible in the Fourier domain motivates the use of compressive sensing (CS) as an efficient acquisition protocol for capturing such signals. In this work, we consider the class of Temporal Higher Order Cyclostationary Statistics (THOCS) estimators when CS is used to acquire the cyclostationary signal assuming compressible cyclic moments in the Fourier domain. We develop a theoretical framework for estimating THOCS using the low-rate nonuniform sampling protocol from CS and illustrate the performance of this framework using simulated data.
\end{abstract}

\section{Introduction}
\subsection{Temporal Higher Order (Cyclostationary) Statistics}
\label{sec:introduction_hos_hocs}
\begin{figure}
\begin{minipage}[b]{.5\linewidth}
\includegraphics[width=0.9\textwidth]{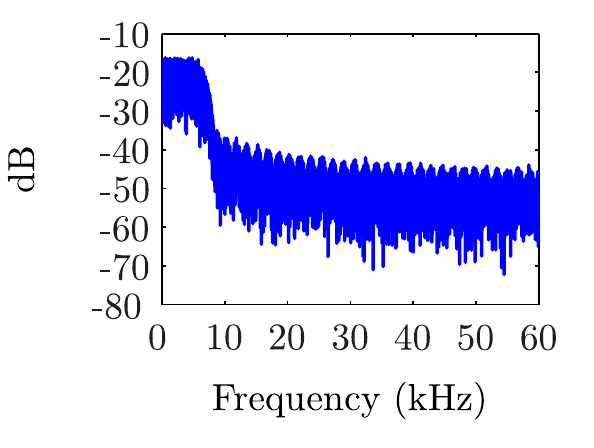}
\subcaption{\small\sl BPSK signal}\label{fig:2psk_1_0_spectrum}
\end{minipage}%
\begin{minipage}[b]{.5\linewidth}
\includegraphics[width=0.9\textwidth]{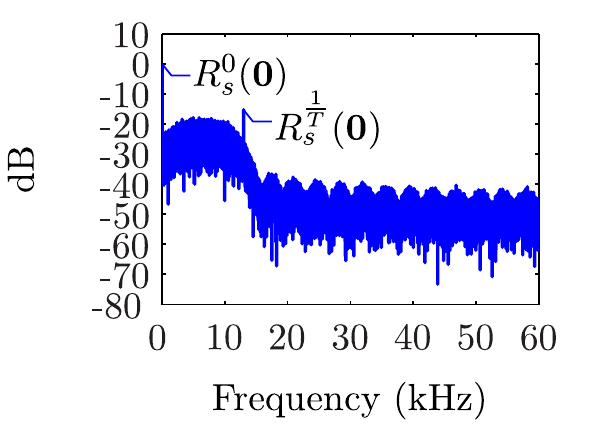}
\subcaption{\small\sl squared BPSK signal}\label{fig:2psk_2_1_spectrum}
\end{minipage}
\caption{\small\sl Periodicities (peaks) are hidden in the spectrum of a BPSK signal but revealed in the spectrum of the squared signal.}
\label{fig:2psk_npt_spectrum}
\end{figure}

A cyclostationary signal (as defined in \cite{Gardner_book}) is one which, after undergoing nonlinear transformations, will contain certain periodicities. Communication signals are not, in general, periodic. However, many communication signals are cyclostationary. For example, it is known~\cite{Reichert92} that by taking a Binary Phase-Shift-Keying (BPSK) signal and multiplying it by itself, one can generate periodicities in the signal at twice its carrier frequency. Figure~\ref{fig:2psk_1_0_spectrum} shows the spectrum of a BPSK signal while Figure~\ref{fig:2psk_2_1_spectrum} shows the spectrum of the squared signal. Likewise, if one raises a BPSK signal to the fourth power, periodicities at four times its carrier frequency can be generated. These ``hidden'' periodicities, referred to as \emph{higher order statistics} (HOS)~\cite{Reichert92}, are generated when such nonlinear transformations (namely, raising the signal to some power) are applied to the signal.
			
The pioneering work of Gardner et al. \cite{Gardner_1_94,Gardner_2_94}, in formulating the theory of \emph{temporal higher order cyclostationary statistics} (THOCS), refers to the hidden periodicities as the \emph{cyclic moments} of the signal. Given the cyclic moments of the signal, one can compute its corresponding \emph{cyclic cumulants} which have been shown to be signal selective and robust to Gaussian noise. The theory of THOCS subsumes that of HOS and defines additional statistical functions such as cyclic cumulants. 

\subsection{Compressive Sensing (CS)}
\label{sec:introduction_cs}

The theory of \emph{compressive sensing} (CS) has in recent years offered the great promise of efficiently capturing essential signal information with low-rate sampling protocols, often below the minimum rate required by the Nyquist sampling theorem. The use of CS can dramatically reduce both the complexity of a signal receiver and the amount of data that must be stored and/or transmitted downstream~\cite{Donoho06,Candes06,Wakin12}.

CS exploits the fact that some high-bandwidth signals are {\em compressible} in that they can be well described by a small number of coefficients when expressed in an appropriate domain. Early work in CS~\cite{Candes06} established the feasibility of collecting a random low-rate stream of nonuniform signal samples (as opposed to a deterministic high-rate stream of uniform signal samples) to acquire a signal that has relatively few large coefficients in its Fourier-domain spectrum. Given a sufficient number of nonuniform samples, subsequent reconstruction of the signal is possible by solving a convex optimization problem. What is remarkable about CS is that the requisite number of samples depends on the compressibility of the signal's spectrum instead of the width of its spectrum (the latter is what is required by the Nyquist sampling theorem).

\subsection{Challenges of THOCS and the Promise of CS}
\label{sec:challenges_hos_hocs}

While THOCS have proven to be useful, there are key challenges that one faces when attempting to estimate THOCS from traditional uniform samples of a cyclostationary signal. The higher the order of THOCS (that is, the higher the order of the nonlinear transformations applied to the signal), the more one must oversample the signal relative to its Nyquist rate to prevent aliasing. To accurately estimate THOCS, one also needs to observe the signal for an adequate (often long) time interval. These restrictions translate to heavy sampling burdens and data storage requirements.

As noted in Section~\ref{sec:introduction_hos_hocs}, a typical cyclostationary signal is not compressible in the frequency domain. However, the signal may be compressible after undergoing a nonlinear transformation, and the dominant peaks in the Fourier spectrum (see, for example, Figure~\ref{fig:2psk_2_1_spectrum}) are in fact the THOCS that one might wish to estimate. This motivates the question of whether CS in general, and low-rate nonuniform sampling in particular, can somehow be used to simplify the data acquisition process in problems involving THOCS. The immediate conundrum that one faces is the following: since the signal coming in to the receiver is not compressible, one cannot simply apply traditional CS sampling and reconstruction protocols to that signal (and then, say, estimate THOCS using conventional methods). While one could consider applying nonlinear transformations to the incoming signal and {\em then} collecting nonuniform samples of the transformed signal, this could significantly increase the complexity of a receiver.

In this paper we consider the problem of estimating THOCS directly from low-rate nonuniform samples of the original cyclostationary signal as well as low-rate nonuniform samples of delayed versions of the signal. In short, rather than collecting nonuniform samples of a nonlinearly transformed signal, one can simply apply these same nonlinear transformations to nonuniform samples of the original cyclostationary signal and nonuniform samples of the delayed versions of it. With these transformed nonuniform samples one could consider applying traditional CS reconstruction protocols and then extracting THOCS features. However, these reconstruction algorithms could result in additional processing latency and require significant computational resources. Therefore, we borrow ideas from the field of compressive signal processing (CSP)~\cite{Davenport10} and propose a very simple alternative method for estimating THOCS directly from the transformed nonuniform samples without the need for full-scale signal reconstruction.

In summary, we propose a traditional CS receiver front end that collects random low-rate streams of nonuniform samples of the incoming signal and delayed versions of it. While the complexity of such a receiver (see, e.g.,~\cite{Wakin12} for a simplified version) is increased due to the additional signal delay paths, it has the potential to dramatically reduce the data acquisition, storage, and transmission burdens compared to conventional THOCS estimators that are based on high-rate uniform sampling. Our back end processing, in which we estimate THOCS from the nonuniform samples, differs from traditional CS and is computationally very simple. We refer to the resulting estimated THOCS as \emph{compressive temporal higher order cyclostationary statistics} (CTHOCS). This paper describes the acquisition/estimation framework underlying CTHOCS and analyzes its effectiveness theoretically and empirically in comparison with conventional THOCS based on high-rate uniform sampling.

Our paper is organized as follows. In Section~\ref{sec:problem_formulation}, we first provide brief reviews of the theories of THOCS and low-rate nonuniform sampling. In Section~\ref{sec:cs_hocs}, we discuss how THOCS can be estimated from low-rate nonuniform samples and subsequently evaluate the performance of such a CTHOCS estimator. We demonstrate the signal selectivity of CTHOCS in Section~\ref{sec:chocs_amr}. We conclude with a final discussion in Section~\ref{sec:conclusions}.

\section{Preliminaries}
\label{sec:problem_formulation}

While the content in this section is intended to make this paper as self-contained as possible, it is not complete. For completeness, the reader may refer to~\cite{Spooner_93,Gardner_1_94,Gardner_2_94,Wakin12,Candes06,Donoho06,Leskow02}.

\subsection{THOCS Preliminaries}
\label{sec:hocs_prelim}

\subsubsection{Fraction-of-Time Probability (FOT)}

Mathematical analysis of HOCS is based on the \emph{fraction-of-time} (FOT) probability framework~\cite{Gardner_1_94,Leskow02,Gardner91}, where statistical parameters or functions are defined based on a single realization of an infinite duration signal. This is in contrast to the stochastic process framework where statistical parameters or functions are defined based on the ensemble average of multiple realizations of the stochastic signal. (For ergodic signals, statistical functions coincide between the two frameworks.) The FOT probability framework is useful in applications where only a single realization of the signal is available. As we will discuss, it facilitates the definitions of expectation, moments and cumulants which are important for the study of cyclostationary time series.

For a given time-series $f(u)$, the multiple sine-wave extraction operator\footnote{This operator is denoted by $\hat{E}^{\{\alpha\}}$ in \cite{Gardner_1_94,Gardner_2_94}.} is defined \cite{Gardner_1_94,Gardner_2_94} as
\begin{align*}
\ealpha\left[f(u)\right](t) 
&\triangleq\sum_{\alpha\in A(f(u))} \langle f(u) e^{-j2\pi\alpha u}\rangle_u e^{j 2\pi \alpha t}
\end{align*}
where $\langle\cdot\rangle_u$ denotes the time averaging operation with respect to the variable $u$ and the set $A(f(u))$ contains all frequencies $\alpha \in \real$ for which $ \langle f(u) e^{-j2\pi\alpha u}\rangle_u \neq 0$; this set is assumed to be countable for the time-series we will deal with. (Equivalently,  $A(f(u))$ contains the set of frequencies $\alpha$ at which the Fourier transform of $f(u)$ contains a Dirac impulse.)

The operator $\ealpha[\cdot]$ is said to extract all additive sine-wave components from $f$. It can be viewed as first analyzing a signal (in the Fourier domain) for all periodic components and then rebuilding the signal only from these same periodic components.\footnote{In general, the signal may contain non periodic components as well.} When $f(u)$ is a time series defined in the FOT probability framework, the $\ealpha[\cdot]$ operator extracts the deterministic component of its argument and hence is analogous to the expectation operator $\mathbb{E}[\cdot]$ in the stochastic process framework.

\subsubsection{THOCS Statistical Functions}
\label{sec:hocs_statistical_functions}

We now provide the definitions of various THOCS statistical functions based on the FOT analysis framework. The THOCS statistical functions, collectively referred to as \emph{moments} and \emph{cumulants}, are analogous to the moments and cumulants defined in the stochastic analysis framework.  We note here that while there exists another set of HOCS in the spectral domain \cite{Gardner_1_94,Gardner_2_94}, THOCS shall be the only set of HOCS considered in this framework.

\begin{enumerate}
\item Time-varying $n$th order moments

The {\em time-varying $n$th order moments} of a signal $x(t)$ are defined\cite{Gardner_1_94,Gardner_2_94} as
\begin{equation}
\label{eqn:tvm}
R_x(t,\btau)_{n,q}\triangleq \ealpha\left[\prod_{i=1}^n x^{(*)_i}(u-\tau_i)\right](t),
\end{equation}
where the $i$th factor $x(u-\tau_i)$\footnote{The use of a minus sign here is to ensure consistency with regards to delaying the signal.} could be (optionally) conjugated and this conjugation is denoted by $(*)_i$. The variable $q$ is the total number of conjugations and $\btau\triangleq [\tau_1\dots \tau_n]^T$ is a vector of lags.\footnote{Following \cite{Gardner_1_94,Gardner_2_94}, \eqref{eqn:tvm} does not indicate which $q$ terms get conjugated, but throughout our paper, we will always assume the first $q$ terms in the product are conjugated.} The argument to $\ealpha[\cdot]$ in \eqref{eqn:tvm} is termed the \emph{lag product} of the signal $x(t)$. In other words, given a signal $x(t)$, its time-varying $n$th order moments can be obtained by first analyzing its lag product for all periodic components and then rebuilding its lag product from only these components.

\item $n$th order cyclic moments

The time-varying $n$th order moments of a signal $x(t)$ can be further expressed in terms of its lag-dependent Fourier coefficients (also known as its {\em $n$th order cyclic moments}) as $R_x(t,\btau)_{n,q}=\sum_{\alpha}R_x^{\alpha}(\btau)_{n,q}e^{j2\pi\alpha t}$, where the $n$th order cyclic moments satisfy
\begin{equation}
\label{eqn:cm}
R_x^{\alpha}(\btau)_{n,q}=\langle R_x(t,\btau)_{n,q}e^{-j2\pi \alpha t}\rangle_t .
\end{equation}
As mentioned in \cite{Gardner_1_94}, due to $R_x^\alpha(\btau)_{n,q}$ being sinusoidal jointly in the $n$ translation variables $\btau$, setting $\tau_1=0$ retains all the information present in $R_x^\alpha(\btau)_{n,q}$. In general, $R_x^\alpha(\btau)_{n,q}$ is not integrable with respect to $\btau$.

\item Time-varying $n$th order cumulants

We first define some key parameters that the time-varying $n$th order cumulants are dependent on. Distinct partitions of the index set $\{1,2,\dots,n\}$ are referred to as $D_n$, $d$ is the number of elements in a partition, and the set of indices belonging to a partition is denoted by $v_i,\;i\in\{1,2,\dots,d\}$. Table~\ref{tab:cumulants_parameters} shows the key parameters with their respective values when $n=3$.

\begin{table}[t]
\setlength{\tabcolsep}{2.2pt}
\centering
\begin{tabular}{|c|c|c|c|c|c|}
\hline
$D_n$& $\{1\}\{2\}\{3\}$& $\{1\;2\}\{3\}$& $\{1\;3\}\{2\}$& $\{2\;3\}\{1\}$& $\{1\;2\;3\}$\\
\hline
$d$&$3$&$2$&$2$&$2$&$1$\\
\hline
\multirow{3}{*}{$v_i$}& $v_1=\{1\}$&$v_1=\{1,2\}$& $v_1=\{1,3\}$& $v_1=\{2,3\}$& $v_1=\{1,2,3\}$\\
& $v_2=\{2\}$&$v_2=\{3\}$& $v_2=\{2\}$& $v_2=\{1\}$& \\
& $v_3=\{3\}$&&&&\\
\hline
\end{tabular}
\caption{\small\sl Partitions for $n=3$. For a given $n$, all distinct partitions $D_n$ can be generated using a recursive formula~\cite{Spooner_93}.
}
\label{tab:cumulants_parameters}
\end{table}

Using the moment-to-cumulant (M-C) conversion formula\footnote{There also exists a cumulant-to-moment conversion formula \cite{Gardner_1_94}.}\cite{Gardner_1_94}, the {\em time-varying $n$th order cumulants} of a signal $x(t)$ are given by
\begin{equation}
\label{eqn:moment_cumulant}
\small C_x(t,\btau)_{n,q}=\sum_{D_n}\left[(-1)^{d-1}(d-1)!\prod_{i=1}^dR_x(t,\btau_{v_i})_{n_i,q_i}\right],
\end{equation}
where $n_i=|v_i|$, $q_i$ is the number of conjugations used to compute the time-varying moments defined below,
\begin{equation}
R_x(t,\btau_{v_i})_{n_i,q_i}\triangleq \ealpha\left[\prod_{k\in v_i}x^{(*)_k}(u-\tau_k)\right](t),
\label{eq:lagvi}
\end{equation}
$\btau_{v_i}$ is the vector of lags in $\btau$ with indices in $v_i$, and $(*)_k$ again denotes optional conjugation of the $k$th factor $x(u-\tau_k)$.

\item $n$th order cyclic cumulants

The {\em $n$th order cyclic cumulants} are the Fourier coefficients of the time-varying $n$th order cumulants, $$C_x^{\beta}(\boldsymbol{\tau})_{n,q}\triangleq \langle C_x(t,\boldsymbol{\tau})_{n,q}e^{-j2\pi\beta t}\rangle_t.$$
It can be further shown\cite{Spooner_93} that
\begin{equation}
C_x^{\beta}(\boldsymbol{\tau})_{n,q} 
=\sum_{D_n}\left[(-1)^{d-1}(d-1)!\sum_{\boldsymbol{{\alpha}^T1}=\beta}\prod_{i=1}^{d}R_x^{\alpha_i}(\boldsymbol{\tau}_{v_i})_{n_i,q_i}\right], \label{eqn:tvcc}
\end{equation}
The inner sum in \eqref{eqn:tvcc} is over all possible vectors $\boldsymbol{\alpha}=[\alpha_1 \dots \alpha_d]^T$ whose entries sum to $\beta$ and satisfy $R_x^{\alpha_i}(\boldsymbol{\tau}_{v_i})_{n_i,q_i} \neq 0$ for $i = 1,2,\dots,d$. (The value of $d$ changes depending on the partition of $D_n$ in the outer sum.) Thus, given the cyclic moments of a signal, one can use \eqref{eqn:tvcc} to compute its corresponding cyclic cumulants. In practice, for a given $n_i$, $q_i$, and $\boldsymbol{\tau}_{v_i}$, one can determine the candidate frequencies for $\alpha_i$ by identifying peaks in the spectrum of the lag product appearing in \eqref{eq:lagvi}.

Given that $C_x^\beta(\btau)_{n,q}$ is also sinusoidal jointly in the $n$ translation variables $\btau$, setting $\tau_1=0$ again retains all the information present in $C_x^\beta(\btau)_{n,q}$. In general, $C_x^\alpha(\btau)_{n,q}$ is integrable with respect to $\btau$ for a large class of cyclostationary signals of practical interest~\cite{Gardner_2_94}.

\end{enumerate}

\subsection{CS Preliminaries}
\label{sec:ncs_preliminaries}

\subsubsection{Sparsity and Compressive Measurements}
\label{sec:cstheory}

CS builds on the framework of sparse (and compressible) signal representations in order to simplify signal acquisition. Suppose we have a finite length-$L$ vector $y$ whose entries are samples of an analog signal acquired at or above the Nyquist rate. If we represent $y$ in terms of an $L\times L$ dictionary (or basis) $\Phi$, where $y=\Phi\eta$, $y$ is said to have an $s$-sparse representation in $\Phi$ if the length $L$-coefficient vector $\eta$ has just $s\ll L$ nonzero entries. If $\eta$ has $s$ significant coefficients (but the rest are very small), $y$ is said to be compressible in the dictionary $\Phi$.

Cand\`{e}s et al.~\cite{Candes06} showed that if one collects just $P < L$ compressive measurements of a sparse signal $y$ via a $P\times L$ sensing matrix $R$ such that $w=Ry=R\Phi\eta=:A\eta$, exact recovery of $\eta$ (and therefore, $y$) from $w$ may be possible. For example, it is possible to solve the under-determined system of equations $w = A \eta$ (having $P$ equations and $L$ unknowns), if the $P\times L$ matrix $A$ satisfies a condition known as the \emph{restricted isometry property} (RIP). A matrix $A$ satisfies the RIP of order $s$ with isometry constant $\delta_s \in(0,1)$ if the following holds for all $s$-sparse vectors $\eta$: $(1-\delta_s)\|\eta\|_2\le \|A\eta\|_2\le (1+\delta_s)\|\eta\|_2$. Exact recovery of $\eta$ from $w$ can be achieved by solving a convex optimization problem (namely, $\ell_1$-minimization~\cite{CandesRIP}) if $A$ satisfies the RIP of order $2s$ and $\delta_{2s}$ is small. A variety of greedy iterative algorithms (e.g., CoSaMP~\cite{Needell10}) have also been proposed. Moreover, these recovery algorithms are robust to noise and provide an approximate solution if $\eta$ is not exactly sparse but compressible.

All of the algorithms mentioned above attempt to find, among all $\eta$ that satisfy the constraint $w = A \eta$, the sparsest possible solution. This recovery problem would be straightforward if $A$ were a square matrix and its columns were orthogonal---one could uniquely identify $\eta$ simply by computing $A^H w$, where $(\cdot)^H$ denotes the conjugate transpose operator. In a CS problem, $A$ is not square (it is underdetermined) and so its columns cannot be orthogonal. However, the RIP property requires that small sets of columns of $A$ be roughly orthogonal, and so $A^H w$ will provide a rough approximation to the true vector $\eta$. Most CS reconstruction algorithms---particularly greedy algorithms---can be interpreted as correcting the estimate $A^H w$ to account for the minor correlations among the columns of $A$. However, it is also worth noting that CSP theory~\cite{Davenport10} supports the fact that the uncorrected estimate $A^H w$ can itself provide a reasonable approximation of $\eta$ when $\eta$ is sparse or compressible, and using this estimate can be computationally attractive because it is much simpler to compute than a full-scale (but more accurate) signal reconstruction.

\subsubsection{Low-rate Nonuniform Sampling}
\label{sec:nusintro}

For a signal that is sparse or compressible in the frequency domain---that is, one whose Fourier spectrum contains relatively few large peaks---Cand\`{e}s et al.~\cite{Candes06} propose a CS acquisition strategy based on low-rate nonuniform sampling. This strategy involves collecting a random low-rate stream of nonuniform samples of an incoming analog signal, as opposed to the deterministic high-rate stream of uniform signal samples acquired by a conventional Nyquist-rate receiver~\cite{Wakin12}. In some implementations of low-rate nonuniform sampling, the random time intervals are integer multiples of a finer sampling interval (at least as fine as Nyquist rate of the signal); we assume this is the case in our subsequent discussions. In practice, one can generate these random time intervals via a pseudorandom number generator.

Low-rate nonuniform sampling is compatible with the formulation and notation defined above. Recall the length-$L$ vector $y$ whose entries are uniform samples of an analog signal acquired at or above the Nyquist rate (denote the sampling interval as $T_s$). A collection of $P$ nonuniform samples of the same analog signal can be represented as $w = Ry$, where $R$ is a $P \times L$ binary matrix containing a single $1$ in a random position in each row. For example, if $P=4$ and $L=10$, one possible $R$ matrix would be
$$\small R=
\begin{bmatrix}
1&0&0&0&0&0&0&0&0&0\\
0&0&0&1&0&0&0&0&0&0\\
0&0&0&0&1&0&0&0&0&0\\
0&0&0&0&0&0&0&0&0&1\\
\end{bmatrix}.
$$
In such an instance, we would have $w = [w(0) ~ w(1) ~ w(2) ~ w(3)]^T = [y(0) ~ y(3T_s) ~ y(4T_s) ~ y(9T_s)]^T$. It was shown in \cite{Rudelson06} that if $\Phi$ is the $L \times L$ DFT matrix and $R$ is a random selection matrix as shown above, then $A = R \Phi$ will satisfy the RIP of order $P/\log^4(L)$ with high probability. (Technically, the RIP holds for the appropriately scaled matrix $A = \sqrt{L/P} \cdot R \Phi$.) Hence, one can recover an $s$-sparse coefficient vector $\eta$ and consequently the Nyquist-rate sample vector $y$ from only a small multiple times $s$ nonuniform samples.

\section{CTHOCS Framework}
\label{sec:cs_hocs}

In this section, we explain our CTHOCS framework, in which THOCS are estimated directly from low-rate nonuniform samples of the original cyclostationary signal as well as low-rate nonuniform samples of delayed versions of the signal.

\subsection{Signal Acquisition}
\label{sec:daq}

In a typical scenario one wishes to estimate THOCS for various orders $n$, optional conjugations $q$, and delay vectors $\btau$; recall from \eqref{eqn:tvm} that each combination of $n$, $q$, and $\btau$ defines a unique lag product of the incoming signal $x(t)$. The question is how to estimate these statistics without excessive sampling or specialized hardware. The idea of using low-rate nonuniform sampling for estimating THOCS is inspired by the compressibility of the lag product.

\begin{figure}[t]
\begin{minipage}[b]{0.5\textwidth}
\includegraphics[width=0.85\textwidth]{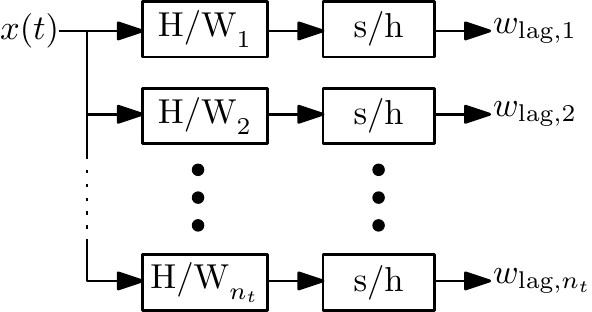}
\subcaption{\small\sl Specialized hardware sampling scheme.}
\label{fig:single_ch_nus_sampling}
\end{minipage}
\begin{minipage}[b]{0.5\textwidth}
\includegraphics[width=0.95\textwidth]{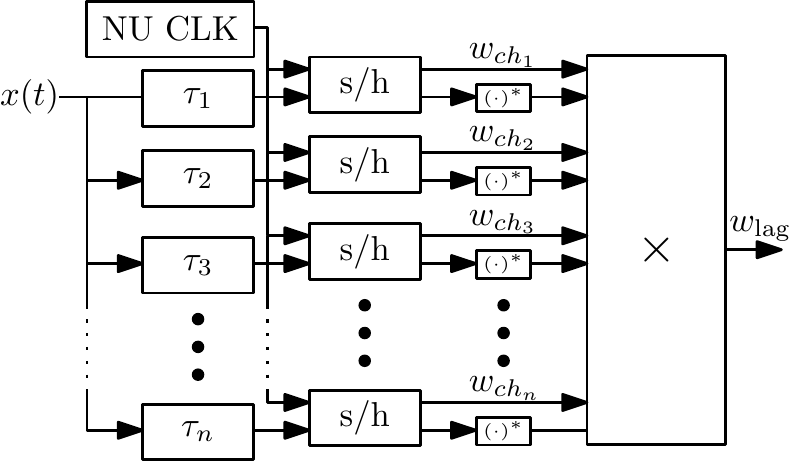}
\subcaption{\small\sl Proposed CS multi-channel sampling scheme.}
\label{fig:multi_ch_nus_sampling}
\end{minipage}
\caption{\small\sl Comparison of two sampling schemes. (a)~Lag products computed with specialized analog hardware. (b)~Proposed scheme using analog signal delays and low-rate nonuniform samplers.}
\end{figure}

For the sake of discussion, Figure~\ref{fig:single_ch_nus_sampling} shows one hypothetical way that low-rate nonuniform sampling could be used assuming specialized analog hardware was available to generate various (compressible) signal lag products. This system consists of $n_t$ channels, where $n_t$ is the total number of unique lag products of interest. In each channel $j \in \{1, 2, \dots, n_t\}$, an analog hardware device $\text{H/W}_j$ computes a lag product of $x(t)$ corresponding to some desired values of $n$, $q$, and $\btau$, all of which may vary from channel to channel. This lag product is then passed through a low-rate nonuniform \emph{sample-and-hold} (s/h) device (to be formally defined in Section \ref{sec:hocsncs}), and the output $w_{\text{lag},j}$ is a nonuniform sample stream of the $j$th lag product. Unfortunately, such an acquisition setup would require a significant amount of specialized hardware and its complexity would scale linearly with the desired number $n_t$ of unique lag products.

As an alternative, we propose the signal acquisition setup shown in Figure~\ref{fig:multi_ch_nus_sampling}. For the sake of simplicity, this figure presents the acquisition scheme for a single lag product as appearing in \eqref{eqn:tvm}, i.e., for a single combination of $n$, $q$, and $\btau$. In each channel $i \in \{1,2,\dots,n\}$, an analog delay produces $x(t-\tau_i)$, where $\tau_i$ is the $i$th entry of the delay vector $\btau$. This delayed signal is then passed through a low-rate nonuniform s/h device, whose output $w_{ch_i}$ is a nonuniform sample stream of the $i$th delayed input signal. In this scheme we assume that all s/h devices are driven by a common nonuniform clock (NU CLK) which dictates the sample times. The output $w_{\text{lag}}$ of the acquisition system is a digital product of the nonuniform sample streams from each channel:
\begin{equation}
w_{\text{lag}}[k] = \prod_{i=1}^n w_{ch_i}^{(*)_i}[k],
\label{eq:mikewlag}
\end{equation}
where the optional conjugations match those in \eqref{eqn:tvm}.

This multi-channel acquisition setup can be easily implemented using only analog signal delays and low-rate nonuniform samplers. At first sight, the proposed sampling scheme may seem at odds with our objective of reducing the overall sampling rate necessary for THOCS estimation due to the use of $n$ channels, each with its own s/h device. However, we note that if the delay vector $\btau$ contains any repeated entries (for example, if $\tau_2 = \tau_3$), the duplicate channel can be removed. Moreover, this system can easily be extended to allow for multiple lag products (e.g., $n_t$ lag products as in Figure~\ref{fig:single_ch_nus_sampling}). The only channels that would need to be added would be those corresponding to delays $\tau$ not already represented in the system. In fact, in most practical cases of interest known to the authors, THOCS estimation is very often restricted to a small set of unique delays $\tau$ of interest. Examples include the \emph{Delay-Reduced Classifier} as proposed in~\cite{910700}, \cite{987051} and the set of classification features proposed in \cite{540605} where all $\tau_i = 0$; in such a case, only {\em one} sampling channel would be needed. Nonetheless, we note here that one cannot expect significant average sampling rate reductions in the case when a wide range of delays $\tau$ are required since the trade-off between average sampling rate reduction and increased number of analog signal delays required no longer justifies such a multi-channel sampling scheme.

\subsection{The Lag Product Measurements}

We now describe how the lag product measurements $w_{\text{lag}}$ (coming from the system in Figure~\ref{fig:multi_ch_nus_sampling}) can be represented mathematically. Although the following hypothetical vector is not physically measured, for each channel $i \in \{1,2,\dots,n\}$, let us define\footnote{Classical windowing techniques could be used in our framework to concentrate main lobe energy and reduce side lobe energy.}
\begin{equation}
\label{eq:highu}
y_{ch_i}\triangleq\begin{bmatrix}x(0-\tau_i)&x(T_s-\tau_i)&\cdots&x((L-1)T_s-\tau_i)\end{bmatrix}^T
\end{equation}
to be a collection of $L$ uniform time samples, having sampling interval $T_s$ (where $\frac{1}{T_s}$ is at least the Nyquist rate), of the analog cyclostationary signal $x(t-\tau_i)$ for the $i$th channel. At the $i$th channel output, the nonuniform sample vector $w_{ch_i}$ can be thought of as being constructed by physically measuring only certain, randomly selected entries of $y_{ch_i}$ for each channel. For $m = 0,1,\dots, L-1$, we define a binary random variable $a_m$ indicating whether entry $m$ of $y_{ch_i}$ is selected for inclusion in $w_{ch_i}$; we assume these $a_m$ are IID Bernoulli random variables with $\mathbb{P}[a_m=1]=\gamma$ and $\mathbb{P}[a_m=0]=1-\gamma$ for some constant $\gamma \in (0,1]$. As indicated in our diagram, we assume that all s/h devices are driven by a common nonuniform clock; that is, we assume that $a_m$ takes on the same value across different channels for the same entry $m$. Thus, on average, the parameter $\gamma$ denotes the fraction of the total Nyquist rate samples retained by the nonuniform sampling protocol per channel. We let $P$ denote the actual total number of nonuniform samples of $x(t)$ collected per channel; we will have $P \approx \gamma L$, and we can write the sample vector as $w_{ch_i} = R y_{ch_i}$, where $R$ is a $P \times L$ selection matrix as explained in Section~\ref{sec:nusintro} and is identical across all $n$ channels.

As noted in Section~\ref{sec:introduction_hos_hocs}, a typical cyclostationary signal is not compressible in the frequency domain. In particular, consider the uniform high-rate sample vector $y_{ch_i}$ defined in~\eqref{eq:highu}. Letting $\Phi$ denote the $L \times L$ DFT matrix, the DFT coefficients of $y_{ch_i}$, given by $\eta_{ch_i} = \Phi^H y_{ch_i}$, will not typically be sparse or compressible. This raises the question of how we can hope to apply CS reconstruction techniques given the nonuniform samples $w_{ch_i} = Ry_{ch_i}$.

The answer comes from the fact that we do not attempt to reconstruct the full rate sample vectors $y_{ch_i}$ in the individual channels. Rather, we note that the overall system output $w_{\text{lag}}$ corresponds to nonuniform samples of the signal's lag product, which is compressible. Defining
$$
x_{\text{lag}}(t) \triangleq \prod_{i=1}^n x^{(*)_i}(t-\tau_i)
$$
to be the lag product appearing in \eqref{eqn:tvm}, we can define $y_{\text{lag}}$ to be a hypothetical collection of $L$ uniform time samples of $x_{\text{lag}}(t)$:
\begin{equation}
y_{\text{lag}}[k] \triangleq x_{\text{lag}}((k-1)T_s) = \prod_{i=1}^n y_{ch_i}^{(*)_i}[k].
\label{eq:mikeylag}
\end{equation}
From \eqref{eq:mikewlag} and \eqref{eq:mikeylag}, it follows that
$$
w_{\text{lag}} = R y_{\text{lag}},
$$
where $R$ is the same $P \times L$ selection matrix used in the individual channels. So, the system output $w_{\text{lag}}$ can be viewed as a subset of the high-rate collection of $L$ uniform time samples from the signal's lag product. We note that such a computation is possible thanks to the commutative relationship of the lag product operation with the nonuniform sampling operator: {\em it is not necessary to compute the analog lag product before sampling the signal}.

For suitable choices of $n$ and $q$, the DFT coefficients of $y_{\text{lag}}$, given by $\eta_{\text{lag}} = \Phi^H y_{\text{lag}}$, may be compressible. Indeed, when dealing with finite data in practice it is common to estimate THOCS directly from $\eta_{\text{lag}}$: the locations and values of peaks in the DFT spectrum provide estimates of the true $\alpha$ values and corresponding cyclic moments $R^\alpha_x(\btau)_{n,q}$, respectively. In situations where $\eta_{\text{lag}}$ is indeed compressible, we should be able to apply CS techniques for reconstructing or estimating $\eta_{\text{lag}}$ from $w_{\text{lag}}$. This is the approach we adopt in the CTHOCS framework.

\subsection{A Simple Estimator of the Lag Product Spectrum}

We can express the collected the nonuniform samples of the lag product as $w_{\text{lag}} = R y_{\text{lag}} = R \Phi \eta_{\text{lag}} = A \eta_{\text{lag}}$, where $A = R \Phi$. As we explained in Section~\ref{sec:cstheory}, when $\eta_{\text{lag}}$ is sparse or compressible, the estimate
\begin{equation}
\widehat{\eta}_{\text{lag}}=A^Hw_{\text{lag}}=\Phi^HR^Hw_ {\text{lag}},
\label{eq:etaLag}
\end{equation}
can provide a suitable approximation to $\eta_{\text{lag}}$ that is very simple to compute. In particular, when $R$ is a random selection matrix and $\Phi$ is a DFT matrix, the estimate $\widehat{\eta}_{\text{lag}}=\Phi^HR^Hw_ {\text{lag}}$ can be computed simply by zero-padding $w_{\text{lag}}$ (creating a length-$L$ vector containing the $P$ entries of $w_{\text{lag}}$ at the nonuniform sample locations) and taking the FFT of this zero-padded vector. This estimation procedure is not only simple (both conceptually and computationally), but as we show in Section~\ref{sec:hocsncs} it is also compatible with the THOCS analytical framework. In particular, we propose to extract statistics from the locations and values of the peaks of $\widehat{\eta}_{\text{lag}}$.

Using the RIP, we can argue that the locations and values of the peaks of $\widehat{\eta}_{\text{lag}}$ should coincide roughly with those of $\eta_{\text{lag}}$.
\begin{lemma}
\label{lm:estimate_error}
Let $s$ denote the number of spectral peaks in $\eta_{\text{lag}}$ and suppose for simplicity that $L = ms$ for some positive integer $m$. Let $\eta_{\text{lag},s}$ denote a length-$L$ vector containing only the $s$ peaks from $\eta_{\text{lag}}$ (and zeros elsewhere), and let $\eta_{\text{lag},t} = \eta_{\text{lag}} - \eta_{\text{lag},s}$ denote the remaining $L-s$ entries of $\eta_{\text{lag}}$. If $A$ satisfies the RIP of order $s+1$ with isometry constant $\delta$, the DFT estimate $\widehat{\eta}_{\text{lag}}=A^H w_ {\text{lag}}$ approximates the true DFT coefficients $\eta_{\text{lag}}$ up to the following accuracy:
\begin{equation}
\Big|\widehat{\eta}_{\text{lag}}[k]-\eta_{\text{lag}}[k]\Big|\le \delta \|\eta_{\text{lag},s}\|_2+\delta (m-1) \sqrt{s}\|\eta_{\text{lag},t}\|_\infty.
\label{eqn:estimate_error}
\end{equation}
\end{lemma}
\noindent \textbf{Proof}: See Appendix \ref{pf:estimate_error}.

~

As noted in Section~\ref{sec:nusintro}, when $\Phi$ is an $L \times L$ DFT matrix and $R$ is a random selection matrix with $P$ rows, then $A = R \Phi$ (appropriately rescaled) will satisfy the RIP of order $P/\log^4(L)$ with high probability. Thus, if $\eta_{\text{lag}}$ is highly concentrated at its $s$ peak values, one can expect $\eta_{\text{lag},t}$ to be small, leading to a rough guarantee in \eqref{eqn:estimate_error} that $\widehat{\eta}_{\text{lag}}[k] \approx \eta_{\text{lag}}[k]$ for each coefficient $k$. While such a simple estimator may not be the most accurate possible---the peak value of $\eta_{\text{lag},t}$ may not be small unless one invokes Lemma 3.1 with a large value of $s$ to account for many large entries in $\eta_{lag}$, and most CS reconstruction algorithms are more sophisticated to account for correlations among the columns of $A$---we opt for this FFT-based estimation technique because of its simplicity and its compatibility with the THOCS analytical framework (see Section~\ref{sec:hocsncs}).

\begin{figure}[t]
\centering
\includegraphics[width=0.35\textwidth]{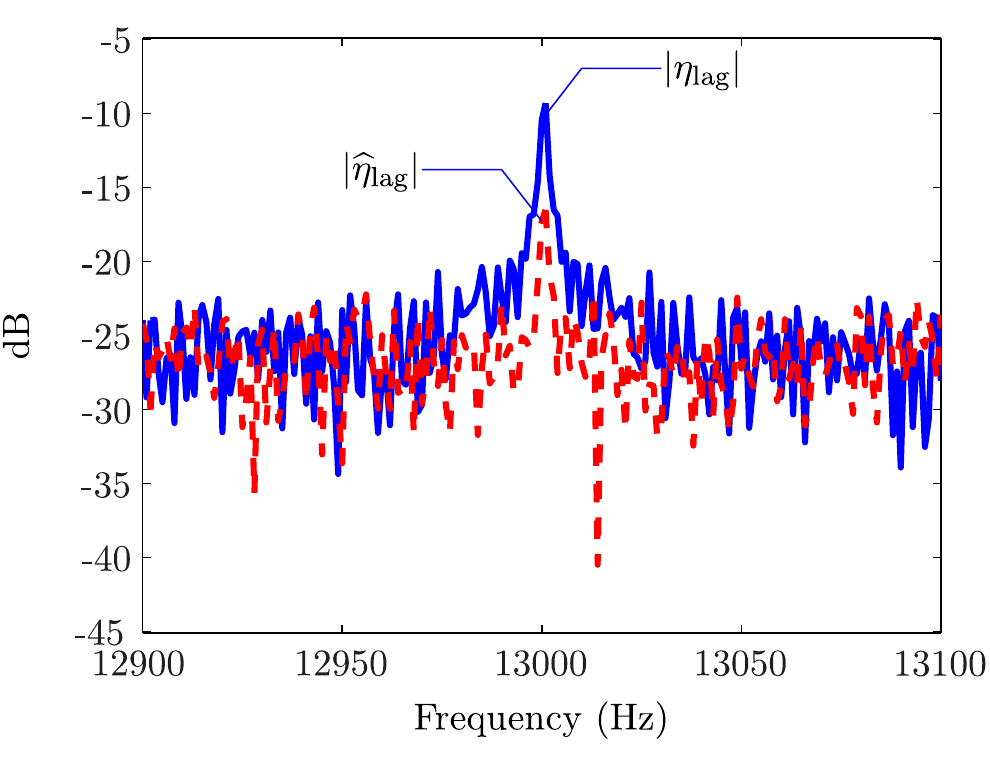}
\caption{\small\sl Lag product DFT (solid blue curve) and estimate obtained from low-rate nonuniform samples (dashed red curve).}
\label{fig:hocs_chocs}
\end{figure}

As an illustration of this technique, the solid blue curve in Figure~\ref{fig:hocs_chocs} shows (a zoomed portion of) the magnitude of the DFT $\eta_{\text{lag}}$ of the lag product of a 2PSK communication signal for $\btau=\boldsymbol{0}$. This curve was generated from uniform samples acquired at the Nyquist rate for this signal. The dashed red curve in this plot shows $\widehat{\eta}_{\text{lag}}$, the zero-padded DFT of the lag product computed from nonuniform samples of the signal with $\gamma = 0.2$ (so that $P \approx 0.2 L$). As we have discussed, the location and value of the peak in the blue curve can be used to infer one of the cyclic moments of the signal. (Figure \ref{fig:hocs_chocs} only shows one of a few peaks.) Conveniently, we see a peak in the red curve that---although weaker---coincides with the position of the peak in the blue curve. This gives some hope that information about the cyclic moments can be recovered from the low-rate nonuniform samples of the signal. We note here that while the weaker peak seems at odds with the approximation error bound given in Lemma~\ref{lm:estimate_error}, the displayed red curve has not been appropriately normalized (rescaled). This shall be discussed in the next section. We shall also explain formally what statistics we extract from $\widehat{\eta}_{\text{lag}}$ and discuss how these relate to THOCS of the original cyclostationary signal in the next section.

\subsection{Definition of CTHOCS}
\label{sec:hocsncs}

Recall that $\eta_{\text{lag}}$ corresponds to the DFT of uniform samples of the lag product of the original signal. Conventionally, one would estimate THOCS directly from $\eta_{\text{lag}}$: the locations and values of peaks in the DFT spectrum provide estimates of the true $\alpha$ values and corresponding cyclic moments $R^\alpha_x(\btau)_{n,q}$, respectively. Although THOCS are formally defined in terms of analog signals and continuous-time Fourier transforms, it is possible to show~\cite{Gardner_2_94} that cyclic moments of the analog signal and cyclic moments of its uniformly sampled counterpart are equivalent up to a periodic repetition in the frequency domain, as occurs with any uniform sampling of an analog signal.

When we compute $\widehat{\eta}_{\text{lag}}$ by taking a zero-padded DFT of $w_{\text{lag}}$, one way to view $\widehat{\eta}_{\text{lag}}$ is simply as an estimate of $\eta_{\text{lag}}$. Another way to view $\widehat{\eta}_{\text{lag}}$, however, is that it is the DFT of zeroed-out uniform samples of the lag product of the original signal (with zeros in locations not sampled in the nonuniform sampling protocol). We adopt this latter viewpoint for the purpose of formally defining CTHOCS: in short, we refer to THOCS of this zeroed-out signal as CTHOCS, we can relate these CTHOCS to THOCS of the original signal, and we can estimate CTHOCS from the locations and values of the peaks in $\widehat{\eta}_{\text{lag}}$.

\subsubsection{Preliminaries}

As explained above, the cyclic moments of the analog cyclostationary signal and cyclic moments of its uniformly sampled counterpart are effectively equivalent~\cite{Gardner_2_94}. In our discussion, to avoid the complexities of dealing with Dirac delta functions (especially when raising these functions to higher powers), we frame our discussion around a s/h version of the delayed analog cyclostationary signal for each channel of the proposed multi-channel sampling scheme as follows. We first assume uniform sampling at each channel of the multi-channel sampler and define the s/h version of the delayed signal $x(t-\tau_i)$ for the $i$th channel to be
\begin{align}
x_{i\:\textnormal{s/h}}(t,\tau_i)&\triangleq
\label{eq:ith_ch_s/h}
 x(t-\tau_i)\left[\sum_{m=-\infty}^{\infty}p_{\textnormal{s/h}}(t-mT_s)\right], 
\end{align}
where $m$ is the sample time index, $T_s$ is the sampling interval (bounded by the Nyquist sampling interval), and $p_{\textnormal{s/h}}(t)$ is the s/h pulse shape used in the proposed multi-channel sampling scheme. For convenience, we will use $p_{\textnormal{s/h}}(t) = \text{rect}\left(\frac{t}{\upsilon}\right)$, where $\upsilon$ is the pulse duration and $\upsilon \ll T_s$. Using \eqref{eq:ith_ch_s/h}, the s/h version of the lag product of signal $x(t)$ for a fixed $\btau$ (corresponding to the $\tau_i$'s in Figure~\ref{fig:multi_ch_nus_sampling}), $x_{\textnormal{lag s/h}}(t,\btau)_{n,q}$ is defined as  $$x_{\textnormal{lag s/h}}(t,\btau)_{n,q}\triangleq\prod_{i=1}^{n}x^{(*)_i}_{i\:\textnormal{s/h}}(t,\tau_i).$$ The Fourier coefficients of $x_{\textnormal{lag s/h}}(t,\btau)_{n,q}$ at the cycle frequencies will approximate the cyclic moments of $x(t)$, and one can make this approximation arbitrarily accurate by letting $\upsilon \rightarrow 0$ and normalizing appropriately.

We next consider nonuniform sampling at each channel of the multi-channel sampler and similarly define the s/h version of the zeroed-out delayed signal for the $i$th channel as follows:
\begin{align}
\label{eqn:ith_ch_ncs}
x_{i\:\textnormal{ncs}}(t,\tau_i)&\triangleq x(t-\tau_i)\left[\sum_{m=-\infty}^{\infty}a_mp_{\textnormal{s/h}}(t-mT_s)\right], 
\end{align}
where $a_m$ are the nonuniform sample indicators as defined in Section~\ref{sec:daq}, and the subscript ``ncs'' refers to nonuniform compressive samples. For use in our analysis, we note that the $a_m$'s are independent, identically distributed (IID) Bernoulli random variables where $\mathbb{E}[a_m^n]=\gamma,\;\forall n\in\mathbb{Z}^+$. In the FOT sense, $a_m$ can be thought of as taking on the value 1 with FOT probability $\gamma$ and taking on the value 0 with FOT probability $1-\gamma$. Again using \eqref{eqn:ith_ch_ncs}, the s/h version of the zeroed-out lag product of signal $x(t)$ for a fixed $\btau$ (corresponding to the $\tau_i$'s in Figure~\ref{fig:multi_ch_nus_sampling}), $x_{\textnormal{lag ncs}}(t,\btau)_{n,q}$ can be easily computed using the various channel outputs as $$x_{\textnormal{lag ncs}}(t,\btau)_{n,q}=\prod_{i=1}^{n}x^{(*)_i}_{i\:\textnormal{ncs}}(t,\tau_i).$$

\subsubsection{CTHOCS Statistical Functions}
\label{sec:chocs_statistical_functions}

We define CTHOCS statistical functions, for a given $\btau,\:n\text{ and } q$, in the context of $x_{\textnormal{lag ncs}}(t,\btau)_{n,q}$, the s/h version of the zeroed-out lag product of the incoming cyclostationary signal, and we compare these to THOCS of $x_{\textnormal{lag s/h}}(t,\btau)_{n,q}$, the s/h version of the lag product of the same signal. Let us remind the reader that the main assumption underlying the definition of CTHOCS is that the lag product of the cyclostationary signal is compressible in the frequency domain, i.e., the signal's lag product only contains a small number of unique dominant cycle frequencies (periodicities).

\begin{enumerate}
\item Time-varying $n$th order compressive moments

The time-varying $n$th order moments of a cyclostationary signal are defined in Section~\ref{sec:hocs_statistical_functions}. We denote the time-varying $n$th order moments computed using the s/h version of the lag products $x_{\textnormal{lag s/h}}(t,\btau)_{n,q}$ and $x_{\textnormal{lag ncs}}(t,\btau)_{n,q}$ by $R_{x_{\textnormal{s/h}}}(t,\btau)_{n,q}$ and $R_{x_{\textnormal{ncs}}}(t,\btau)_{n,q}$, respectively. Specifically,
\begin{equation*}
R_{x_{\textnormal{s/h}}}(t,\btau)_{n,q}\triangleq\ealpha \left[ x_{\textnormal{lag s/h}}(t,\btau)_{n,q}\right]
\end{equation*}
and \begin{equation*}
R_{x_{\textnormal{ncs}}}(t,\btau)_{n,q}\triangleq\ealpha\left[x_{\textnormal{lag ncs}}(t,\btau)_{n,q}\right].
\end{equation*}

Our CTHOCS framework relies heavily on the fact that these statistics are strongly related.

\begin{lemma}
\label{lm:comp_moments} The time-varying $n$th order moments obtained using the nonuniformly sampled lag product signal $x_{\textnormal{lag ncs}}(t,\btau)_{n,q}$ for a fixed $\btau$, $n$ and $q$ satisfy
\begin{equation}
R_{x_{\textnormal{ncs}}}(t,\btau)_{n,q} = \gamma R_{x_{\textnormal{s/h}}}(t,\btau)_{n,q}.
\label{eqn:tvm_ncs_1}
\end{equation}
\end{lemma}
\textbf{Proof}: See Appendix \ref{pf:comp_moments}.

In other words, the time-varying $n$th order moments computed using the collection of nonuniformly sampled signals \{$x_{i\:\textnormal{ncs}}(t,\tau_i)$\} are simply scaled by a factor of $\gamma$ compared to the time-varying $n$th order moments computed using the collection of uniformly sampled signals \{$x_{i\:\textnormal{s/h}}(t,\tau_i)$\} for a given $\btau,\: n\text{ and }q$. In light of \eqref{eqn:tvm_ncs_1}, we define our first CTHOCS quantity: the {\em time-varying $n$th order compressive moments} of our signal are defined to be
\begin{equation}
R^c_{x_{\textnormal{ncs}}}(t,\btau)_{n,q} \triangleq \frac{1}{\gamma} R_{x_{\textnormal{ncs}}}(t,\btau)_{n,q}.
\label{eqn:tvm_ncs_2}
\end{equation}
(All CTHOCS quantities will be denoted with a superscript ``$c$''.) Thanks to \eqref{eqn:tvm_ncs_1}, it follows that the ``compressive moments'' $R^c_{x_{\textnormal{ncs}}}(t,\btau)_{n,q}$ we define are actually equal to the traditional moments $R_{x_{\textnormal{s/h}}}(t,\btau)_{n,q}$ of the original uniformly sampled lag product of the signal for a given $\btau$, $n$ and $q$.

In practice, one must approximate $R^c_{x_{\textnormal{ncs}}}(t,\btau)_{n,q}$ based on a finite number of nonuniform samples of the signal at the output of each channel in Figure~\ref{fig:multi_ch_nus_sampling}. We propose to do this by computing the zero-padded DFT $\widehat{\eta}_{\text{lag}}$ of the vector $w_{\text{lag}}$, preserving just the peaks in $\widehat{\eta}_{\text{lag}}$, rescaling those peaks by $\frac{1}{\gamma}$, and then inverting the DFT. The resulting estimate of $R^c_{x_{\textnormal{ncs}}}(t,\btau)_{n,q}$ doubles as an approximation to the true moments $R_{x_{\textnormal{s/h}}}(t,\btau)_{n,q}$.

It is important to note that the strength of the peaks of the zero-padded DFT reduces proportionally with $\gamma$ (recall Figure~\ref{fig:hocs_chocs}), and so when dealing with a finite set of data, a peak detection scheme will be necessary to identify the locations of the peaks. When dealing with noisy data and high compression ratios ($\gamma \ll 1$) some peaks may actually fall below the noise floor and will thus be missed. The implications of dealing with finite data are discussed more thoroughly in Section~\ref{sec:chocs_measurement}.

\item $n$th order compressive cyclic moments

We define the {\em $n$th order compressive cyclic moments} of our signal by applying the cyclic moment formula \eqref{eqn:cm} to the time-varying $n$th order compressive moments we defined in \eqref{eqn:tvm_ncs_2}:
\begin{equation}
\label{eqn:tvcm_ncs}
R_{x_{\text{ncs}}}^{\alpha,c}(\btau)_{n,q} \triangleq \langle R^c_{x_{\text{ncs}}}(t,\btau)_{n,q}e^{-j2\pi \alpha t}\rangle_t. \end{equation}
Again, thanks to \eqref{eqn:tvm_ncs_1} and \eqref{eqn:tvm_ncs_2}, these ``compressive cyclic moments'' $R_{x_{\text{ncs}}}^{\alpha,c}(\btau)_{n,q}$ are equal to the cyclic moments $R_{x_{\textnormal{s/h}}}^\alpha(\btau)_{n,q}$ of the uniformly sampled lag product of the signal for a given $\btau$, $n$ and $q$.

In practice, when dealing with finite data one can estimate these compressive cyclic moments simply by using the locations and rescaled values of the peaks identified in the estimation of the compressive moments $R^c_{x_{\textnormal{ncs}}}(t,\btau)_{n,q}$ above.

By setting $\tau_1=0$ here again, all the information present in $R_{x_{\text{ncs}}}^{\alpha,c}(\btau)_{n,q}$ is retained. 

\item Time-varying $n$th order compressive cumulants

We define the {\em time-varying $n$th order compressive cumulants} of our signal by applying the moment-to-cumulant conversion formula \eqref{eqn:moment_cumulant} to the time-varying $n$th order compressive moments we defined in \eqref{eqn:tvm_ncs_2}:
\begin{equation}
C^c_{x_{\text{ncs}}}(t,\btau)_{n,q} \triangleq \sum_{D_n}\left[(-1)^{d-1}(d-1)!\prod_{i=1}^dR^c_{x_{\text{ncs}}}(t,\btau)_{n_i,q_i}\right], 
\label{eqn:tvc_ncs}
\end{equation}
where each term $R^c_{x_{\textnormal{ncs}}}(t,\btau)_{n_i,q_i}$ can be computed from the channel outputs in Figure~\ref{fig:multi_ch_nus_sampling}. Again, because of \eqref{eqn:tvm_ncs_1} and \eqref{eqn:tvm_ncs_2}, these ``compressive cumulants'' $C^c_{x_{\text{ncs}}}(t,\btau)_{n,q}$ are equal to the traditional cumulants $C_{x_{\textnormal{s/h}}}(t,\btau)_{n,q}$ of the uniformly sampled lag product of the signal for a given $\btau$, $n$ and $q$.\footnote{Because the product in \eqref{eqn:tvc_ncs} contains a variable number of terms $d$, the ``compressive cumulants'' $C^c_{x_{\textnormal{ncs}}}(t,\btau)_{n,q}$ will {\em not} have a simple linear relationship to the traditional cumulants $C_{x_{\text{ncs}}}(t,\btau)_{n,q}$ of the zeroed-out lag product signal.}

When dealing with finite data one can approximate the compressive moments $R^c_{x_{\textnormal{ncs}}}(t,\btau)_{n,q}$ as discussed above and then plug these estimates into \eqref{eqn:tvc_ncs} to obtain an estimate of the compressive cumulants since cumulants are not directly measurable quantities. If some peaks were missed in the zero-padded DFT spectrum, the estimated compressive cumulants could differ substantially from the original signal cumulants $C_{x_{\textnormal{s/h}}}(t,\btau)_{n,q}$.

Again, by setting $\tau_1=0$, all the information present in $C_{x_{\text{ncs}}}^{c}(t,\btau)_{n,q}$ is retained.

\item $n$th order compressive cyclic cumulants

We define the {\em $n$th order compressive cyclic cumulants} of our signal by applying the moment-to-cumulant conversion formula \eqref{eqn:tvcc} to the $n$th order compressive cyclic moments we defined in \eqref{eqn:tvcm_ncs}:
\begin{equation}
C_{x_{\text{ncs}}}^{\beta,c}(\btau)_{n,q}  
\triangleq \sum_{D_n}\left[(-1)^{d-1}(d-1)!\sum_{\boldsymbol{{\alpha}^T1}=\beta}\prod_{i=1}^{d}  R_{x_{\text{ncs}}}^{\alpha_i,c}(\btau)_{n_i,q_i}\right]. \label{eqn:ccc_1}
\end{equation}
Again, thanks to \eqref{eqn:tvm_ncs_1} and \eqref{eqn:tvm_ncs_2}, these ``compressive cyclic cumulants'' are equal to the traditional cyclic cumulants $C_{x_{\textnormal{s/h}}}^{\beta}(n,q)$ of the uniformly sampled signal for a given $\btau$, $n$ and $q$.

In practice, when dealing with finite data one can approximate the compressive cyclic moments $R_{x_{\text{ncs}}}^{\alpha,c}(n,q)$ as discussed above and then plug these estimates into \eqref{eqn:ccc_1} to obtain an estimate of the compressive cyclic cumulants. Again, if some peaks were missed, the estimated compressive cyclic cumulants could differ substantially from the original signal's cyclic cumulants $C_{x_{\textnormal{s/h}}}^{\beta}(n,q)$. We will demonstrate in Section~\ref{sec:chocs_amr}, however, that the estimated compressive cyclic cumulants still retain their signal selective property as compared to their conventional cyclic cumulants.

\end{enumerate}

\subsection{Implications of Finite Observations}
\label{sec:chocs_measurement}

We conclude this section with a few additional remarks concerning the estimates of THOCS and CTHOCS from finite, rather than infinite, data streams. Using finite observations of the uniformly sampled channel outputs in Figure~\ref{fig:multi_ch_nus_sampling}, an estimator of the cyclic moment $R_{x_{\textnormal{s/h}}}^\alpha(\btau)_{n,q}$ referred to as $\widehat{R}_{x_{\textnormal{s/h}}}^\alpha(t,\btau)_{n,q}$ can be defined as
\begin{equation}
\widehat{R}_{x_{\textnormal{s/h}}}^\alpha(t,\btau)_{n,q} \triangleq \frac{1}{\widehat{T}}\int_{t-\frac{\widehat{T}}{2}+\frac{\tau_0^*}{2}}^{t+\frac{\widehat{T}}{2}-\frac{\tau_0^*}{2}}x_{\textnormal{lag s/h}}(u,\btau)_{n,q}e^{-j2\pi\alpha u}\;du,
\label{eq:cm_estimator_1}
\end{equation}
where $\widehat{T}$ is the duration of a finite time window centered at time $t$ applied to the incoming analog signal $x(u)$ before its being fed into the multi-channel sampler of Figure~\ref{fig:multi_ch_nus_sampling}, $\tau_l\triangleq \text{max}\{\tau_1,\cdots,\tau_n\}$, $\tau_r\triangleq \text{min}\{\tau_1,\cdots,\tau_n\}$ and $\tau_0^*\triangleq \tau_r-\tau_l$. The time-varying mean of this estimator is equivalent to that proposed in \cite{Spooner_93} and (from \cite{Spooner_93}) is given by
\begin{equation}
\frac{1}{\widehat{T}}R_{x_{\textnormal{s/h}}}^\alpha(\btau)_{n,q}(\tau_0^*+\widehat{T}) +\frac{1}{\widehat{T}}\sum_{b\neq \alpha}R_{x_{\textnormal{s/h}}}^b(\btau)_{n,q}e^{-i\pi(\alpha-b)(\tau_l+\tau_r)} (\tau_0^*+\widehat{T}) \text{sinc}{\left[(\alpha-b)(\tau_0^*+\widehat{T})\right]e^{-i2\pi (\alpha-b)t}}.
\label{eq:meanBias}
\end{equation}
This equation reveals a bias in the estimator due to contributions by other (cycle) frequencies. This bias vanishes as $\widehat{T} \rightarrow \infty$. By extending this analysis, one can show that when estimating a compressive cyclic moment $R_{x_{\textnormal{ncs}}}^{\alpha,c}(\btau)_{n,q}$ from observations of the nonuniformly sampled channel outputs in Figure~\ref{fig:multi_ch_nus_sampling} with finite duration $\widehat{T}$ centered at time $t$, e.g., by computing the estimator
\begin{equation}
\widehat{R}_{x_{\textnormal{ncs}}}^{\alpha,c}(t,\btau)_{n,q}\triangleq\frac{1}{\gamma \widehat{T}} \int_{t-\frac{\widehat{T}}{2}+\frac{\tau_0^*}{2}}^{t+\frac{\widehat{T}}{2}-\frac{\tau_0^*}{2}}x_{\textnormal{lag ncs}}(u,\btau)_{n,q}e^{-j2\pi\alpha u}\;du,
\label{eq:ccm_estimator_1}
\end{equation}
the mean of the estimator is again given by~\eqref{eq:meanBias}.

The time-averaged variance of a cyclostationary time series $x(t)$ is defined \cite{Gardner_2_94} as
\begin{equation}
\text{Var}_0\{x(t)\}\triangleq\langle|x(t)|^2\rangle_t-\langle|\ealpha\{x(t)\}|^2\rangle_t.
\label{eqn:ta_var}
\end{equation}
Hence, the time-averaged variance of the compressive cyclic moment estimator $\widehat{R}_{x_{\textnormal{ncs}}}^{\alpha,c}(t,\btau)_{n,q}$ is given by
\begin{equation}
\text{Var}_0\left\{\widehat{R}_{x_{\textnormal{ncs}}}^{\alpha,c}(t,\btau)_{n,q}\right\}=\left\langle\left| \widehat{R}_{x_{\textnormal{ncs}}}^{\alpha,c}(t,\btau)_{n,q}\right|^2\right\rangle_t-\left\langle\left|\ealpha\left\{\widehat{R}_{x_{\text{ncs}}}^{\alpha,c}(t,\btau)_{n,q}\right\}\right|^2\right\rangle_t,
\label{eqn:ccm_var_1}
\end{equation}
where
\begin{equation}
\left\langle\left| \widehat{R}_{x_{\textnormal{ncs}}}^{\alpha,c}(t,\btau)_{n,q}\right|^2\right\rangle_t=\left\langle\frac{1}{\gamma^2 \widehat{T}^2} \int_{t-\frac{\widehat{T}}{2}+\frac{\tau_0^*}{2}}^{t+\frac{\widehat{T}}{2}-\frac{\tau_0^*}{2}}\int_{t-\frac{\widehat{T}}{2}+\frac{\tau_0^*}{2}}^{t+\frac{\widehat{T}}{2}-\frac{\tau_0^*}{2}} x_{\textnormal{lag ncs}}(r,\btau)_{n,q}\Big(x_{\textnormal{lag ncs}}(s,\btau)_{n,q}\Big)^* e^{-j 2 \pi \alpha (r-s)}drds\right\rangle_t
\label{eqn:ms_av_estimator}
\end{equation}
is the mean square absolute value of the estimator and $\ealpha\left\{\widehat{R}_{x_{\textnormal{ncs}}}^{\alpha,c}(t,\btau)_{n,q}\right\}$ is given by \eqref{eq:meanBias}.

\begin{lemma}
\label{lm:ccm_msav}The mean square absolute value of the estimator $\left\langle\left| \widehat{R}_{x_{\textnormal{ncs}}}^{\alpha,c}(t,\btau)_{n,q}\right|^2\right\rangle_t$ for a fixed $\btau$ can be shown to satisfy
\begin{align}
&\left\langle\left| \widehat{R}_{x_{\textnormal{ncs}}}^{\alpha,c}(t,\btau)_{n,q}\right|^2\right\rangle_t\nonumber\\
&=\frac{1}{\widehat{T}^2} \int_{-\frac{\widehat{T}}{2}+\frac{\tau_0^*}{2}}^{\frac{\widehat{T}}{2}-\frac{\tau_0^*}{2}}\int_{-\frac{\widehat{T}}{2}+\frac{\tau_0^*}{2}}^{\frac{\widehat{T}}{2}-\frac{\tau_0^*}{2}} \left\langle\prod_{i=1}^{n}x^{(*)_i}(r+t-\tau_i)\left(x^{(*)_i}(s+t-\tau_i)\right)^*\right\rangle_t\nonumber\\
&\times\left[\frac{1}{\gamma}\left\langle\sum_{m=-\infty}^{\infty} p^{n}_{\textnormal{s/h}}(r+t-mT_s)p^{n}_{\textnormal{s/h}}(s+t-mT_s)\right\rangle_t+\left\langle\sum_{u\ne v} p^n_{\textnormal{s/h}}(r+t-uT_s)p^n_{\textnormal{s/h}}(s+t-vT_s)\right\rangle_t\right]\nonumber\\
&\times e^{-j 2 \pi \alpha (r-s)}drds.
\label{eqn:ccm_var}
\end{align}
\end{lemma}
\textbf{Proof}: See Appendix \ref{pf:ccm_msav}.

We note that when $\gamma=1$, $\left\langle\left| \widehat{R}_{x_{\textnormal{ncs}}}^{\alpha,c}(t,\btau)_{n,q}\right|^2\right\rangle_t$ equals $\left\langle\left| \widehat{R}_{x_{\textnormal{s/h}}}^{\alpha,c}(t,\btau)_{n,q}\right|^2\right\rangle_t$ which results in identical variance when comparing both estimators. Since \eqref{eqn:ccm_var} can be split into two integrals, which must add to a real, nonnegative number for all $\gamma\in(0,1]$, it follows that $\left\langle\left| \widehat{R}_{x_{\textnormal{ncs}}}^{\alpha,c}(t,\btau)_{n,q}\right|^2\right\rangle_t$ must increase as $\gamma$ decreases. This, in turn, causes $\text{Var}_0\left\{\widehat{R}_{x_{\textnormal{ncs}}}^{\alpha,c}(t,\btau)_{n,q}\right\}$ to increase if $\widehat{T}$ is fixed and $\gamma$ decreases; however, this increase in variance can be mitgated by increasing $\widehat{T}$. While our main goal in this section is to establish the variance of our compressive cyclic moment estimator in relation to its full rate uniform sampling cyclic moment estimator counterpart, the interested reader is referred to  \cite{370106} and \cite{Spooner_93} for more detailed discussions on the variance of the full rate uniform sampling cyclic moment estimator as well as the cyclic cumulant estimator. In the following section, we focus our discussions on the signal selectivity aspect of CTHOCS and validate our theorems with simulations.

\section{CTHOCS Signal Selectivity}
\label{sec:chocs_amr}

As highlighted in our introduction and in the seminal papers of Gardner et al.~\cite{Gardner_1_94,Gardner_2_94}, two highly desirable properties of cyclic cumulants are signal selectivity and tolerance to Gaussian noise (when $n>2$). Therefore, we devote this section to empirically validating the signal selectivity of our proposed compressive cyclic cumulants.

Our proposed CTHOCS only requires the lag product of the cyclostationary signal to be compressible in the frequency domain. One example of such a cyclostationary signal is a baseband digital Quadrature Amplitude Modulation (Digital QAM) communication signal, as we explain in the next section.

\subsection{Baseband Digital QAM Signal Model}
\label{sec:signal_model}

We assume the cyclostationary signal to be acquired is a baseband digital QAM communication signal of the form
\begin{equation}
\label{eqn:signal_model}
s(t)
=\sum_{k=-\infty}^{\infty}as_kp(t-kT-t_0)e^{j(2\pi\Delta f_ct+\theta_c)},
\end{equation}
where $a$ is the signal amplitude, $s_k$ is the $k$th transmitted symbol, $p(t)$ is the signaling pulse shape, $T$ is the symbol period, $\Delta f_c$ is the residual carrier frequency offset, and $\theta_c$ is the carrier phase. We make the common assumption that $\Delta f_c \ll \frac{1}{T}$, and we also assume that symbols $s_k$ are IID, uniformly chosen from some finite dictionary depending on the modulation type of the signal. In the sections that follow, the notation $s(t)$ is reserved specifically for an incoming cyclostationary signal that obeys the model above. Henceforth, in the subsequent sections, we shall restrict ourselves to signals belonging to the signal model of \eqref{eqn:signal_model}.

\subsection{Signal Selectivity of Cyclic Cumulants}
\label{sec:hocs_amr_review}

We now provide a brief literature review of previous works focused on the signal selectivity of cyclic cumulants for the Digital QAM signal model \eqref{eqn:signal_model}.  We note here that while other papers may focus on the signal selectivity of other signal statistics such as non-cyclic cumulants or spectral HOCS for our signal model \eqref{eqn:signal_model}, our aim here is to establish the validity of compressive cyclic cumulants using cyclic cumulants as a benchmark.

Spooner considered the classification of both single and multiple carrier (co-channel) signals belonging to \eqref{eqn:signal_model} using cyclic cumulants in \cite{540605}. The signal selective features proposed were
\begin{equation}
F_s=\left\{\operatorname*{max}_{\btau} \left|C_s^\beta(\btau)_{n,q}\right|\right\},
\end{equation}
for $n=2,4,\dots,N$, $q=0,1,\dots,n$, $\btau=[\tau_1\dots \tau_n]$, $\beta=(n-2q)\Delta f_c +\frac{k}{T}$, $k=0,\dots,K$, and
\begin{equation}
C_s^{\beta}(\boldsymbol{\tau})_{n,q}
=\frac{a^n}{T}C_{s\: n,q}e^{j\theta_c(n-2q)}e^{j2\pi\Delta f_c\sum_{m=1}^{n}(-)_m\tau_m}e^{-j2\pi\frac{k}{T} t_0}
\left[\int_{-\infty}^{\infty}\left(\prod_{\ell=1}^{n}p(t+\tau_\ell)\right)e^{-j2\pi \frac{k}{T} t}dt\right].
\label{eqn:tvcc_2}
\end{equation}
Here, $C_{s\: n,q}=\sum_{D_n}(-1)^{d-1}(d-1)!\prod_{i=1}^dR_{s,v_i}$ is known as the symbol cumulants parameter, $R_{s,v_i}$ are known as the symbol moments, and $(-)_m$ denotes an optional minus sign depending on the optional conjugation. The symbol dependent parameters $R_{s,v_i}$ can be computed from the symbols $s_k$ in \eqref{eqn:signal_model} using $R_{s,v_i}=\mathbb{E}\left[s_k^{|v_i|-q_i}(s_k^*)^{q_i}\right]$, where $q_i$ is the number of conjugated indices in the set $v_i$. The proposed features from the incoming signal were subsequently estimated assuming known cycle frequencies (values of $\beta$), normalized, grouped in vector form and classified using a minimum Euclidean distance approach with theoretical feature vectors. Classification results were obtained for scenarios involving six (single carrier) and ten classes (four co-channel carriers).

We note here that for each $n$, only the maximum value of \eqref{eqn:tvcc_2} was used and this would occur for the case when $\btau=\boldsymbol{0}$ (the length-$n$ vector of zeros) based on inspection of \eqref{eqn:tvcc_2} and given the decaying nature of typical communication signal pulse shapes $p(t)$ \cite{Dobre03}.

As a revision to \cite{540605}, Spooner et al.\ demonstrated the validity of cyclic cumulants, for both single carrier and multiple co-channel carriers, when estimated with no prior information about the signals in \cite{910700} using warped cyclic cumulants (a variant of the cyclic cumulant features proposed in \cite{540605}). A costly maximum likelihood (ML) classifier was derived and subsequently approximated with two realizable alternatives, namely, the order-reduced classifier (ORC) and the delay-reduced classifier (DRC). Classification results were obtained for scenarios involving five classes (single carrier) and up to three co-channel carriers.

In \cite{987051}, Spooner extended the use of the fourth order cyclic cumulants (used in \cite{540605,910700}) to the sixth order and demonstrated that classification performance can be further improved with its use. Classification results were obtained for up to four class scenarios involving two co-channel signals.

Marchand et al.\ considered the use of a linear combination of cyclic cumulants containing second and fourth order cyclic cumulants for classification of digital QAM signals in \cite{613485,681573}.

Finally, Dobre et al.\ investigated the use of cyclic cumulant features for classification of digital QAM signals up to the eighth order in the single carrier scenario in \cite{Dobre03} assuming all parameters of the signal are known.

\subsection{Signal Selectivity of Compressive Cyclic Cumulants}
\label{sec:chocs_features}

As mentioned in the previous section, our goal here is to establish the validity of compressive cyclic cumulants using cyclic cumulants as a benchmark. Therefore, we shall be using the signal selective cyclic cumulant features as proposed in \cite{540605} as well as the minimum distance classification scheme proposed in \cite{540605}. We note here that in subsequent simulations, a four class scenario comprising of 2PSK, 4PSK, 8PSK and 16QAM single carrier classification will be considered. Therefore, a single feature minimum classification scheme, namely using
\begin{equation}
\label{eqn:c_4_0_feature}
F_s=\left|C^{\beta,c}_{s_{\text{ncs}}}(\boldsymbol{0})_{4,0}\right|,
\end{equation}
where $\beta=4\Delta f_c$ suffices. In this instance, using \eqref{eqn:tvcc_2}, the theoretical features can be computed as
\begin{equation}
F_s=\frac{a^4}{T}\left|C_{s\: 4,0}\right|
\left|\int_{-\infty}^{\infty}p(t)^4dt\right|,
\label{eqn:tvcc_3}
\end{equation}
where $\left|C_{s\: 4,0}\right|$ takes the values $2$, $1$, $0$, and $0.68$ for 2PSK, 4PSK, 8PSK and 16QAM respectively. In our simulations, we assume the signal amplitude $a$ is known and the interested reader is referred to \cite{540605} for a detailed discussion on feature normalization when various signal parameters of \eqref{eqn:signal_model} are unknown.

\subsection{Walkthrough CTHOCS Estimation Example}
\label{sec:chocs_estimation}

We now present an example of how $\widehat{C}_{s_{\text{ncs}}}^{\beta,c}(\btau)_{4,2}$, an estimate of $C_{s_{\text{ncs}}}^{\beta,c}(\btau)_{4,2}$ can be obtained from finite sets of nonuniform samples of a cyclostationary signal belonging to \eqref{eqn:signal_model} obtained from our proposed multi-channel sampling scheme. 
\begin{enumerate}
\item Using the collection of $w_{ch_i}$ low-rate nonuniform sample vectors, compute its lag product $w_{\text{lag}}$ using \eqref{eq:mikewlag} for $n=4$ and $q\in\{0,1,2\}$. For each $q\in\{0,1,2\}$, compute $\widehat{\eta}_{\text{lag}}$, the zero-padded DFT of $w_{\text{lag}}$, using \eqref{eq:etaLag}.

\item For each $q\in\{0,1,2\}$, detect the peaks in $\widehat{\eta}_{\text{lag}}$, rescale these peaks by $\frac{1}{\gamma}$, and denote these rescaled values as the estimated compressive cyclic moments $\widehat{R}_{s_{\text{ncs}}}^{\alpha,c}(\btau)_{n,q}$. For each peak detected, $\alpha$ denotes the cycle frequency at which the peak was detected.

\item Form an estimate of the compressive cyclic cumulant $C_{s_{\text{ncs}}}^{\beta,c}(\btau)_{4,2}$ (denote it as $\widehat{C}_{s_{\text{ncs}}}^{\beta,c}(\btau)_{4,2}$) by plugging the estimated compressive cyclic moments $\widehat{R}_{s_{\text{ncs}}}^{\alpha,c}(\btau)_{n,q}$ into \eqref{eqn:ccc_1}. For the sake of clarity, we provide the formula for computing $\widehat{C}_{s_{\text{ncs}}}^{\beta,c}(\btau)_{4,2}$:
    \begin{equation*}
\widehat{C}_{s_{\text{ncs}}}^{\beta,c}(\btau)_{4,2}= \sum_{D_4}\left[(-1)^{d-1}(d-1)!\sum_{\boldsymbol{{\alpha}^T1}=\beta}\prod_{i=1}^{d}  \widehat{R}_{s_{\text{ncs}}}^{\alpha_i,c}(\btau)_{n_i,q_i}\right].
    \end{equation*}    Before proceeding further, it may be helpful to examine the set $D_4$, which is the set containing all possible partitions of the index set \{1 2 3 4\}, as well as the $\sum$ term inside the square brackets. Since we consider the case where $n=4$ and $q=2$, we shall assume the first two indices are the conjugated terms and denote the index set as \{$1^*$ $2^*$ 3 4\}. Then $D_4$ contains the following partitions of \{$1^*$ $2^*$ 3 4\}:
\begin{enumerate}
\item \{$1^*$ $2^*$ 3 4\}: For this partition, $d=1$ (since there is only one factor) and the only compressive cyclic moment estimate we require is $\widehat{R}_{s_{\text{ncs}}}^{\beta,c}(\btau)_{4,2}$.
\item \{$1^*$ $2^*$\}\{3 4\}: For this partition, $d=2$ (since there are two factors) and the $\sum$ term has the form $\sum \widehat{R}_{s_{\text{ncs}}}^{\alpha_i,c}(\btau)_{2,2}\widehat{R}_{s_{\text{ncs}}}^{\alpha_j,c}(\btau)_{2,0}$ over all $\alpha_i$ and $\alpha_j$ such that $\alpha_i+\alpha_j=\beta$. We require the compressive cyclic moment estimates $\widehat{R}_{s_{\text{ncs}}}^{\alpha_i,c}(\btau)_{2,2}$ and $\widehat{R}_{s_{\text{ncs}}}^{\alpha_j,c}(\btau)_{2,0}$ due to the factors \{$1^*$ $2^*$\} and \{3 4\}, respectively.
\item \{$1^*$ 3\}\{$2^*$ 4\}: For this partition, $d=2$ (since there are two factors) and the $\sum$ term has the form $\sum \widehat{R}_{s_{\text{ncs}}}^{\alpha_i,c}(\btau)_{2,1}\widehat{R}_{s_{\text{ncs}}}^{\alpha_j,c}(\btau)_{2,1}$ over all $\alpha_i$ and $\alpha_j$ such that $\alpha_i+\alpha_j=\beta$.
\item \{$1^*$ 4\}\{$2^*$ 3\}: For this partition, $d=2$ and the $\sum$ term has the form $\sum \widehat{R}_{s_{\text{ncs}}}^{\alpha_i,c}(\btau)_{2,1}\widehat{R}_{s_{\text{ncs}}}^{\alpha_j,c}(\btau)_{2,1}$ over all $\alpha_i$ and $\alpha_j$ such that $\alpha_i+\alpha_j=\beta$.
\end{enumerate}
Recall that THOCS (and thus CTHOCS) are nonzero only for even values of $n$ and this holds for signals belonging to the model of \eqref{eqn:signal_model}, hence we have eliminated partitions containing factors with odd number of indices.

\end{enumerate}

\subsection{CTHOCS Validation}
\label{sec:rhocs_validation}

\subsubsection{Simulation}
\label{simulation}

To validate the inference quality and signal selectivity quality of CTHOCS, Monte Carlo (MC) simulations were performed for 2PSK, 4PSK, 8PSK and 16QAM signal modulation types. The following signal parameters were kept constant for all simulations: signal amplitude ($a=1$), residual carrier frequency offset ($\Delta f_c=23.0625$Hz), symbol rate ($\frac{1}{T}=12999.5625$ Hz, $6499.5625$ Hz, $3249.5625$ Hz, $1624.5625$ Hz, $799.5625$ Hz and $399.5625$ Hz for 13000, 6500, 3250, 1625, 800, and 400 processing symbols respectively), RC pulse shape with roll-off factor of $0.3$, and signal duration ($\widehat{T}=1$ second), and all of these parameters were assumed to be known. While prior knowledge of the parameters simplified the estimation of the CTHOCS features, a peak detection algorithm was necessary (and used in the simulations) since the actual number of peaks (in the lag spectrum of the signal) to be detected for a specific signal parameter configuration was not known beforehand. Specifically, each lag product spectrum was tested for the presence of a peak up to the 6th harmonic frequency (a total of 12 including the negative frequencies). The absolute value at each harmonic frequency was then compared against those of its neighboring frequencies with a peak detected if the absolute value was greater than an empirically pre-determined threshold.

Due to the various symbol rates considered, sampling frequencies $\frac{1}{T_s}=131072$~Hz, $65536$~Hz, $32768$~Hz, $16384$~Hz, $8192$~Hz, and $4096$~Hz were chosen for symbol rates $12999.5625$ Hz, $6499.5625$ Hz, $3249.5625$ Hz, $1624.5625$ Hz, $799.5625$ Hz, and $399.5625$ Hz for all simulations. For each signal type, 50 trials were performed.

\subsubsection{Compressive Cyclic Cumulants Inference Quality}
\label{sec:chocs_recon_quality}

\begin{figure*}[tp]
\vspace{-5ex}
\begin{minipage}[b]{0.31\textwidth}
\includegraphics[width=1\textwidth]{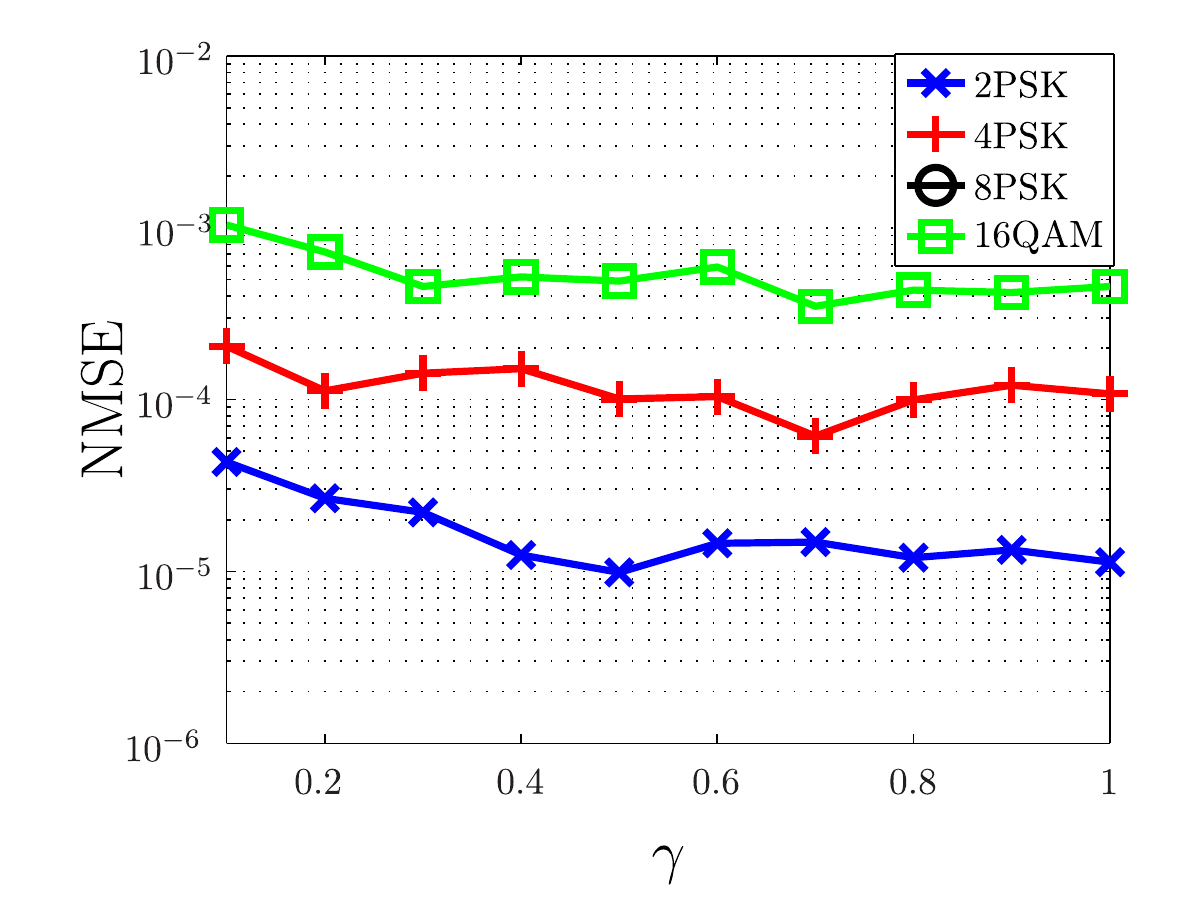}
\subcaption{\small\sl 13000 Symbols: $n=4$, $q=0$}\label{fig:compare_nmse_c4_0_500_13000}
\end{minipage}
\begin{minipage}[b]{0.31\textwidth}
\includegraphics[width=1\textwidth]{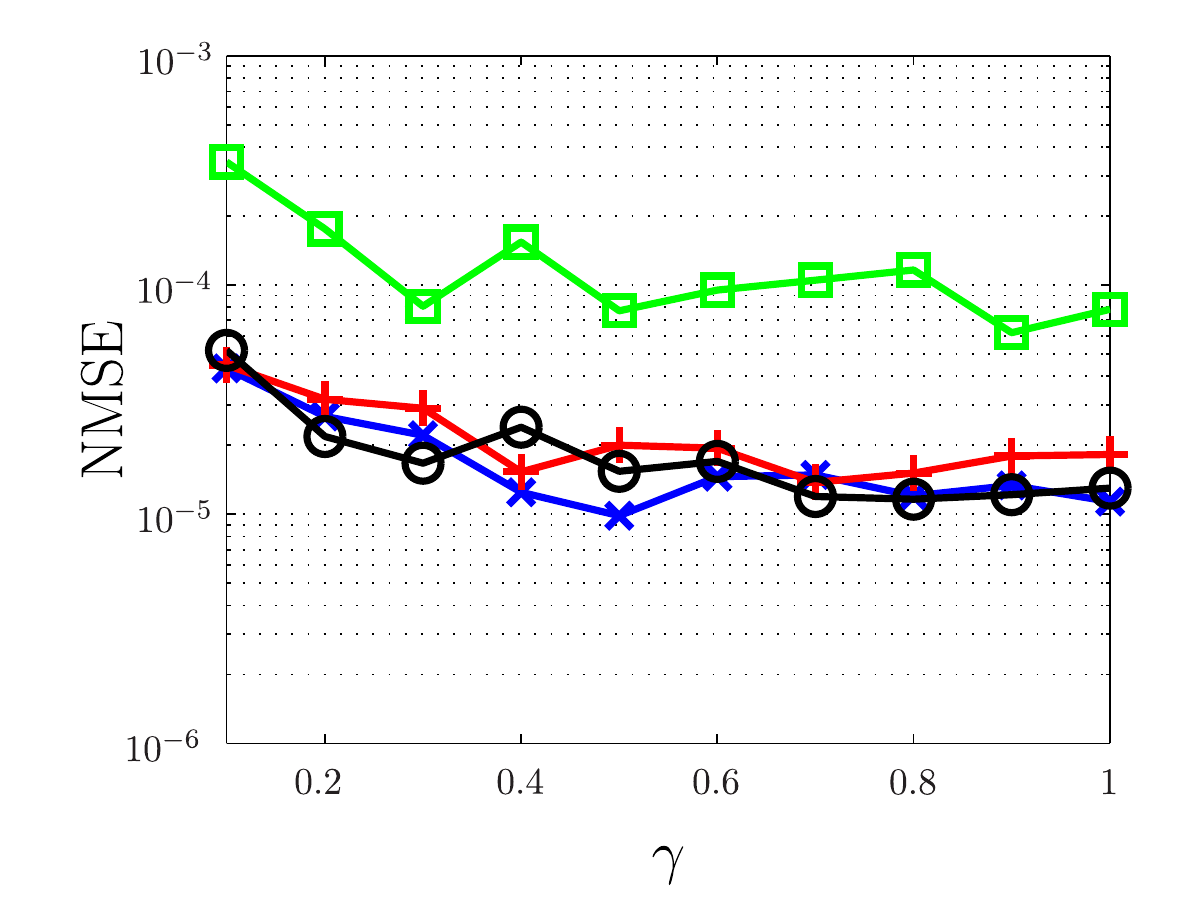}
\subcaption{\small\sl 13000 Symbols: $n=4$, $q=2$}\label{fig:compare_nmse_c4_2_500_13000}
\end{minipage}
\begin{minipage}[b]{0.31\textwidth}
\includegraphics[width=1\textwidth]{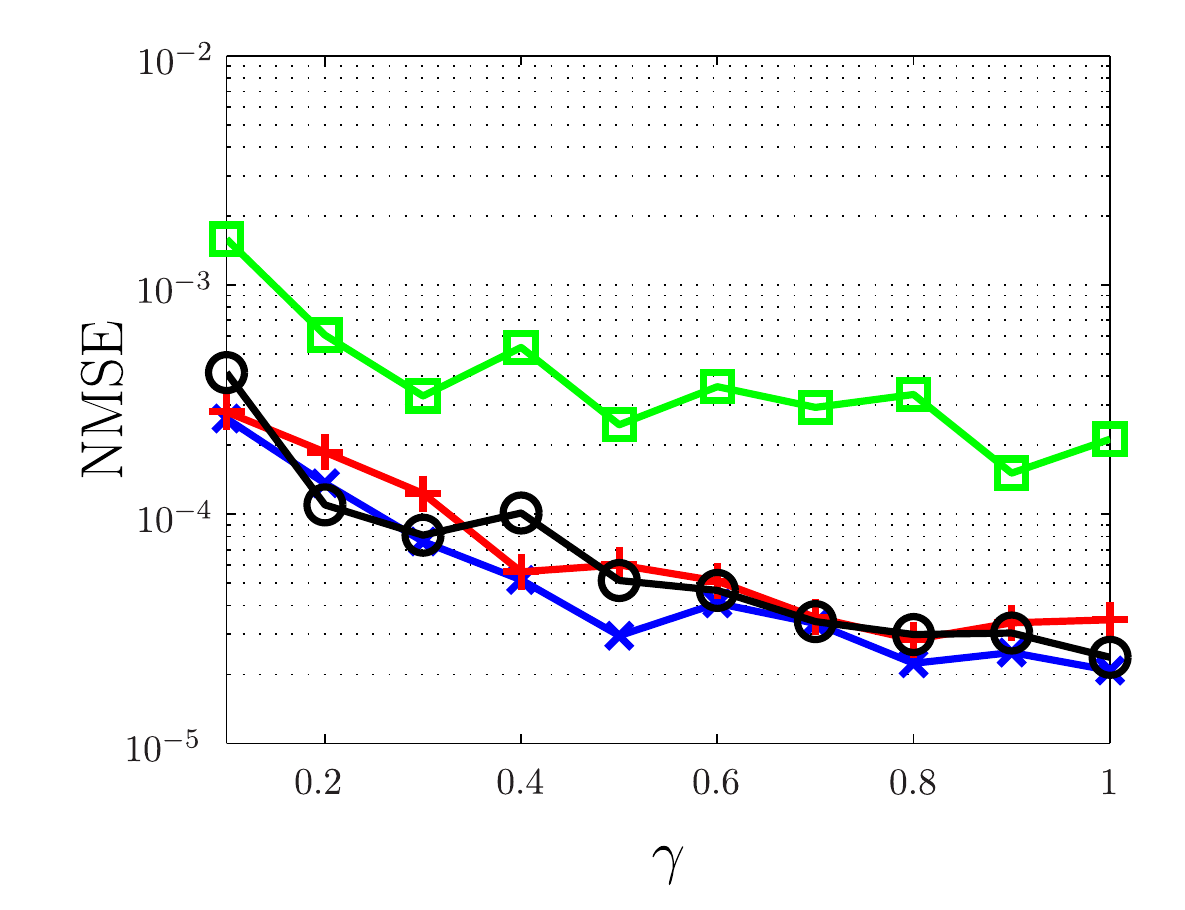}
\subcaption{\small\sl 13000 Symbols: $n=6$, $q=3$}\label{fig:compare_nmse_c6_3_500_13000}
\end{minipage}\\
\begin{minipage}[b]{0.31\textwidth}
\includegraphics[width=1\textwidth]{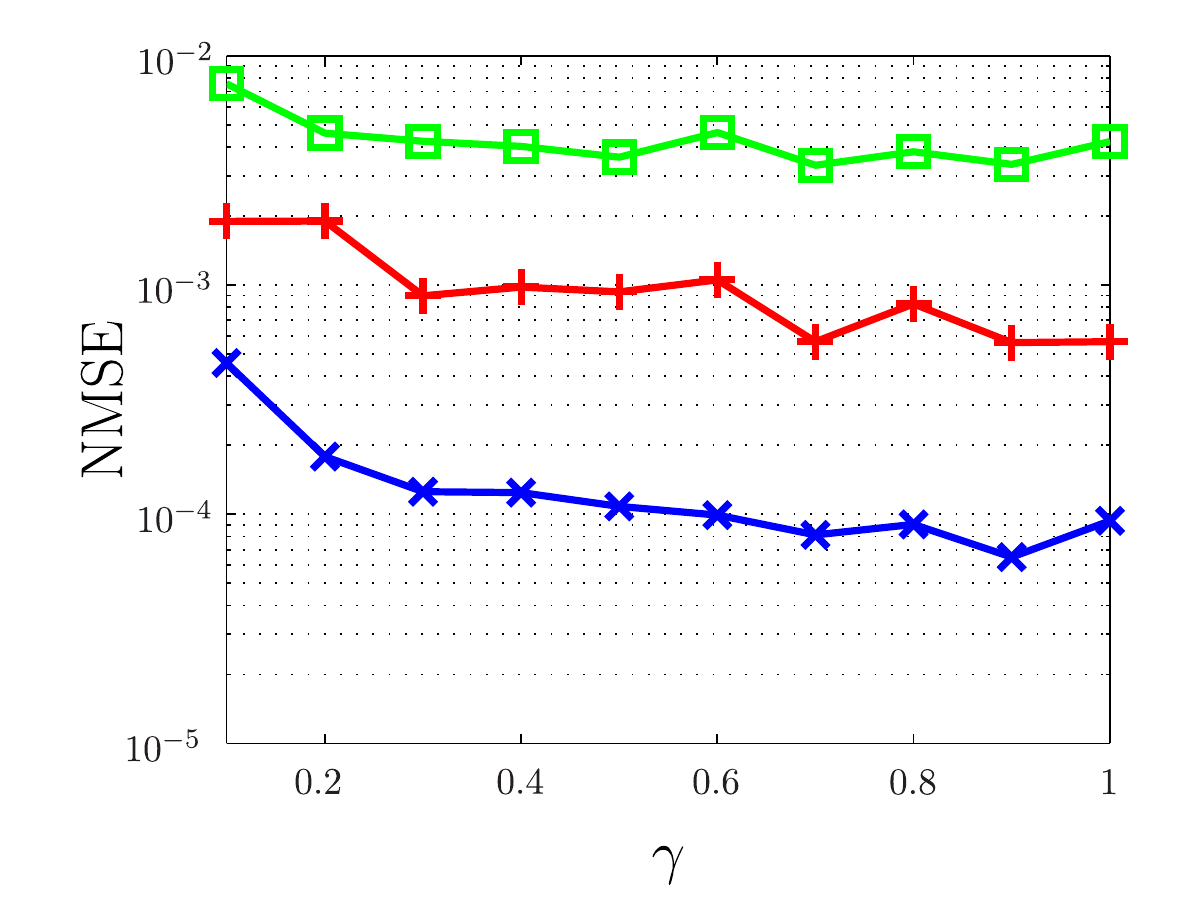}
\subcaption{\small\sl 1625 Symbols: $n=4$, $q=0$}\label{fig:compare_nmse_c4_0_500_1625}
\end{minipage}
\begin{minipage}[b]{0.31\textwidth}
\includegraphics[width=1\textwidth]{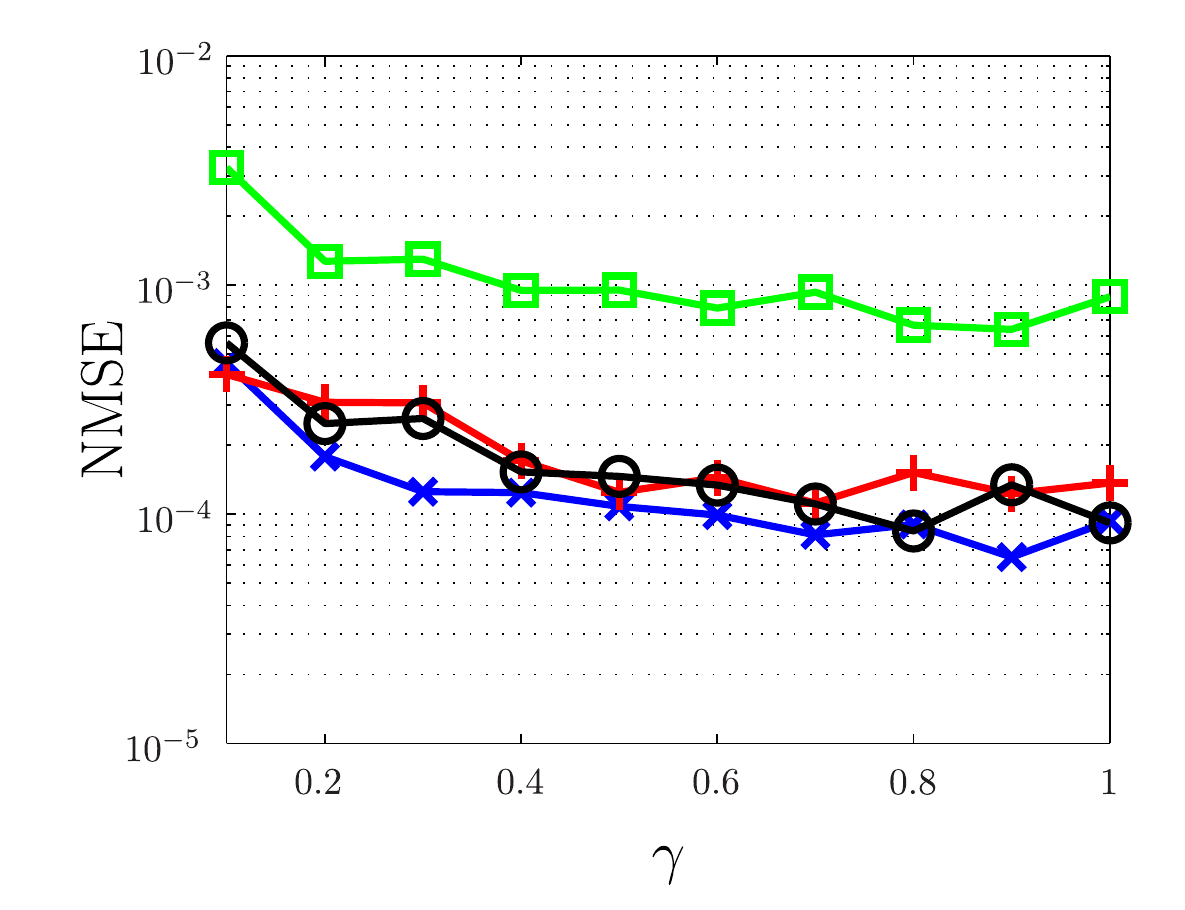}
\subcaption{\small\sl 1625 Symbols: $n=4$, $q=2$}\label{fig:compare_nmse_c4_2_500_1625}
\end{minipage}
\begin{minipage}[b]{0.31\textwidth}
\includegraphics[width=1\textwidth]{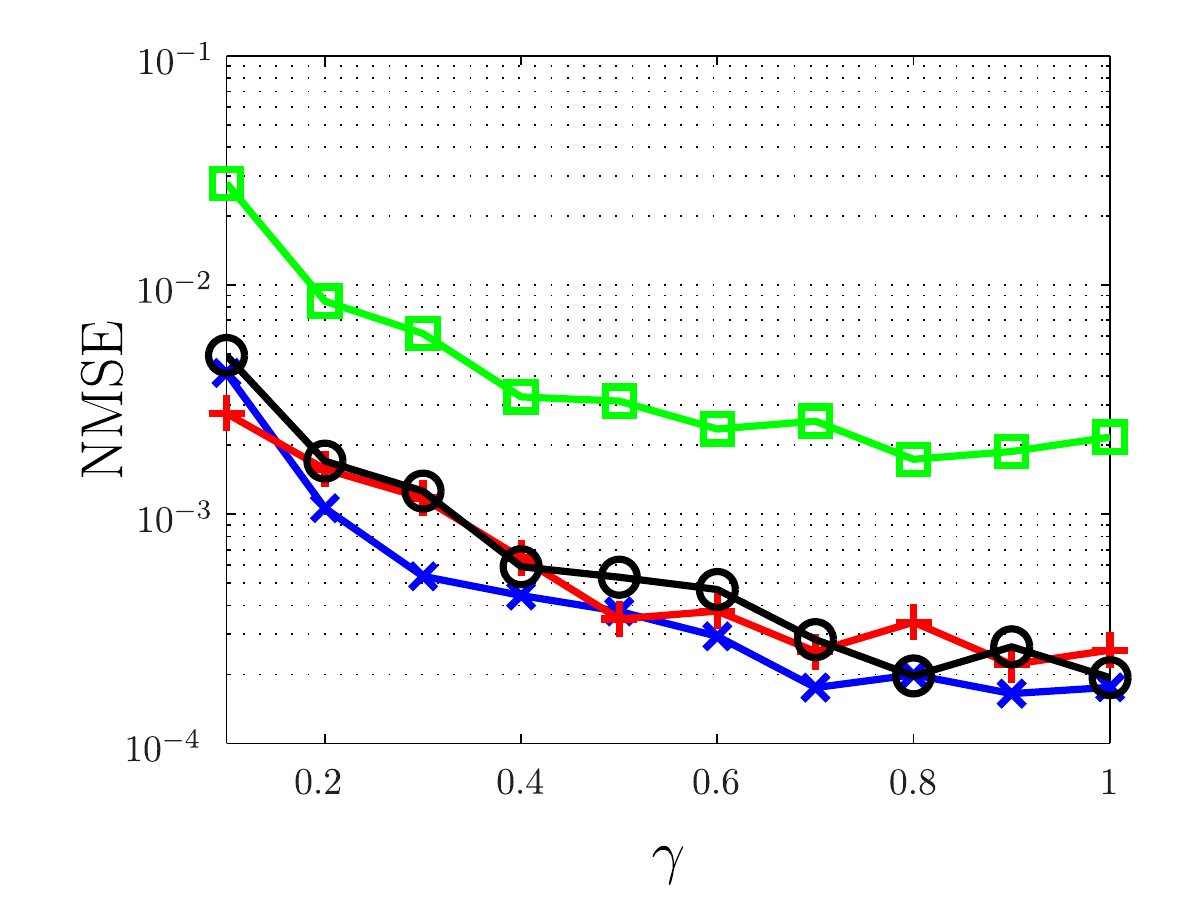}
\subcaption{\small\sl 1625 Symbols: $n=6$, $q=3$}\label{fig:compare_nmse_c6_3_500_1625}
\end{minipage}\\
\begin{minipage}[b]{0.31\textwidth}
\includegraphics[width=1\textwidth]{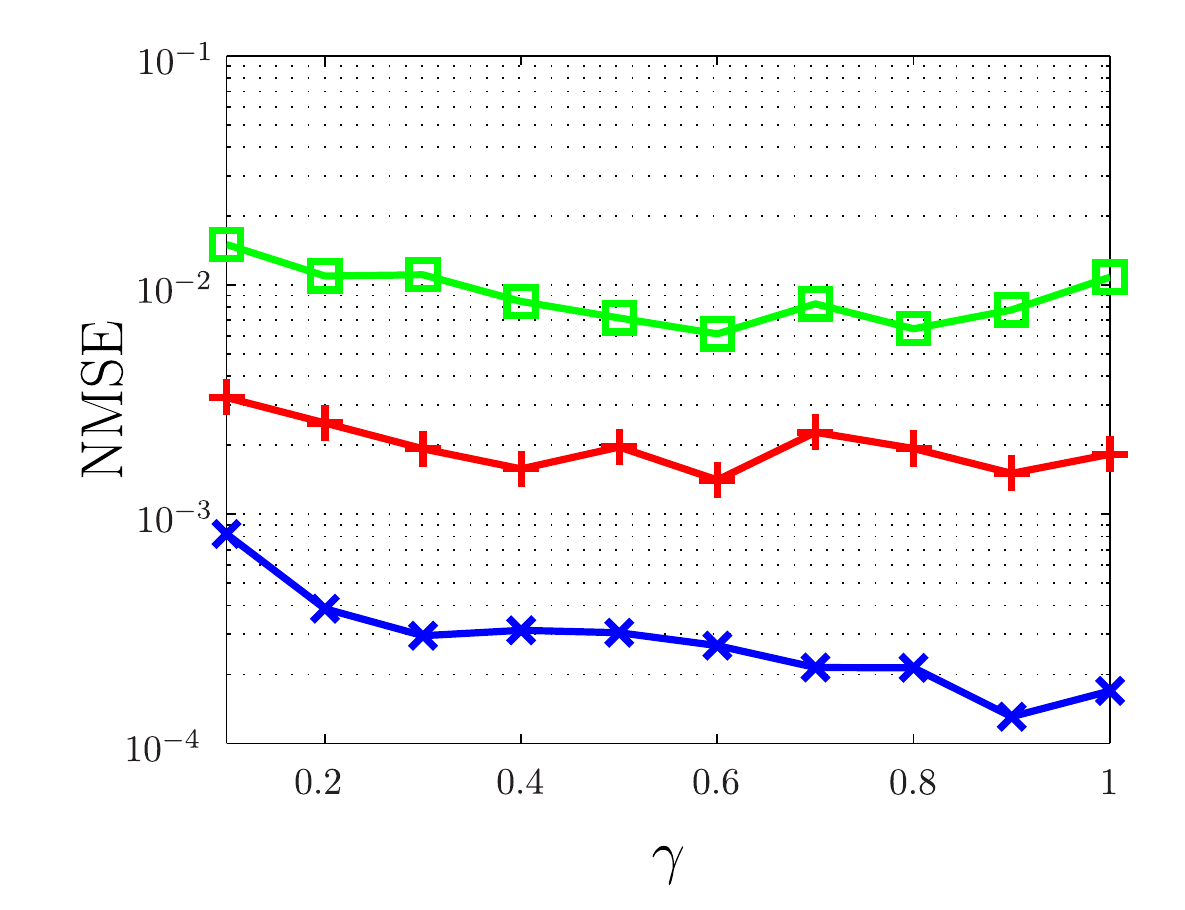}
\subcaption{\small\sl 800 Symbols: $n=4$, $q=0$}\label{fig:compare_nmse_c4_0_500_800}
\end{minipage}
\begin{minipage}[b]{0.31\textwidth}
\includegraphics[width=1\textwidth]{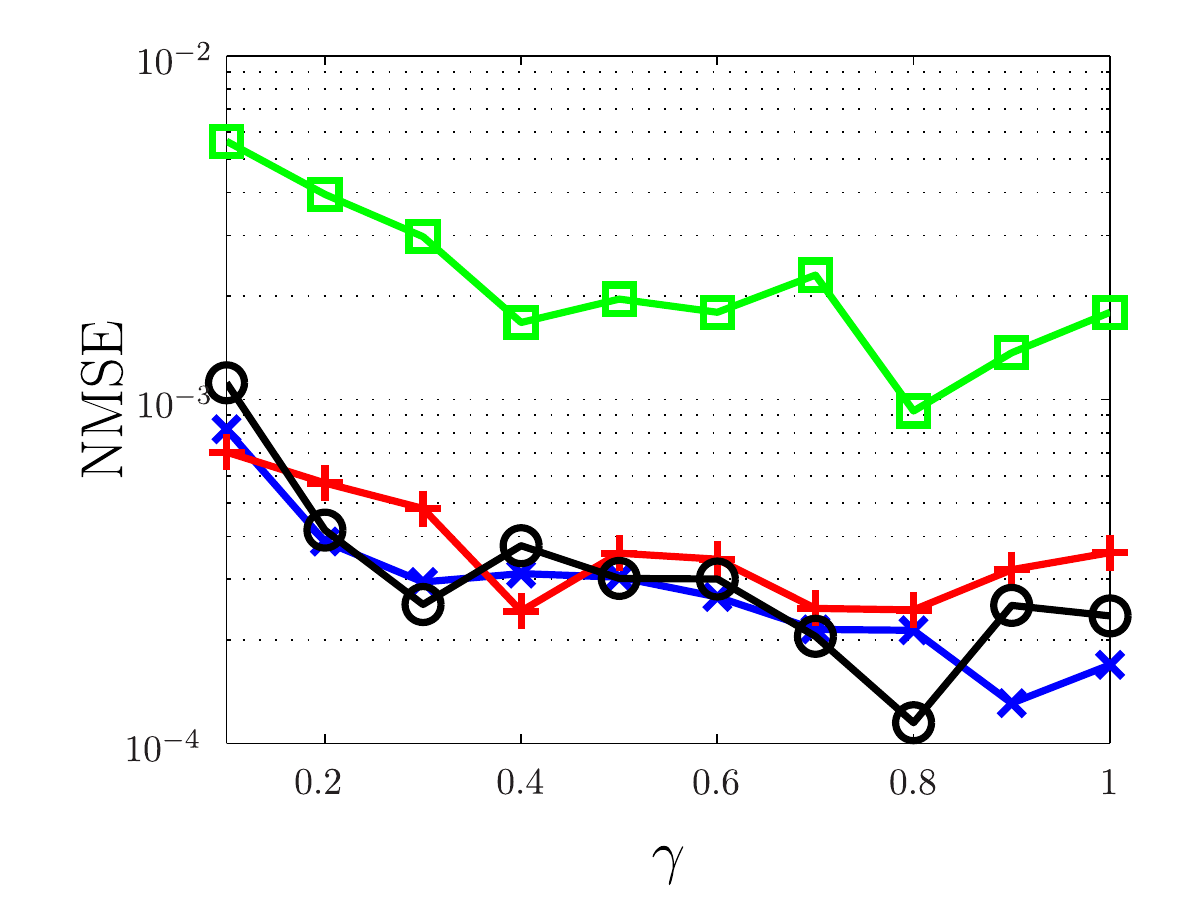}
\subcaption{\small\sl 800 Symbols: $n=4$, $q=2$}\label{fig:compare_nmse_c4_2_500_800}
\end{minipage}
\begin{minipage}[b]{0.31\textwidth}
\includegraphics[width=1\textwidth]{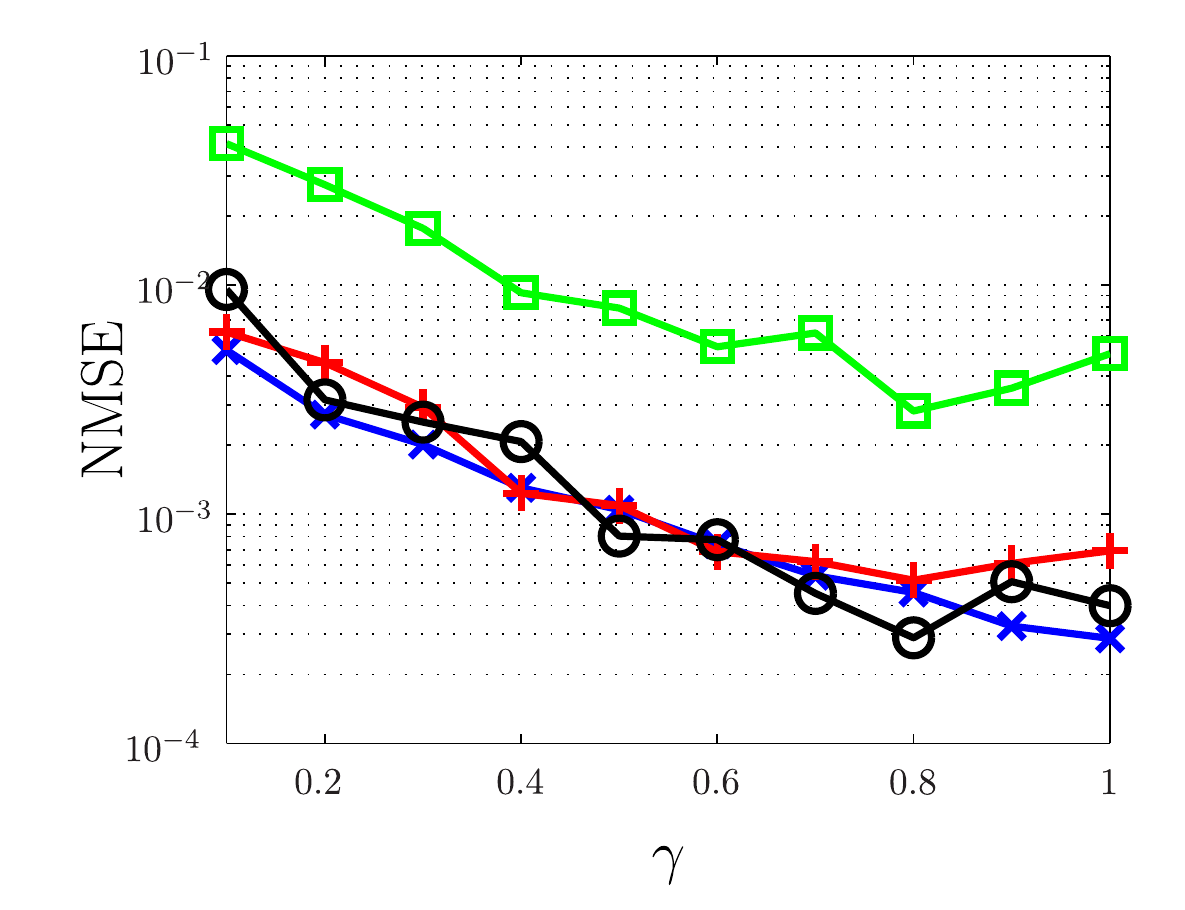}
\subcaption{\small\sl 800 Symbols: $n=6$, $q=3$}\label{fig:compare_nmse_c6_3_500_800}
\end{minipage}\\
\begin{minipage}[b]{0.31\textwidth}
\includegraphics[width=1\textwidth]{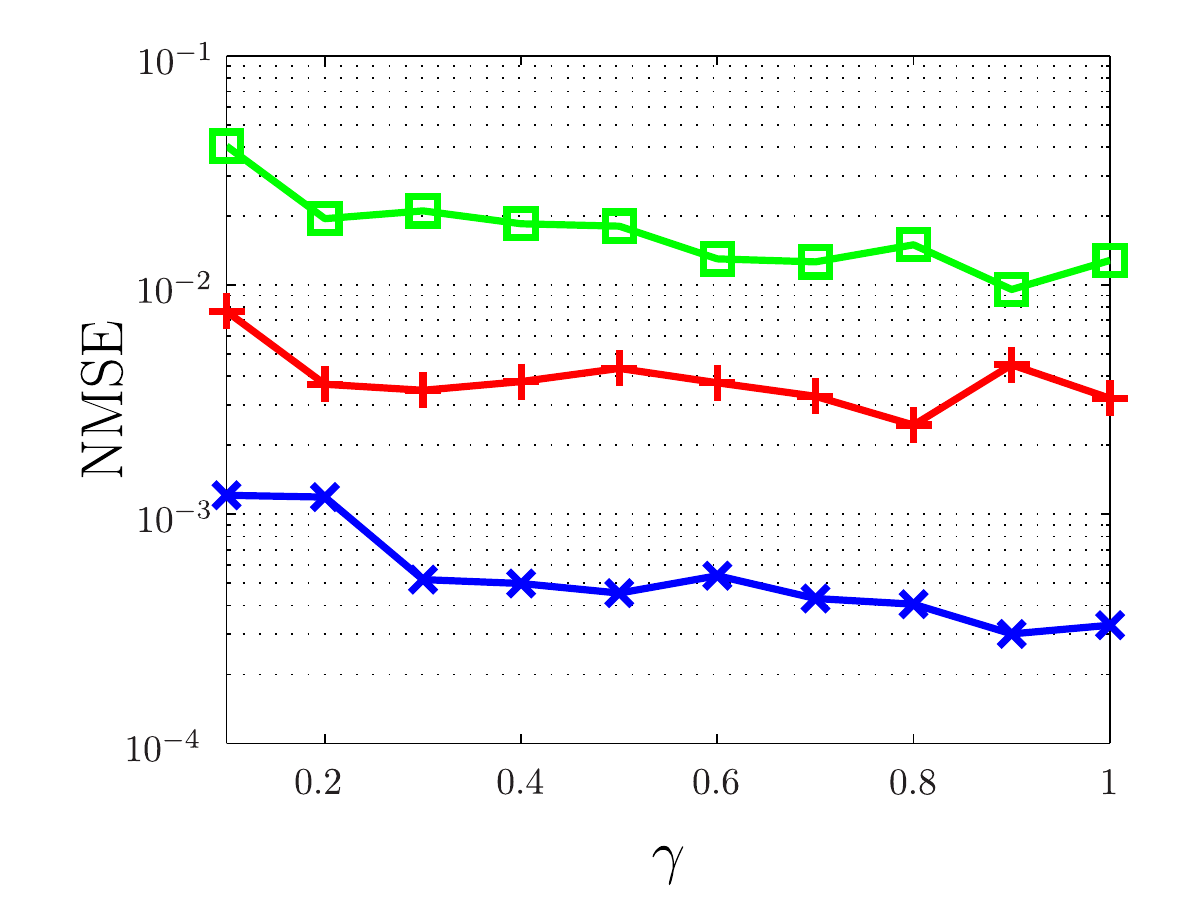}
\subcaption{\small\sl 400 Symbols: $n=4$, $q=0$}\label{fig:compare_nmse_c4_0_500_400}
\end{minipage}
\begin{minipage}[b]{0.31\textwidth}
\includegraphics[width=1\textwidth]{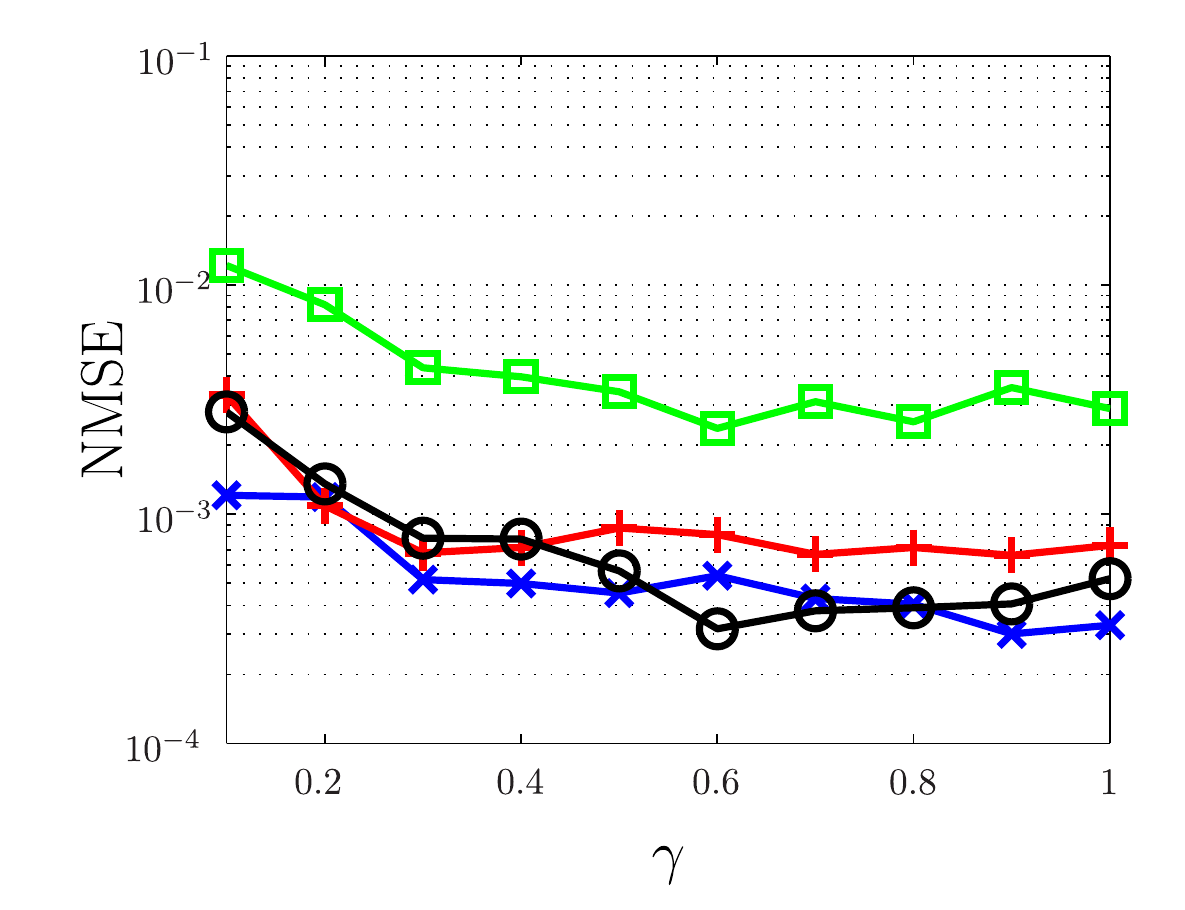}
\subcaption{\small\sl 400 Symbols: $n=4$, $q=2$}\label{fig:compare_nmse_c4_2_500_400}
\end{minipage}
\begin{minipage}[b]{0.31\textwidth}
\includegraphics[width=1\textwidth]{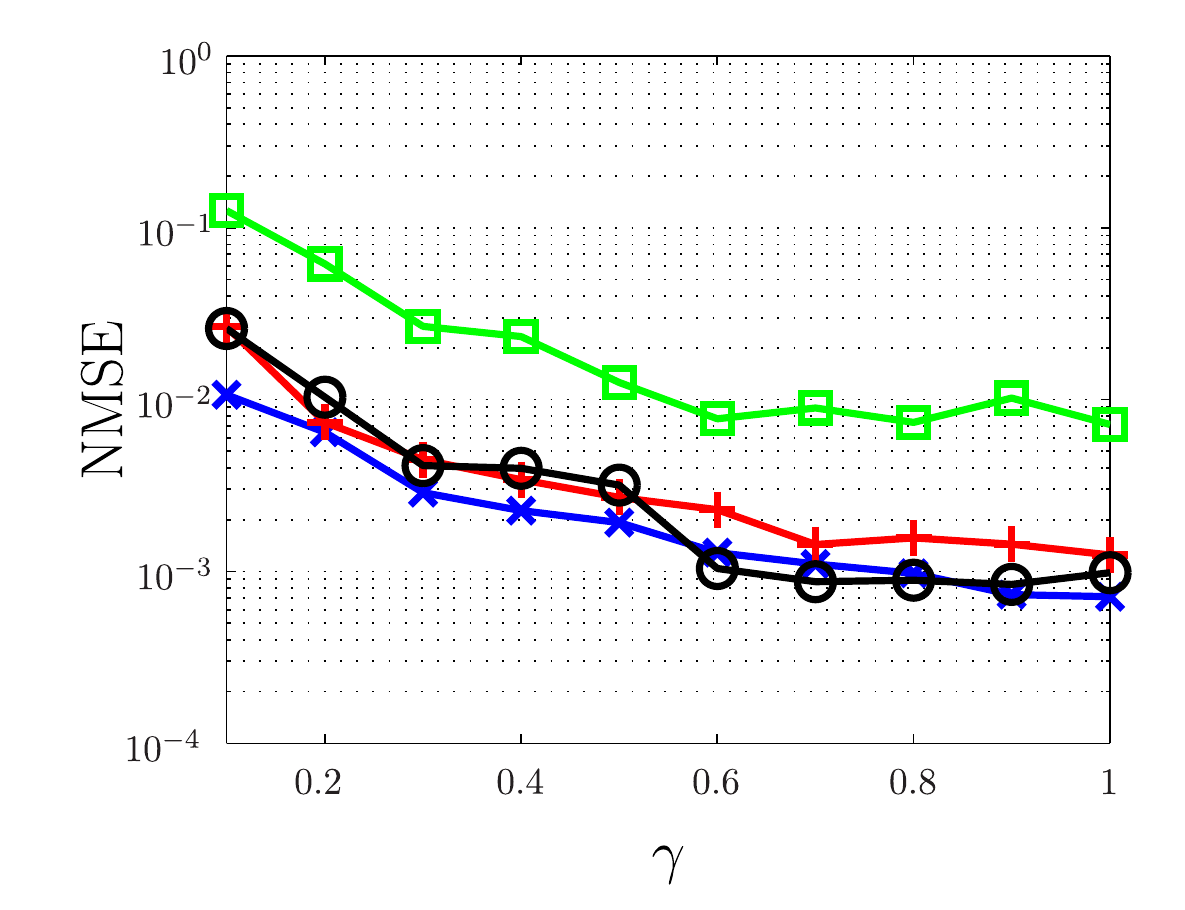}
\subcaption{\small\sl 400 Symbols: $n=6$, $q=3$}\label{fig:compare_nmse_c6_3_500_400}
\end{minipage}\\
\caption{\small\sl NMSE of compressive cyclic cumulants against cyclic cumulants. Across each row, the plots show NMSE of compressive cyclic cumulants versus cyclic cumulants for select values of $n$ and $q$ as a function of $\gamma$ for select processed data length under noiseless conditions.}\label{fig:compare_chocs_mse_noiseless}
\end{figure*}

\begin{figure*}[tp]
\vspace{-5ex}
\begin{minipage}[b]{0.31\textwidth}
\includegraphics[width=1\textwidth]{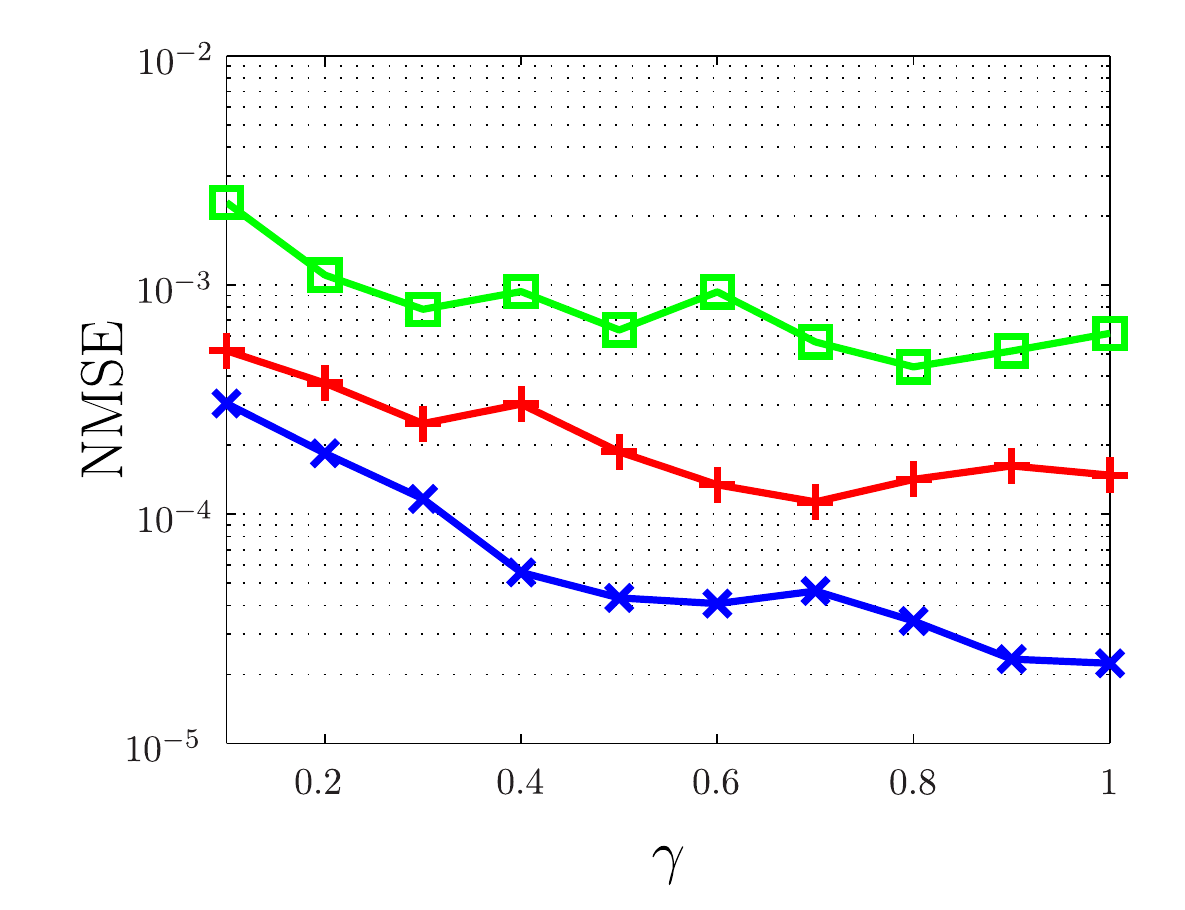}
\subcaption{\small\sl 13000 Symbols: $n=4$, $q=0$}\label{fig:compare_nmse_c4_0_9_13000}
\end{minipage}
\begin{minipage}[b]{0.31\textwidth}
\includegraphics[width=1\textwidth]{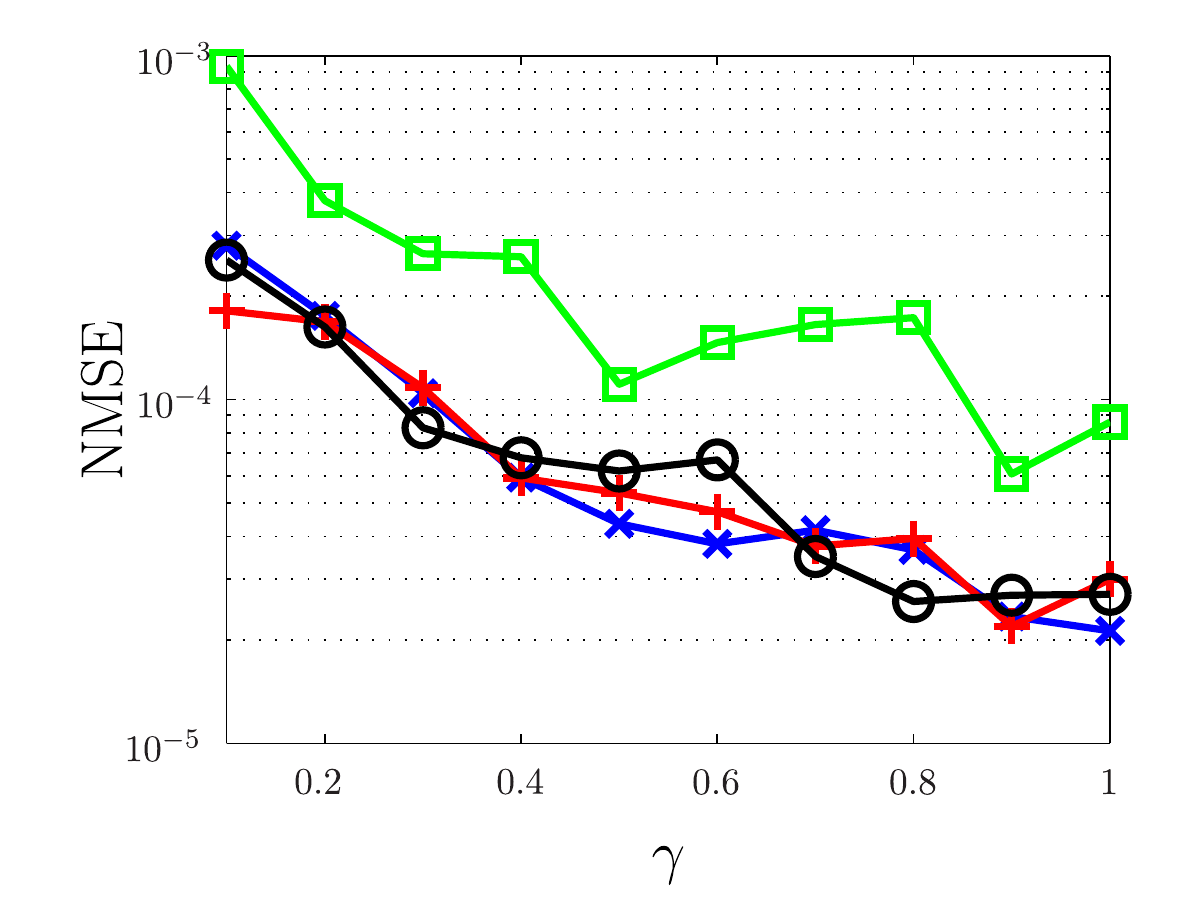}
\subcaption{\small\sl 13000 Symbols: $n=4$, $q=2$}\label{fig:compare_nmse_c4_2_9_13000}
\end{minipage}
\begin{minipage}[b]{0.31\textwidth}
\includegraphics[width=1\textwidth]{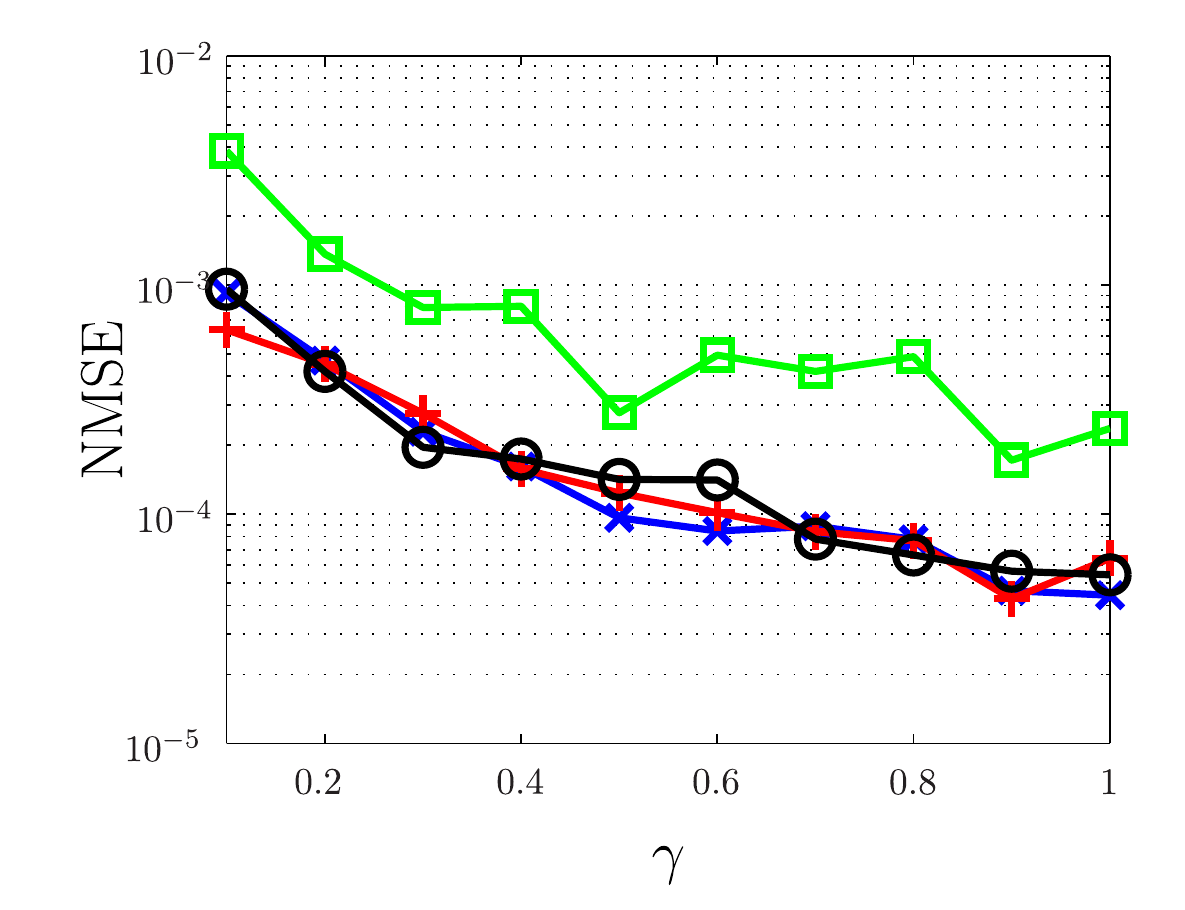}
\subcaption{\small\sl 13000 Symbols: $n=6$, $q=3$}\label{fig:compare_nmse_c6_3_9_13000}
\end{minipage}\\
\begin{minipage}[b]{0.31\textwidth}
\includegraphics[width=1\textwidth]{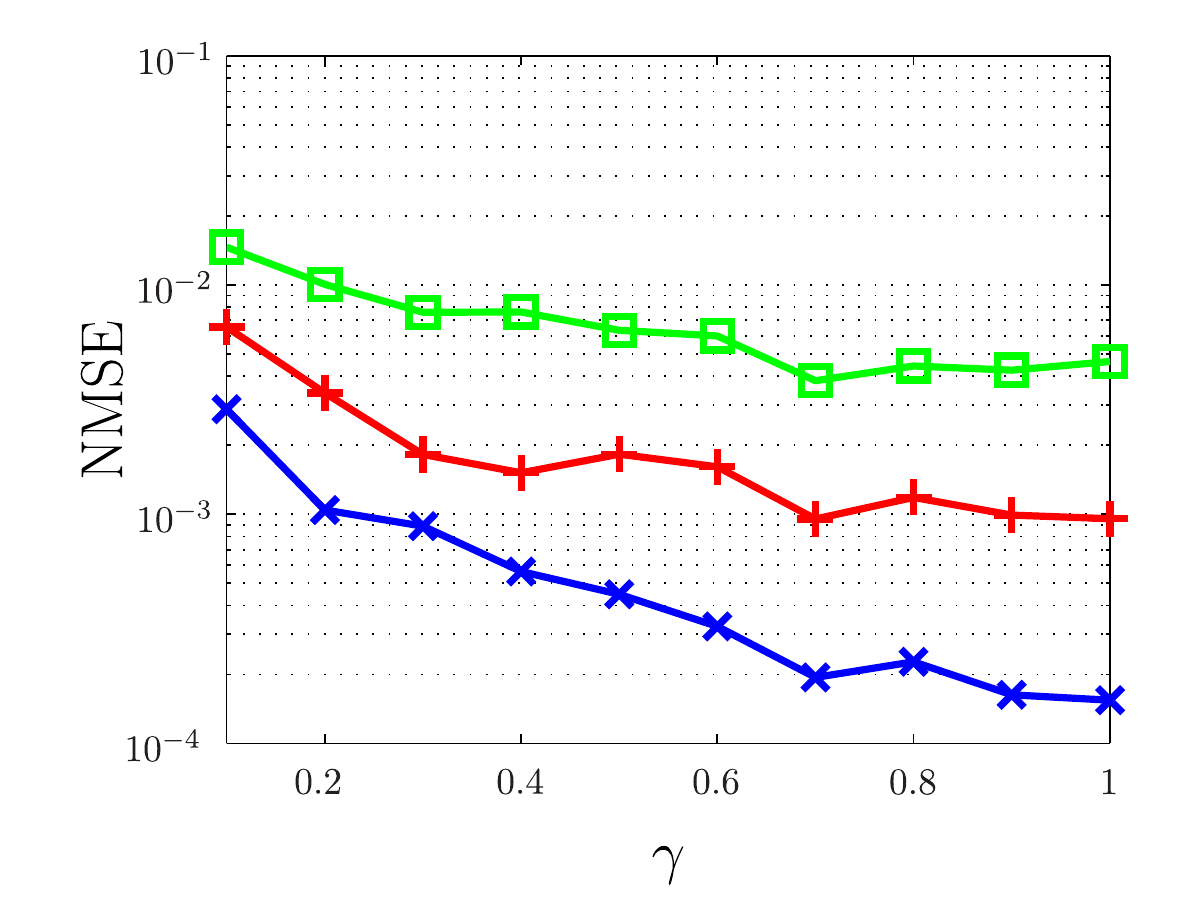}
\subcaption{\small\sl 1625 Symbols: $n=4$, $q=0$}\label{fig:compare_nmse_c4_0_9_1625}
\end{minipage}
\begin{minipage}[b]{0.31\textwidth}
\includegraphics[width=1\textwidth]{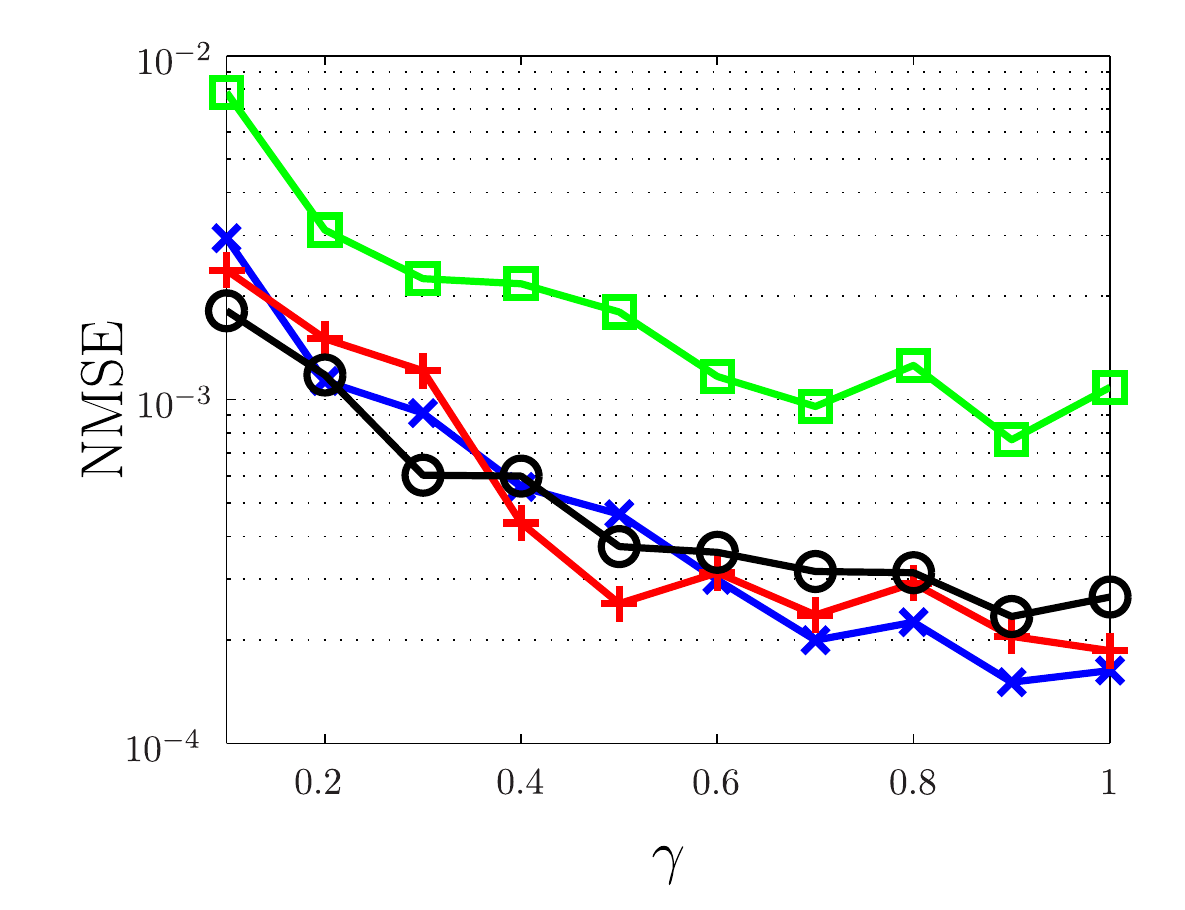}
\subcaption{\small\sl 1625 Symbols: $n=4$, $q=2$}\label{fig:compare_nmse_c4_2_9_1625}
\end{minipage}
\begin{minipage}[b]{0.31\textwidth}
\includegraphics[width=1\textwidth]{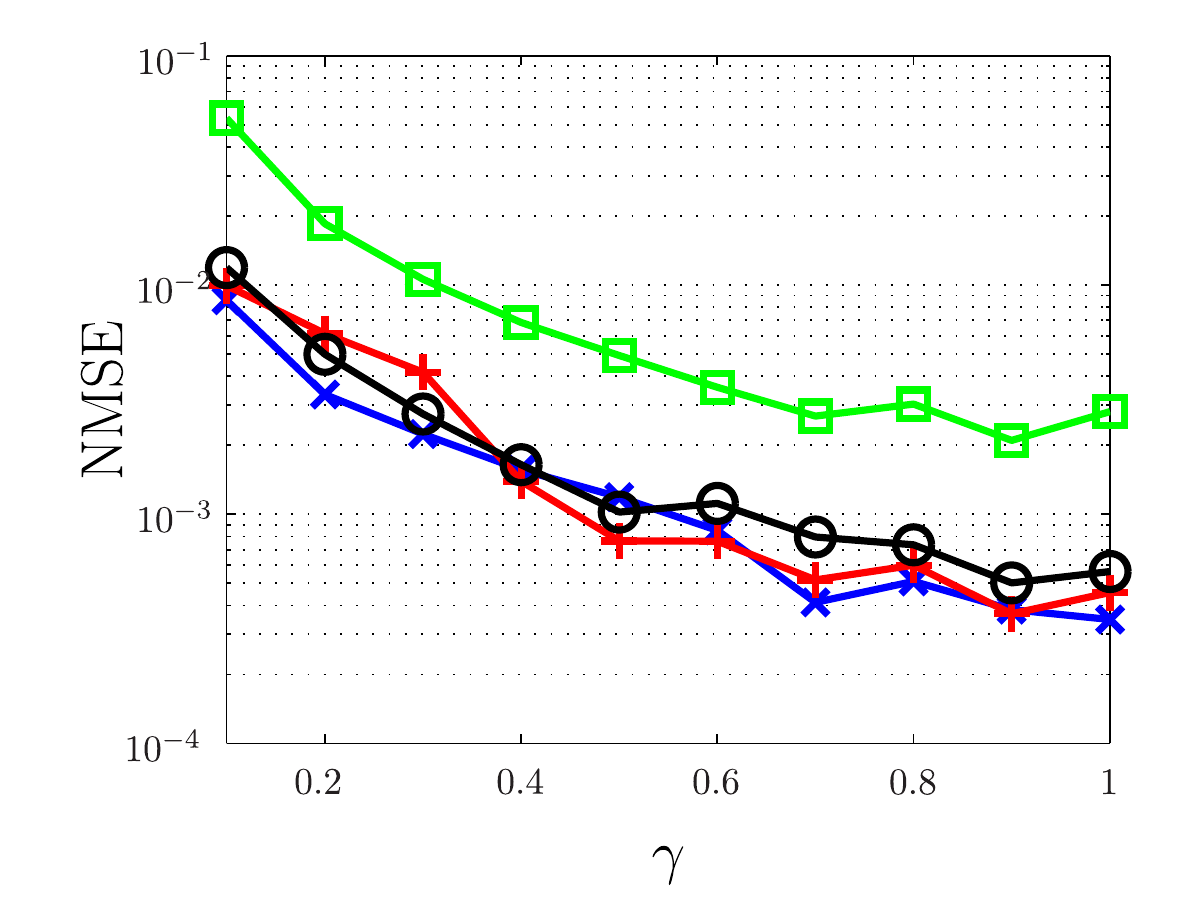}
\subcaption{\small\sl 1625 Symbols: $n=6$, $q=3$}\label{fig:compare_nmse_c6_3_9_1625}
\end{minipage}\\
\begin{minipage}[b]{0.31\textwidth}
\includegraphics[width=1\textwidth]{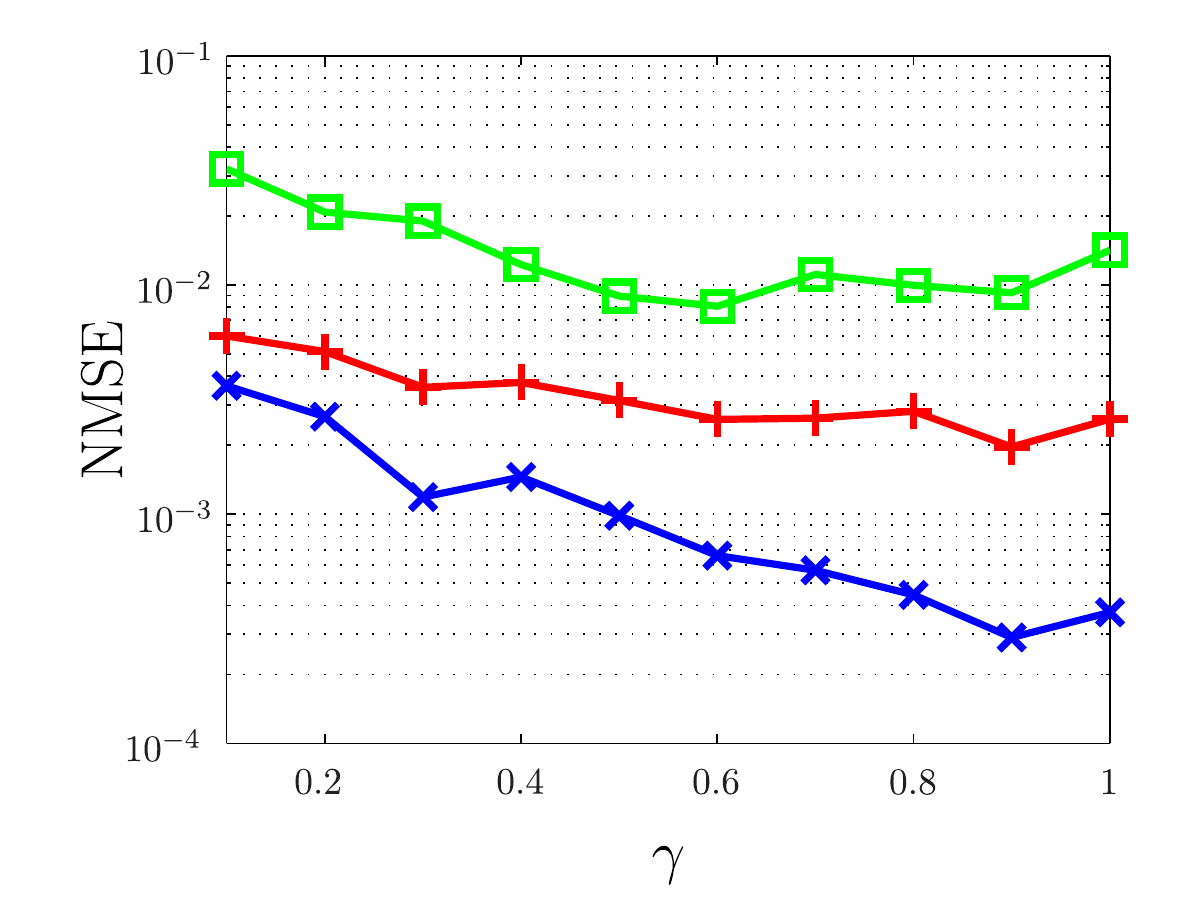}
\subcaption{\small\sl 800 Symbols: $n=4$, $q=0$}\label{fig:compare_nmse_c4_0_9_800}
\end{minipage}
\begin{minipage}[b]{0.31\textwidth}
\includegraphics[width=1\textwidth]{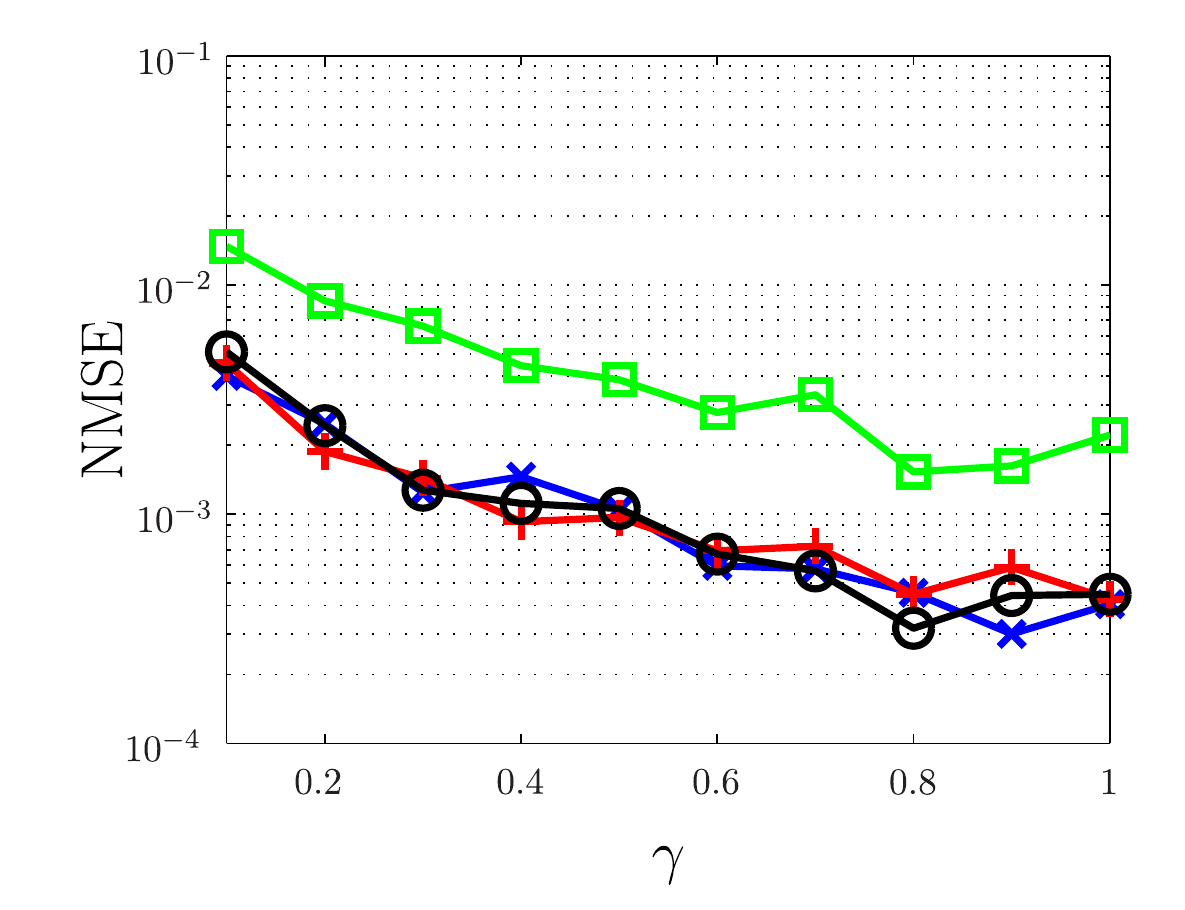}
\subcaption{\small\sl 800 Symbols: $n=4$, $q=2$}\label{fig:compare_nmse_c4_2_9_800}
\end{minipage}
\begin{minipage}[b]{0.31\textwidth}
\includegraphics[width=1\textwidth]{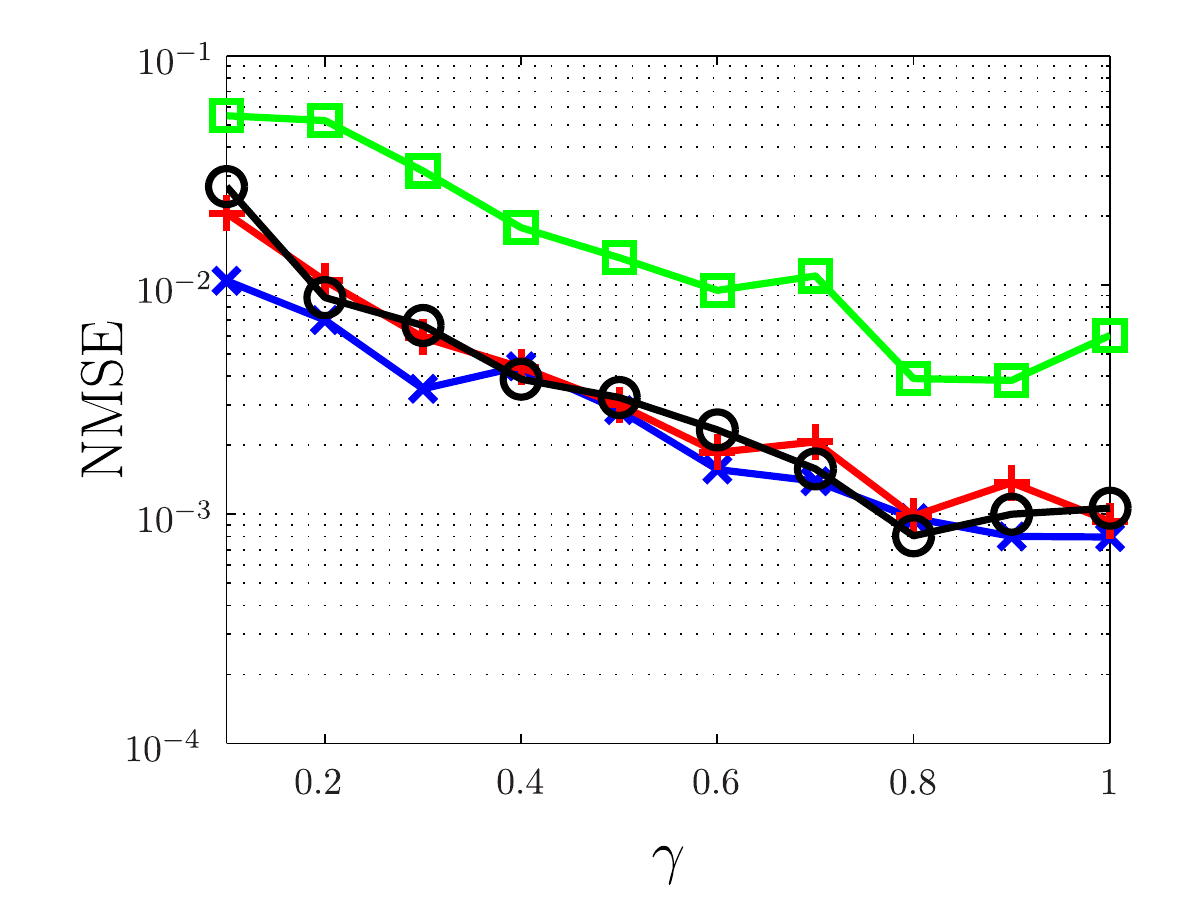}
\subcaption{\small\sl 800 Symbols: $n=6$, $q=3$}\label{fig:compare_nmse_c6_3_9_800}
\end{minipage}\\
\begin{minipage}[b]{0.31\textwidth}
\includegraphics[width=1\textwidth]{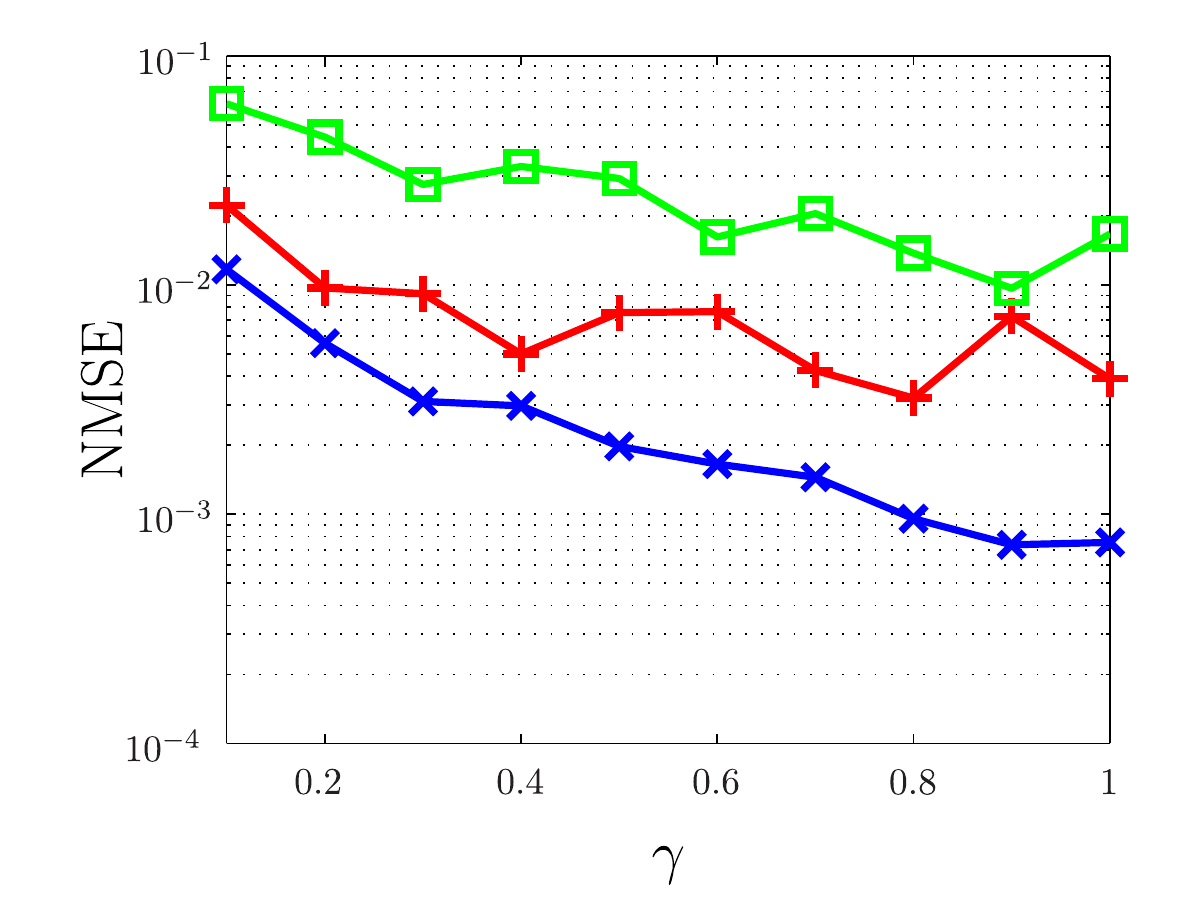}
\subcaption{\small\sl 400 Symbols: $n=4$, $q=0$}\label{fig:compare_nmse_c4_0_9_400}
\end{minipage}
\begin{minipage}[b]{0.31\textwidth}
\includegraphics[width=1\textwidth]{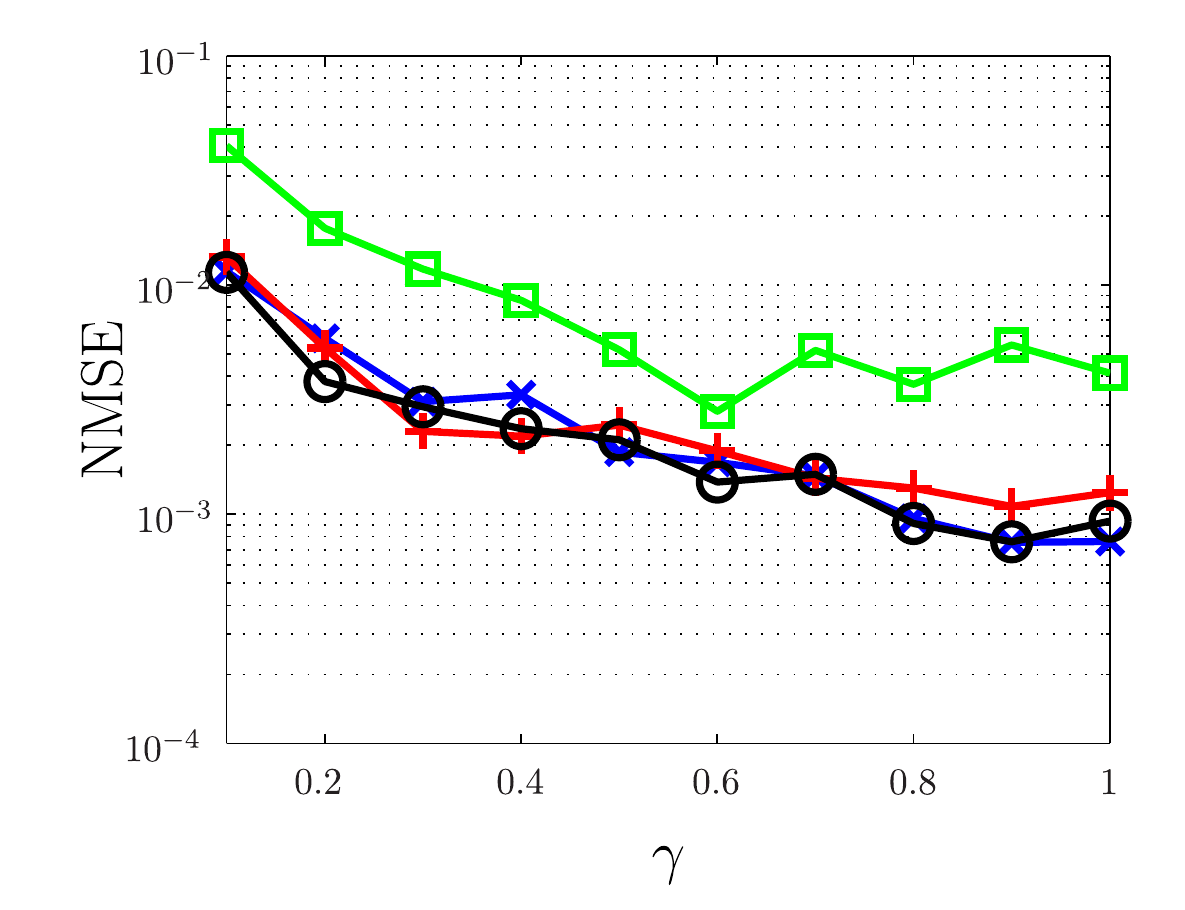}
\subcaption{\small\sl 400 Symbols: $n=4$, $q=2$}\label{fig:compare_nmse_c4_2_9_400}
\end{minipage}
\begin{minipage}[b]{0.31\textwidth}
\includegraphics[width=1\textwidth]{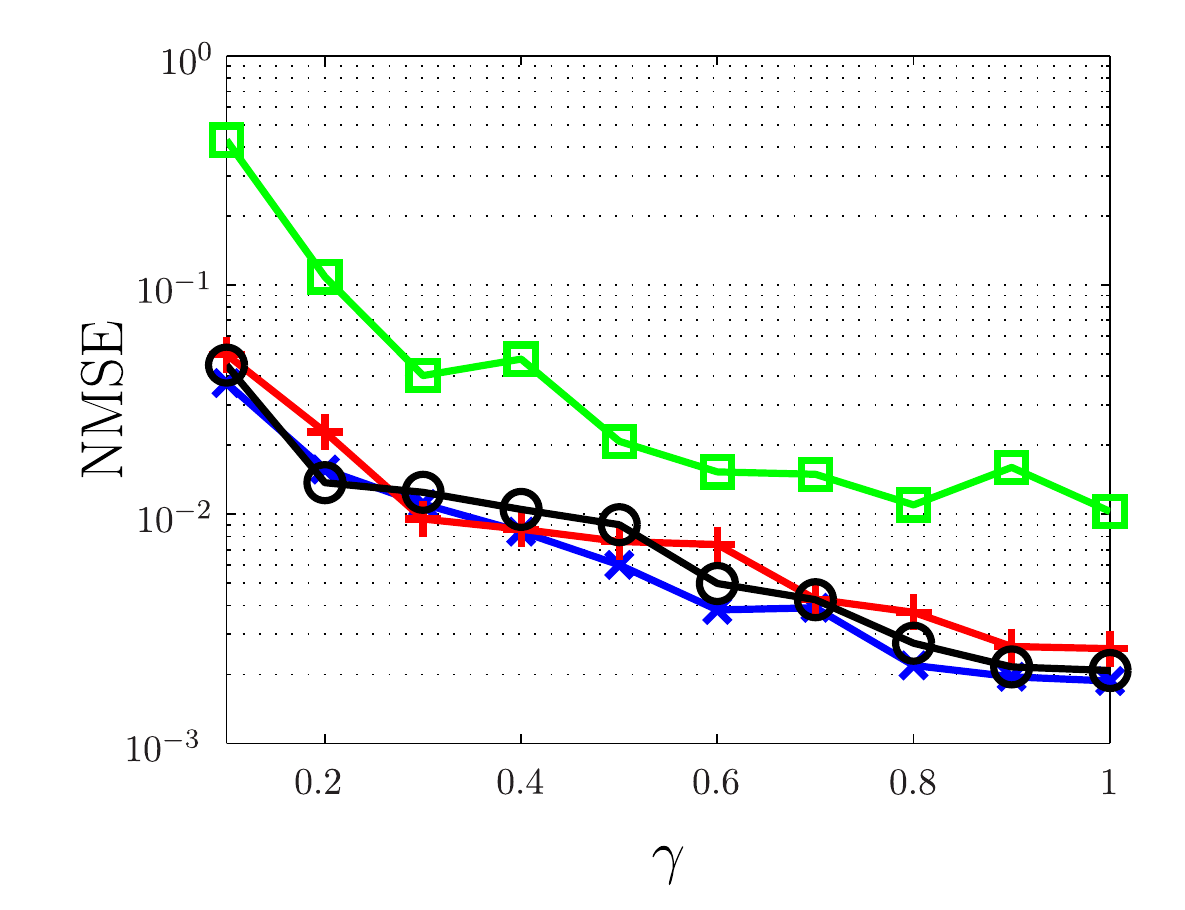}
\subcaption{\small\sl 400 Symbols: $n=6$, $q=3$}\label{fig:compare_nmse_c6_3_9_400}
\end{minipage}\\
\caption{\small\sl NMSE of compressive cyclic cumulants against cyclic cumulants. Across each row, the plots show NMSE of compressive cyclic cumulants versus cyclic cumulants for select values of $n$ and $q$ as a function of $\gamma$ for select processed data length when CNR=9dB.}\label{fig:compare_chocs_mse_9dB}
\end{figure*}

\begin{figure*}[tp]
\vspace{-5ex}
\begin{minipage}[b]{0.31\textwidth}
\includegraphics[width=1\textwidth]{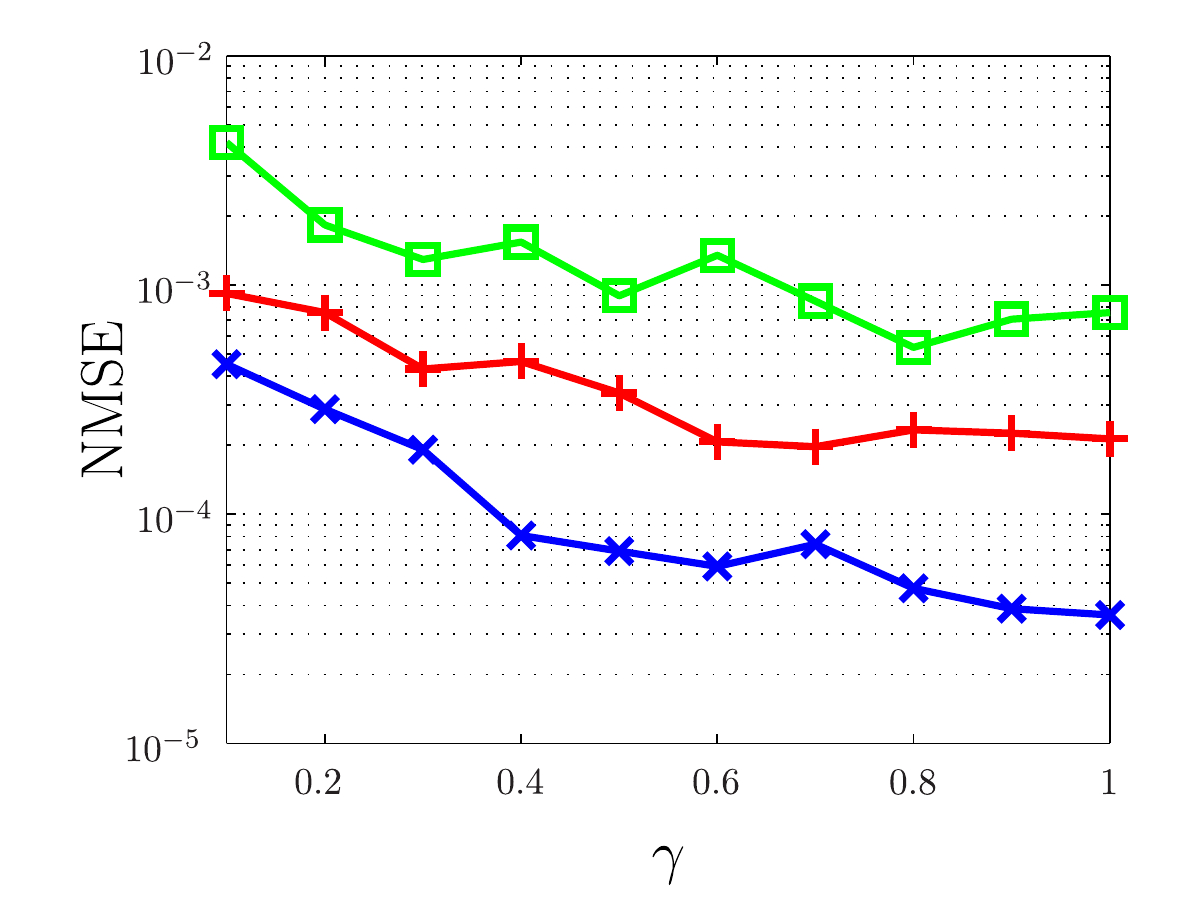}
\subcaption{\small\sl 13000 Symbols: $n=4$, $q=0$}\label{fig:compare_nmse_c4_0_6_13000}
\end{minipage}
\begin{minipage}[b]{0.31\textwidth}
\includegraphics[width=1\textwidth]{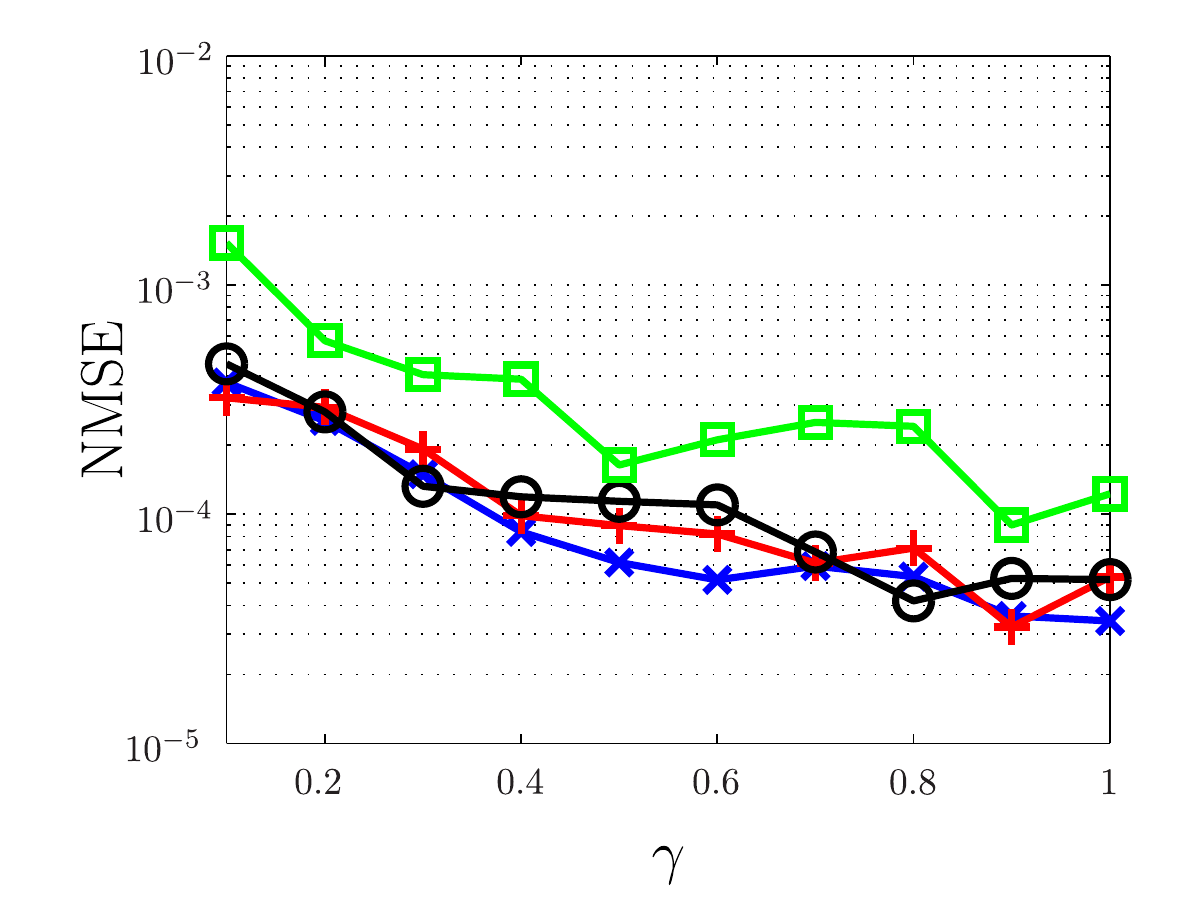}
\subcaption{\small\sl 13000 Symbols: $n=4$, $q=2$}\label{fig:compare_nmse_c4_2_6_13000}
\end{minipage}
\begin{minipage}[b]{0.31\textwidth}
\includegraphics[width=1\textwidth]{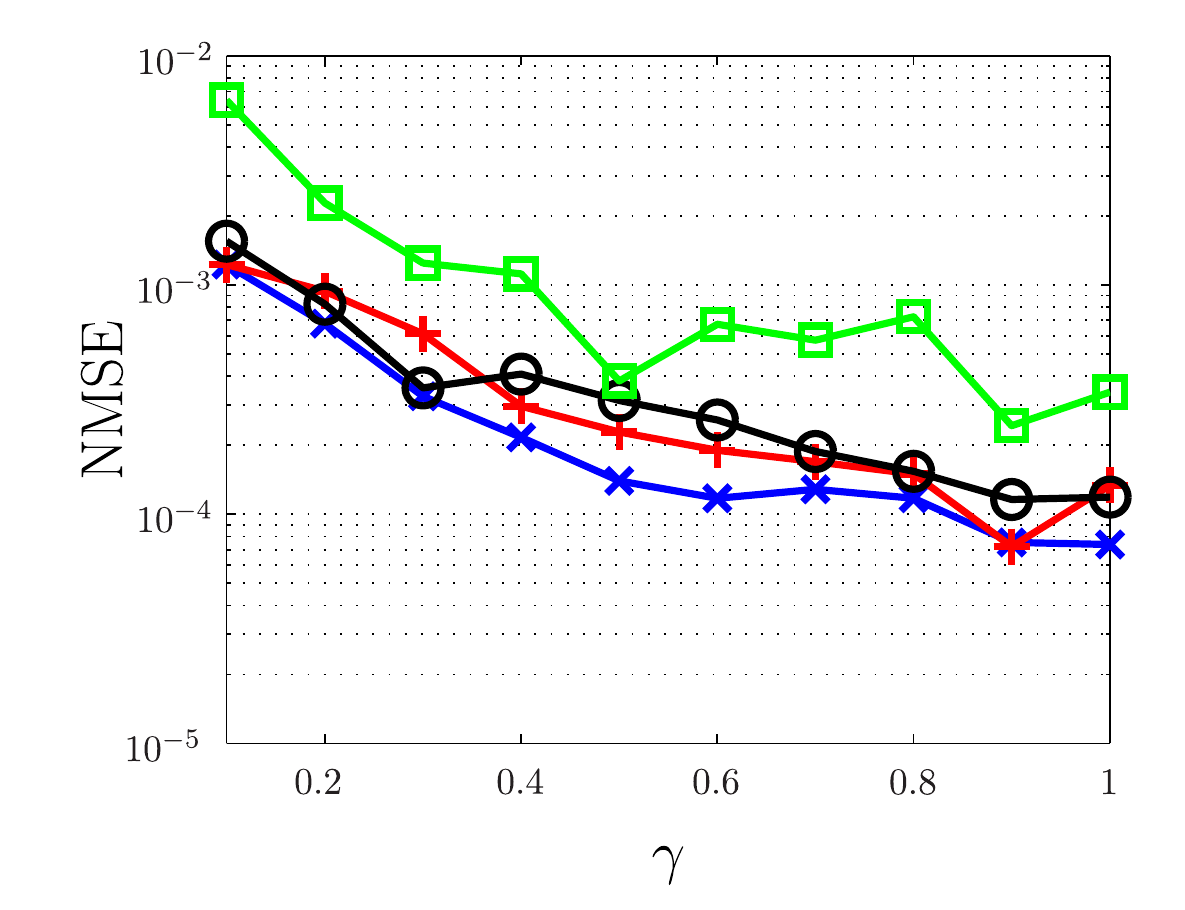}
\subcaption{\small\sl 13000 Symbols: $n=6$, $q=3$}\label{fig:compare_nmse_c6_3_6_13000}
\end{minipage}\\
\begin{minipage}[b]{0.31\textwidth}
\includegraphics[width=1\textwidth]{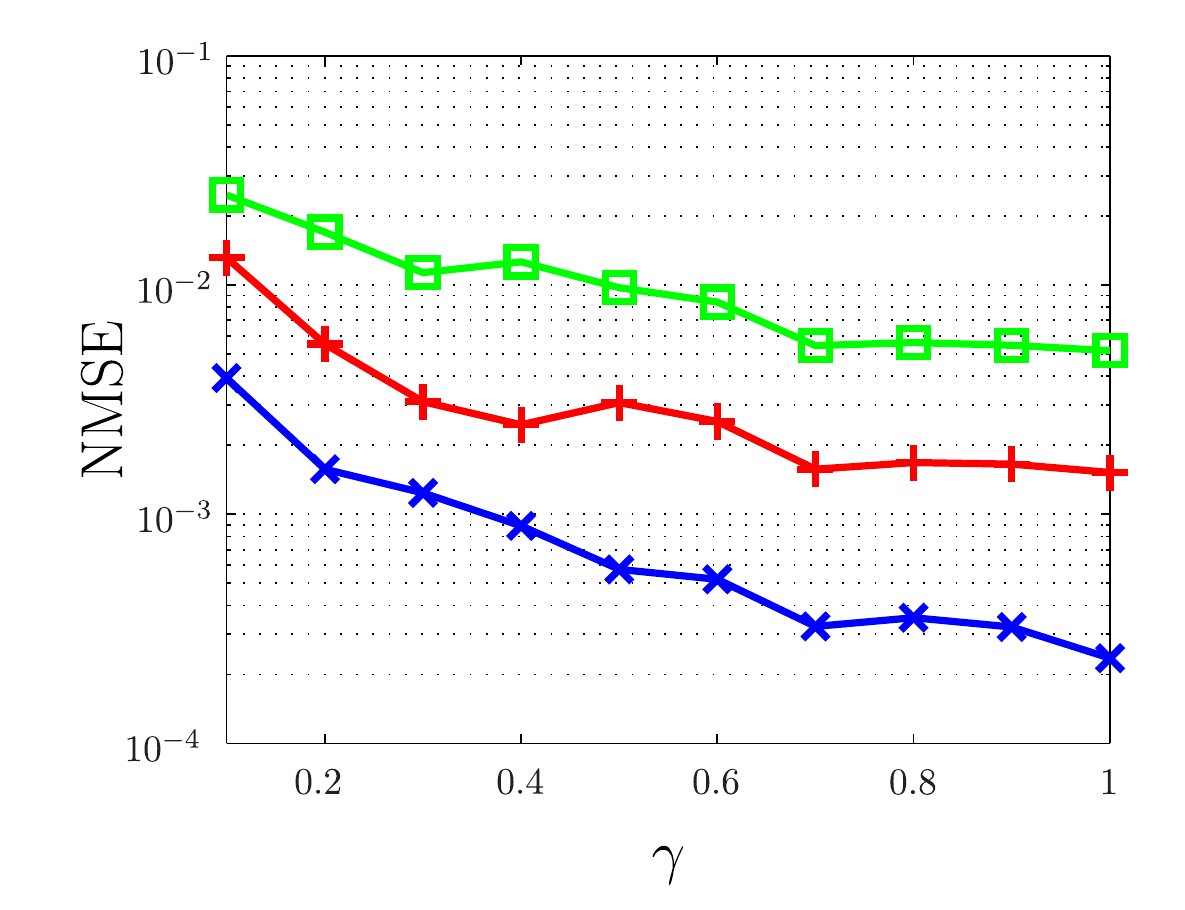}
\subcaption{\small\sl 1625 Symbols: $n=4$, $q=0$}\label{fig:compare_nmse_c4_0_6_1625}
\end{minipage}
\begin{minipage}[b]{0.31\textwidth}
\includegraphics[width=1\textwidth]{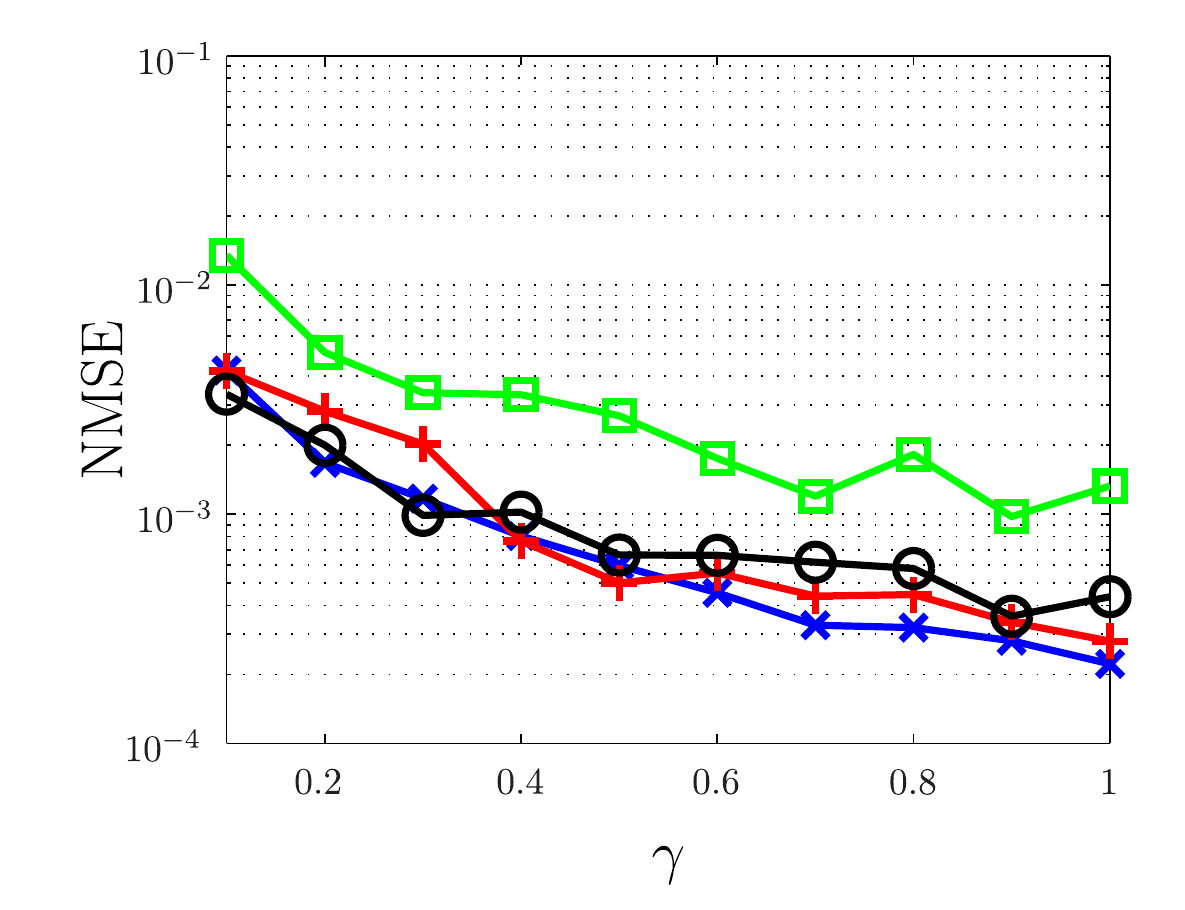}
\subcaption{\small\sl 1625 Symbols: $n=4$, $q=2$}\label{fig:compare_nmse_c4_2_6_1625}
\end{minipage}
\begin{minipage}[b]{0.31\textwidth}
\includegraphics[width=1\textwidth]{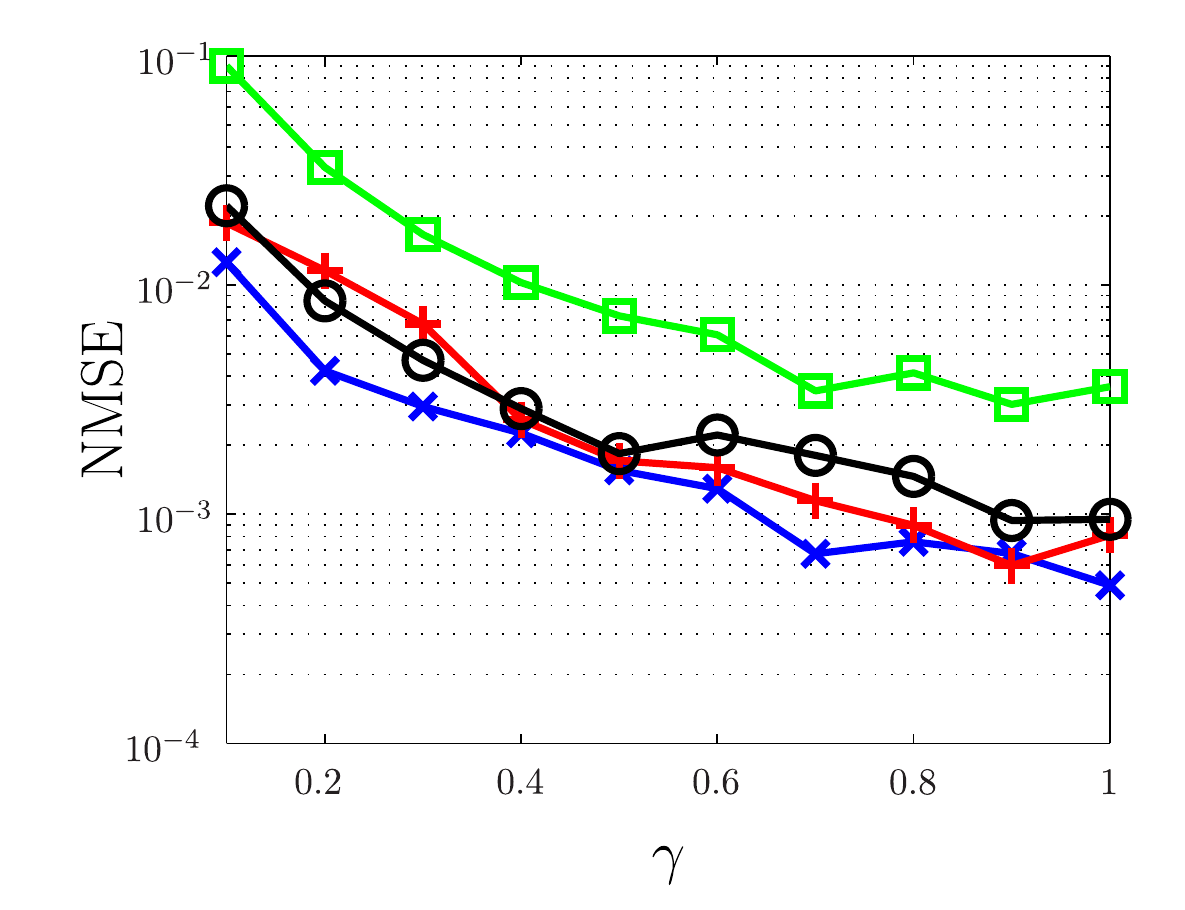}
\subcaption{\small\sl 1625 Symbols: $n=6$, $q=3$}\label{fig:compare_nmse_c6_3_6_1625}
\end{minipage}\\
\begin{minipage}[b]{0.31\textwidth}
\includegraphics[width=1\textwidth]{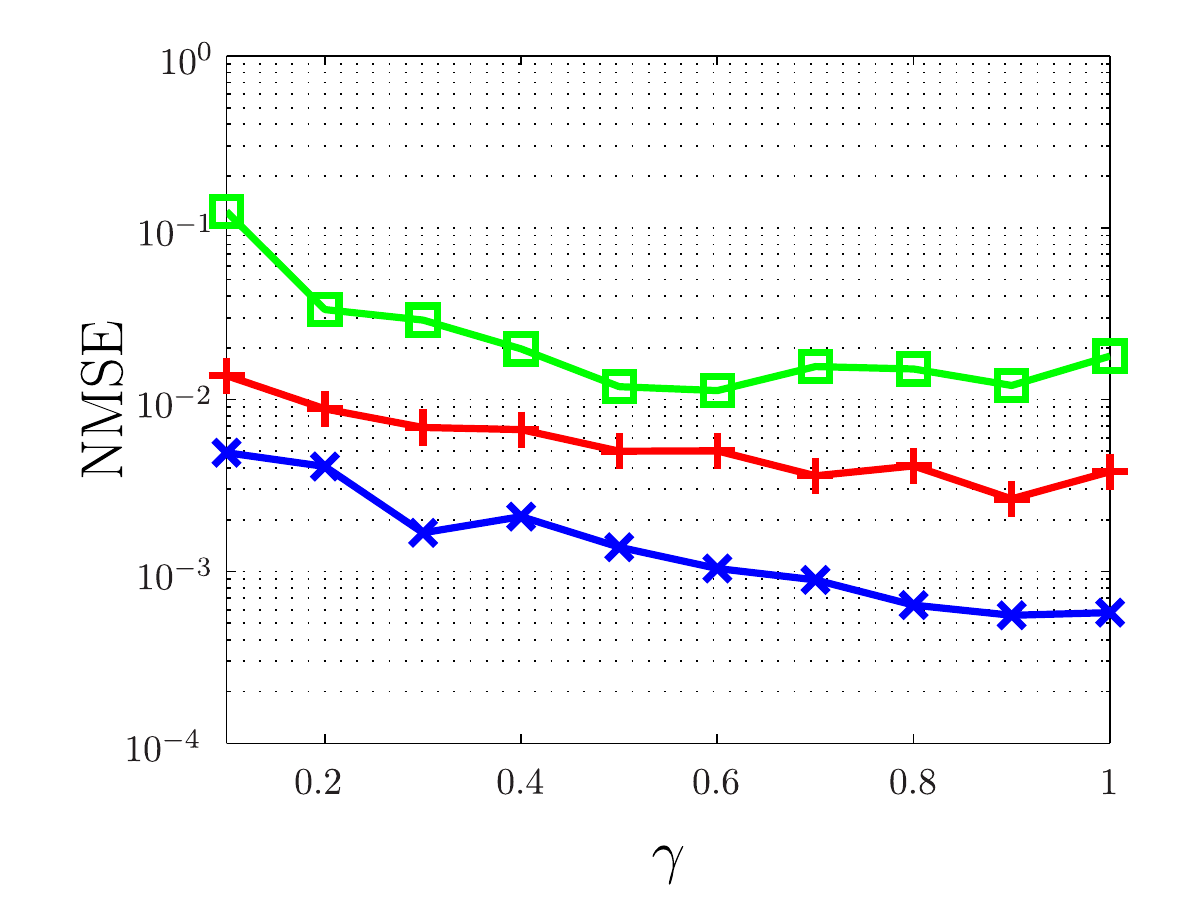}
\subcaption{\small\sl 800 Symbols: $n=4$, $q=0$}\label{fig:compare_nmse_c4_0_6_800}
\end{minipage}
\begin{minipage}[b]{0.31\textwidth}
\includegraphics[width=1\textwidth]{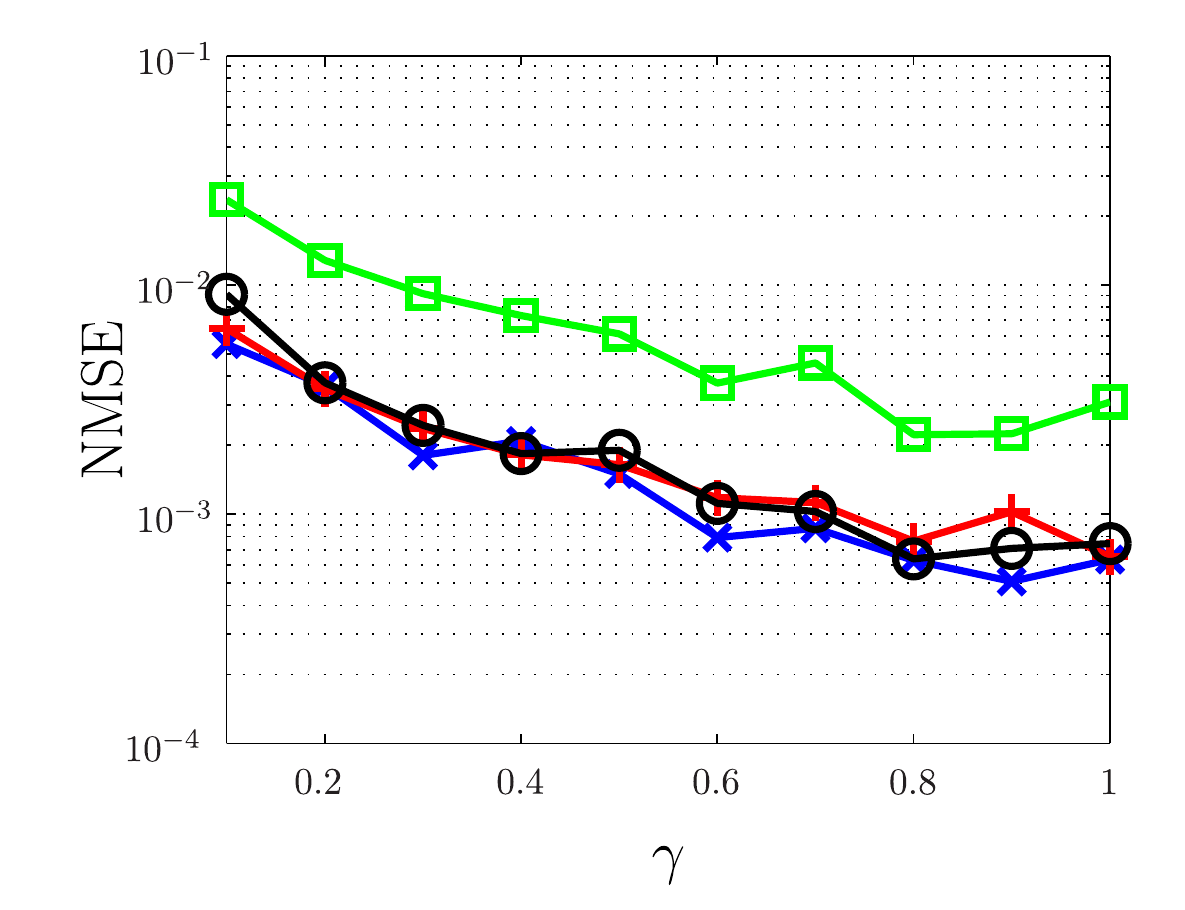}
\subcaption{\small\sl 800 Symbols: $n=4$, $q=2$}\label{fig:compare_nmse_c4_2_6_800}
\end{minipage}
\begin{minipage}[b]{0.31\textwidth}
\includegraphics[width=1\textwidth]{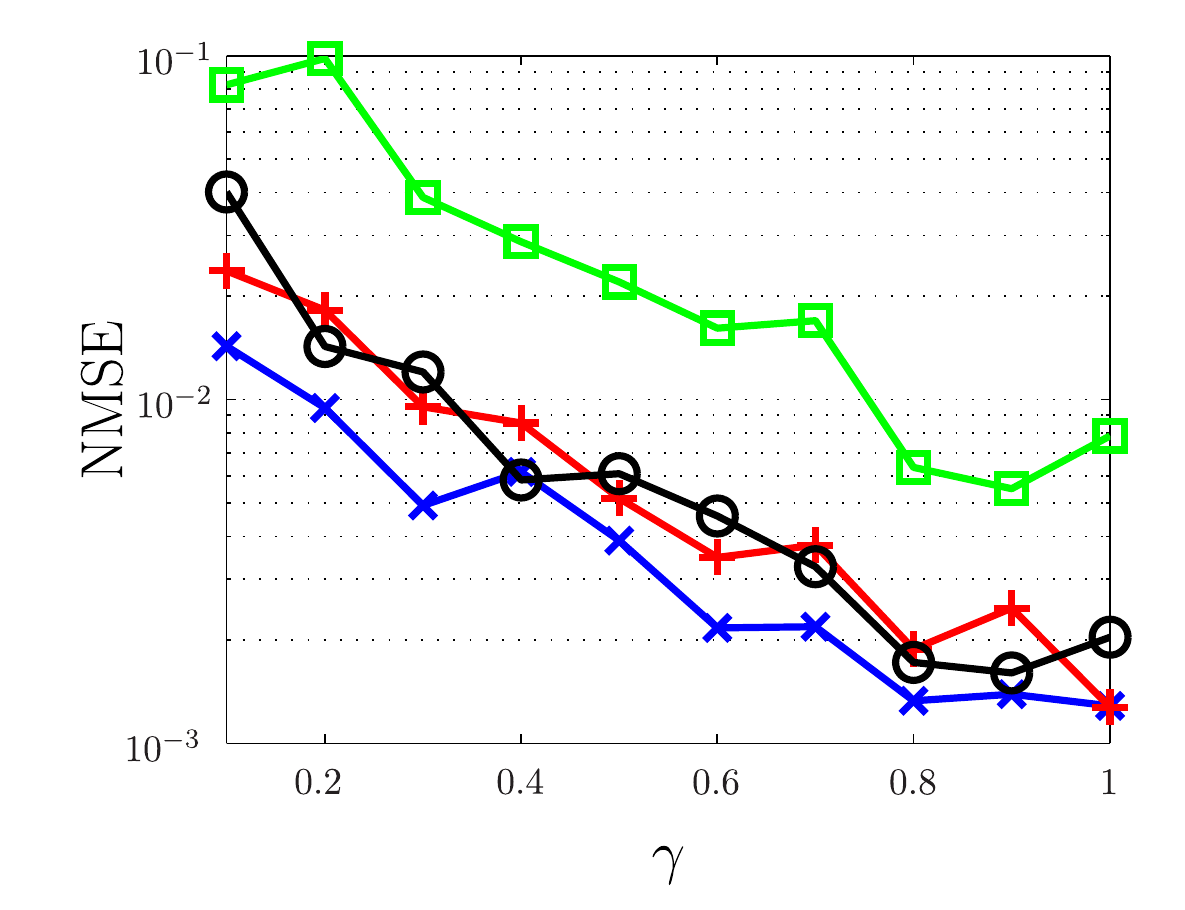}
\subcaption{\small\sl 800 Symbols: $n=6$, $q=3$}\label{fig:compare_nmse_c6_3_6_800}
\end{minipage}\\
\begin{minipage}[b]{0.31\textwidth}
\includegraphics[width=1\textwidth]{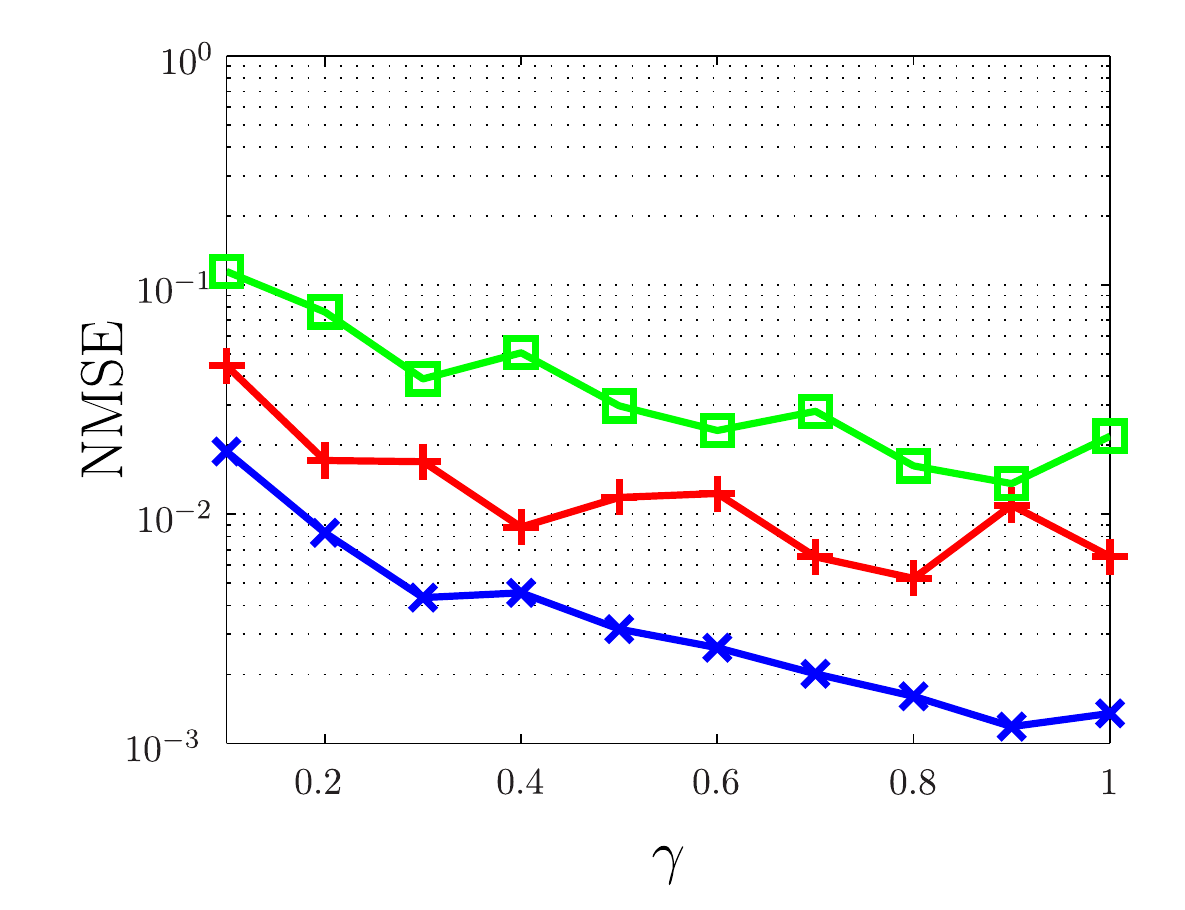}
\subcaption{\small\sl 400 Symbols: $n=4$, $q=0$}\label{fig:compare_nmse_c4_0_6_400}
\end{minipage}
\begin{minipage}[b]{0.31\textwidth}
\includegraphics[width=1\textwidth]{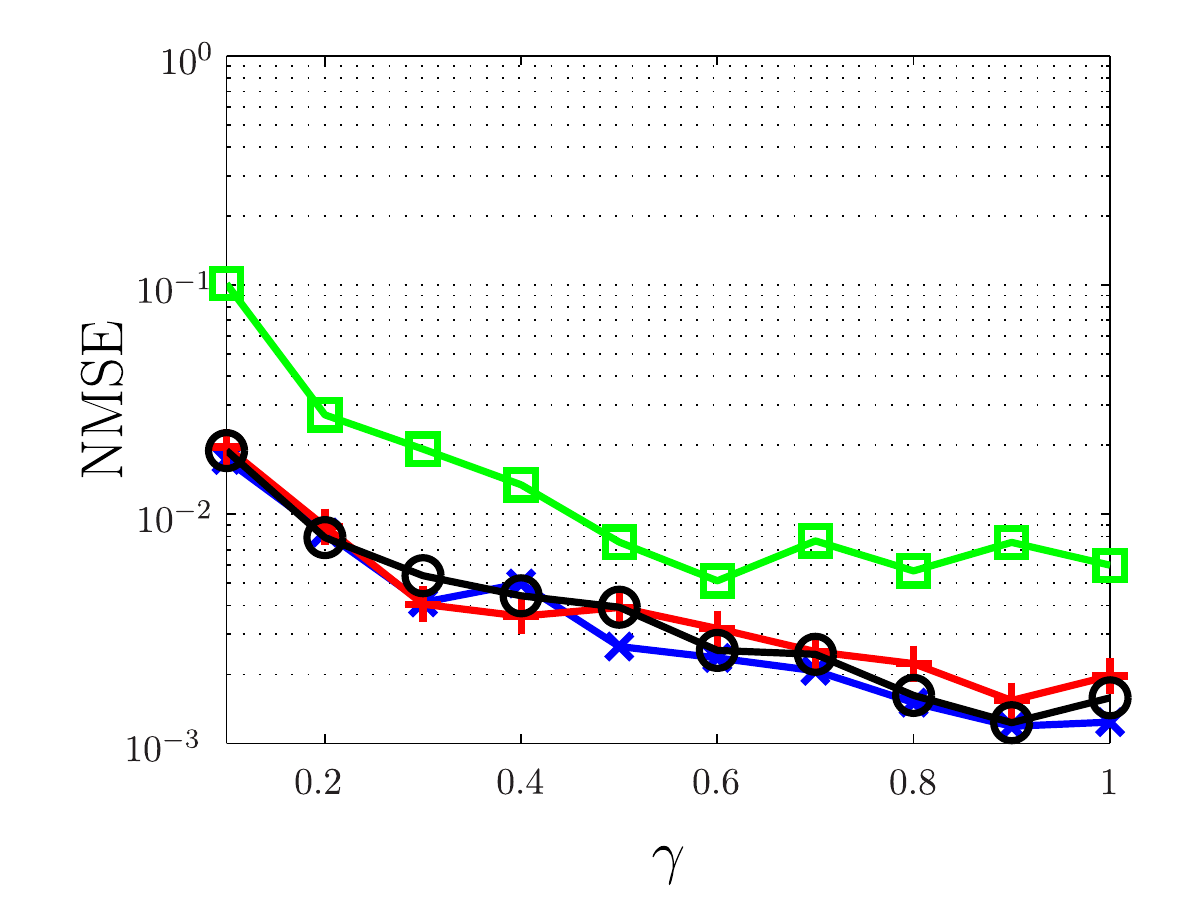}
\subcaption{\small\sl 400 Symbols: $n=4$, $q=2$}\label{fig:compare_nmse_c4_2_6_400}
\end{minipage}
\begin{minipage}[b]{0.31\textwidth}
\includegraphics[width=1\textwidth]{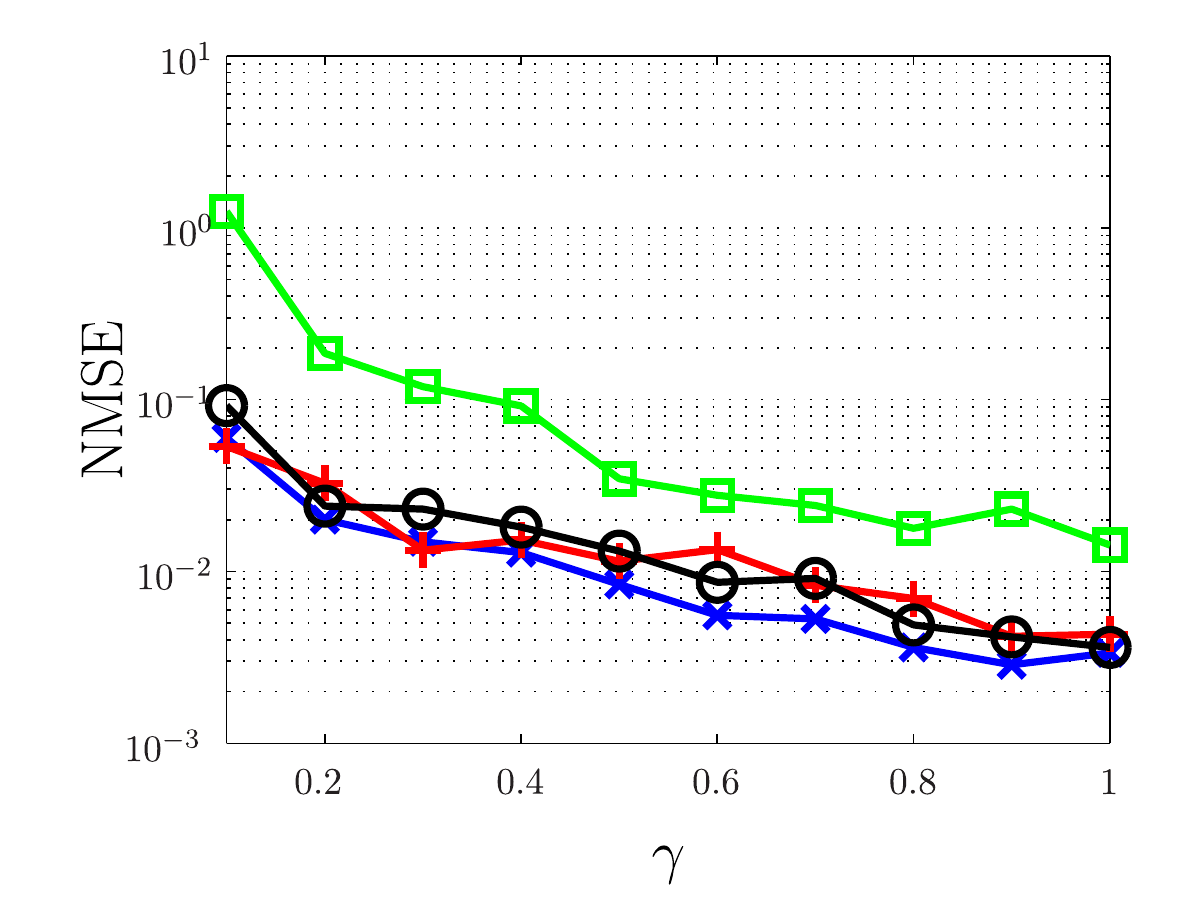}
\subcaption{\small\sl 400 Symbols: $n=6$, $q=3$}\label{fig:compare_nmse_c6_3_6_400}
\end{minipage}\\
\caption{\small\sl NMSE of compressive cyclic cumulants against cyclic cumulants. Across each row, the plots show NMSE of compressive cyclic cumulants versus cyclic cumulants for select values of $n$ and $q$ as a function of $\gamma$ for select processed data length when CNR=6dB.}\label{fig:compare_chocs_mse_6dB}
\end{figure*}

\begin{figure*}[tp]
\vspace{-5ex}
\begin{minipage}[b]{0.31\textwidth}
\includegraphics[width=1\textwidth]{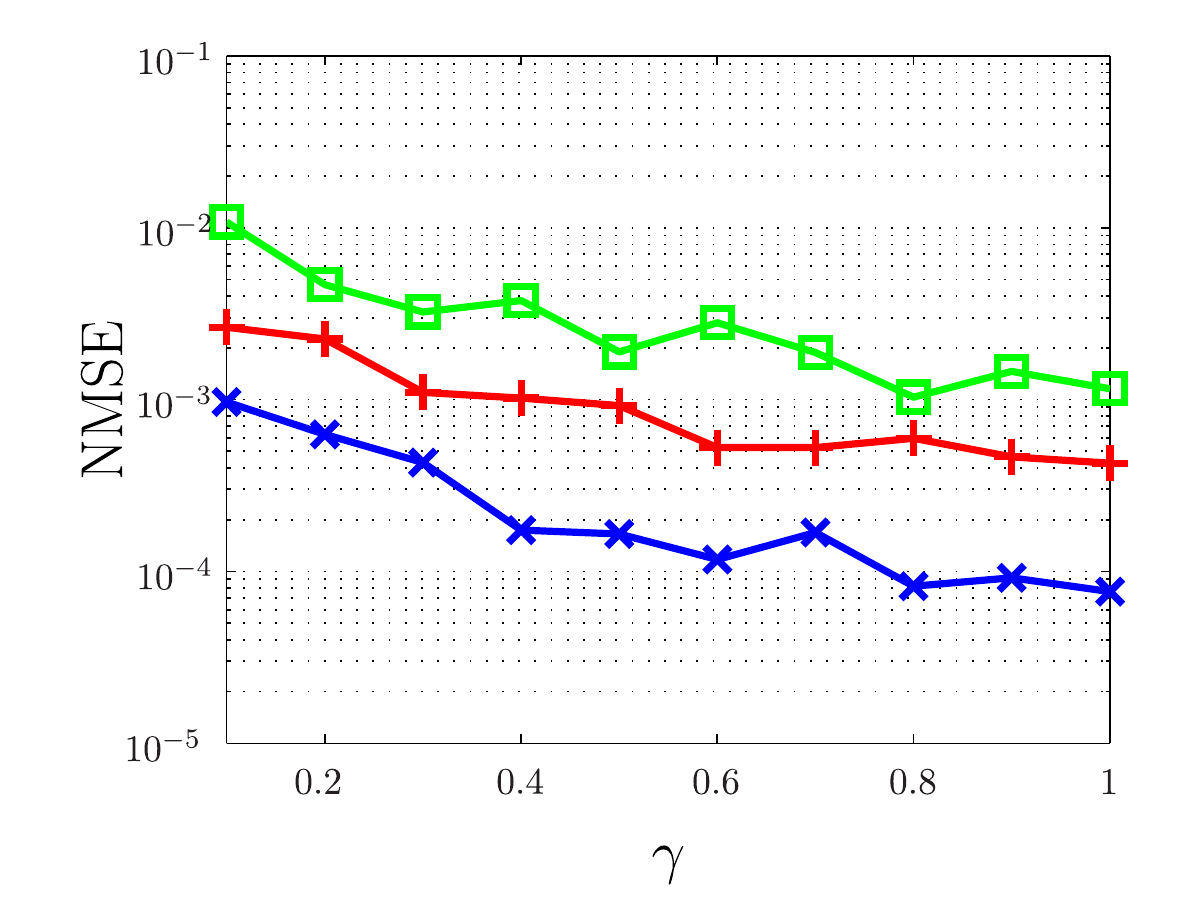}
\subcaption{\small\sl 13000 Symbols: $n=4$, $q=0$}\label{fig:compare_nmse_c4_0_3_13000}
\end{minipage}
\begin{minipage}[b]{0.31\textwidth}
\includegraphics[width=1\textwidth]{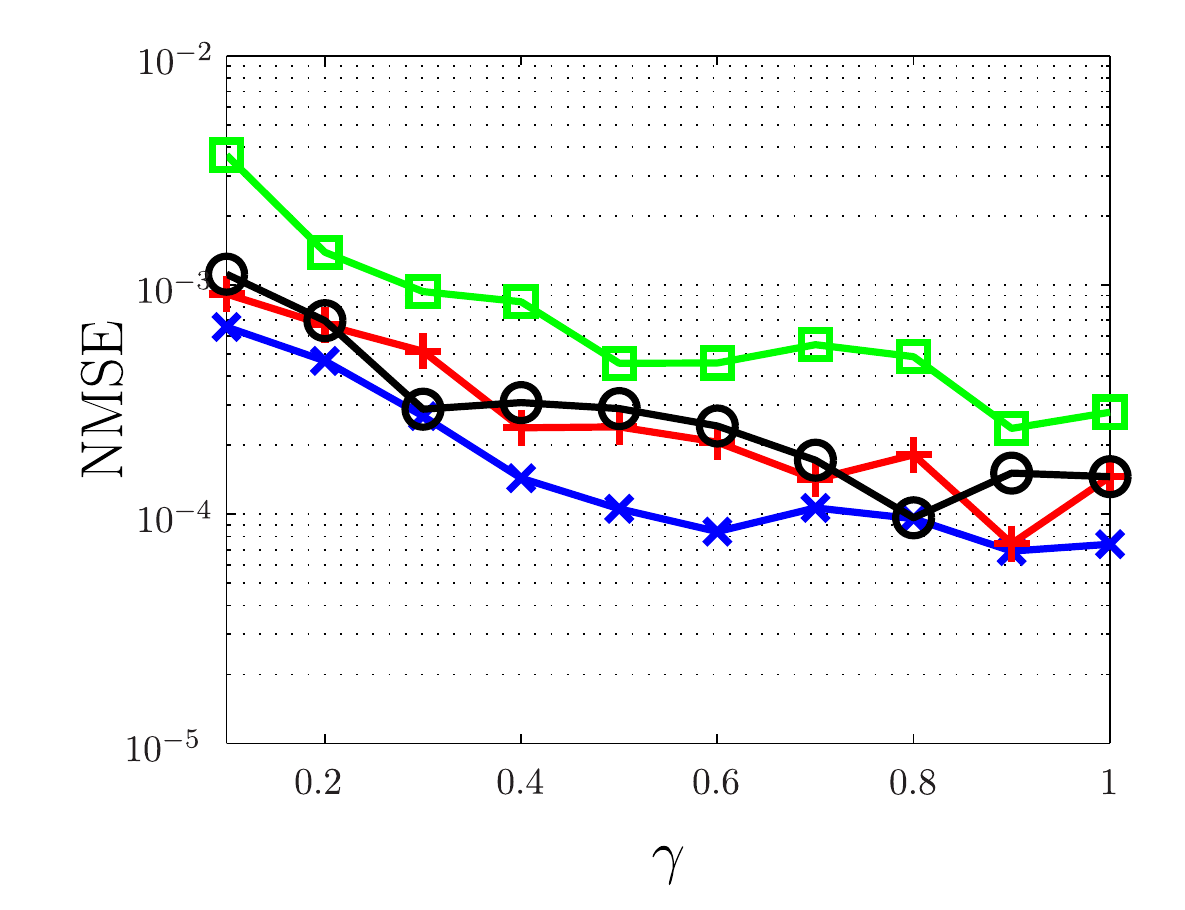}
\subcaption{\small\sl 13000 Symbols: $n=4$, $q=2$}\label{fig:compare_nmse_c4_2_3_13000}
\end{minipage}
\begin{minipage}[b]{0.31\textwidth}
\includegraphics[width=1\textwidth]{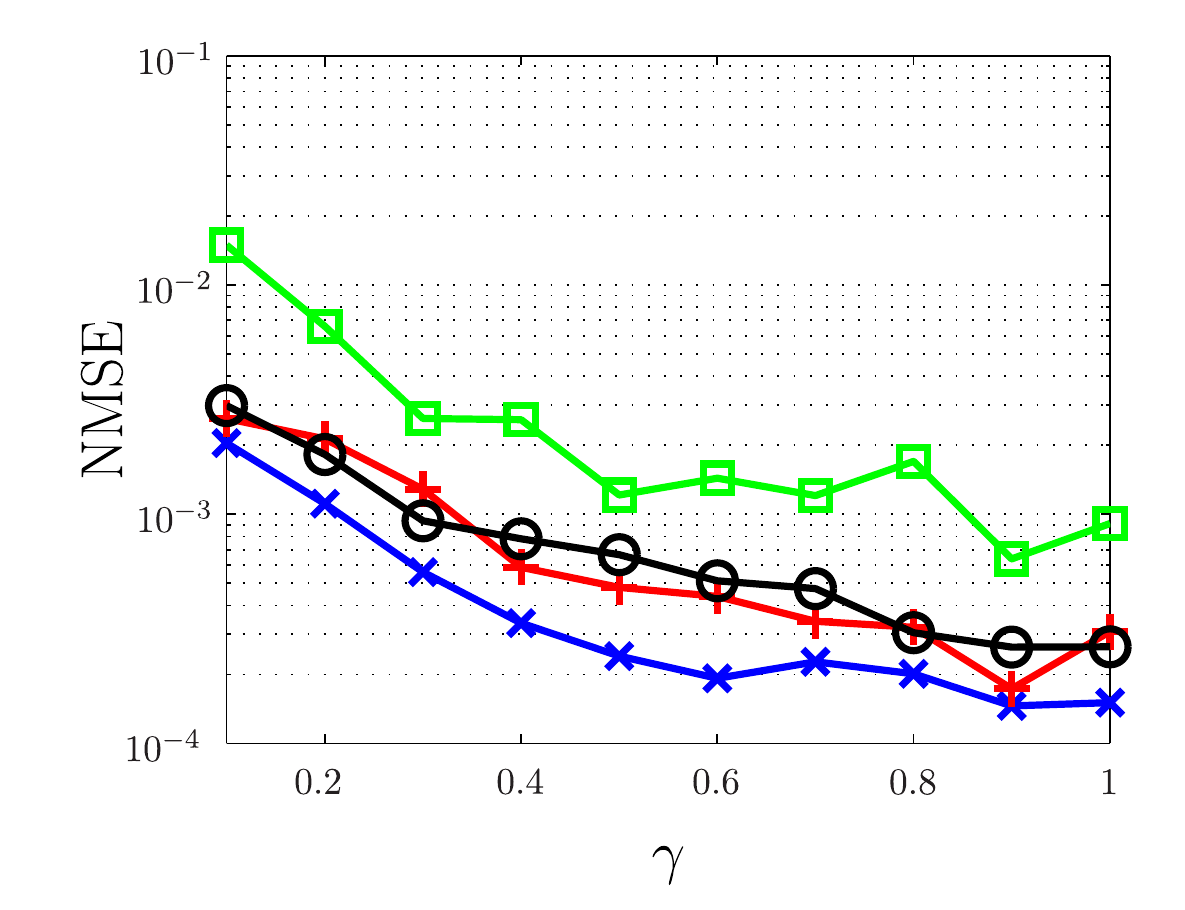}
\subcaption{\small\sl 13000 Symbols: $n=6$, $q=3$}\label{fig:compare_nmse_c6_3_3_13000}
\end{minipage}\\
\begin{minipage}[b]{0.31\textwidth}
\includegraphics[width=1\textwidth]{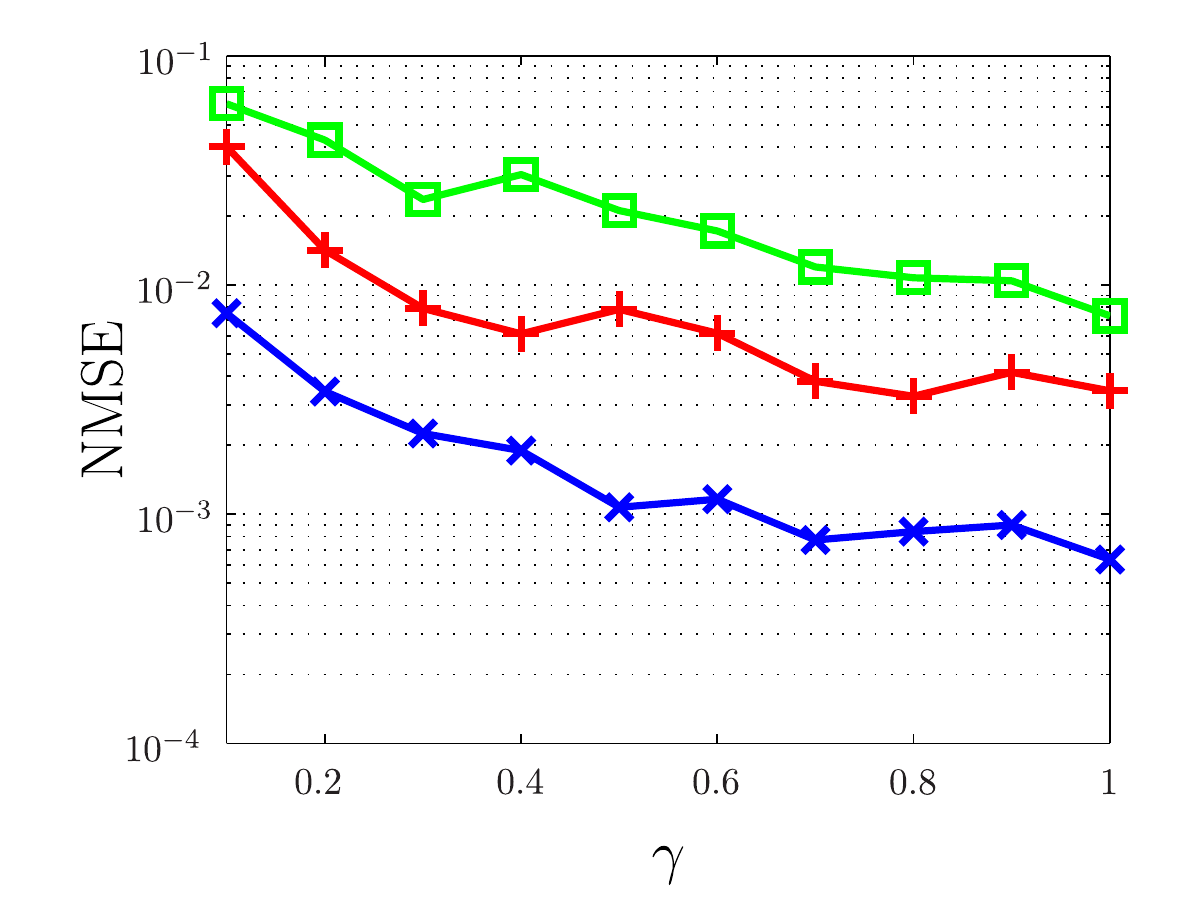}
\subcaption{\small\sl 1625 Symbols: $n=4$, $q=0$}\label{fig:compare_nmse_c4_0_3_1625}
\end{minipage}
\begin{minipage}[b]{0.31\textwidth}
\includegraphics[width=1\textwidth]{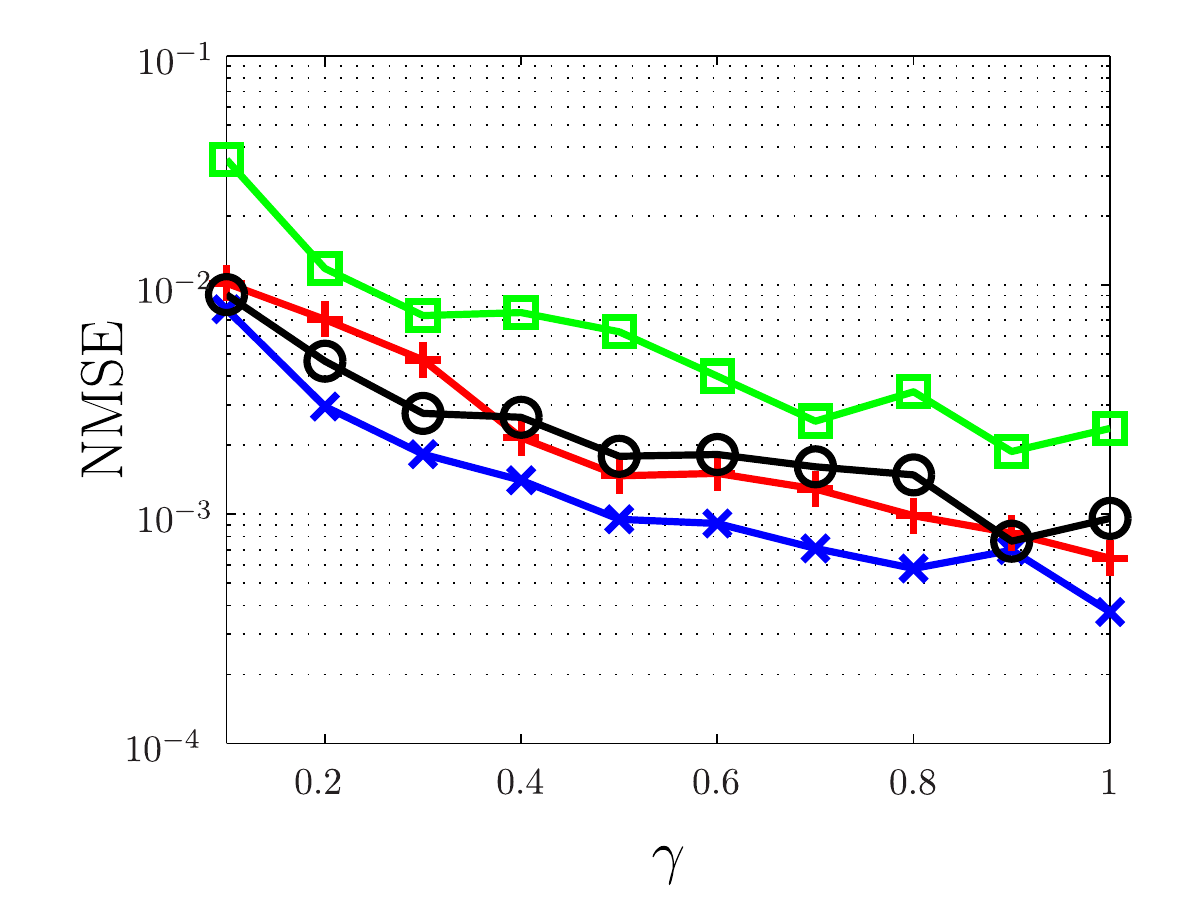}
\subcaption{\small\sl 1625 Symbols: $n=4$, $q=2$}\label{fig:compare_nmse_c4_2_3_1625}
\end{minipage}
\begin{minipage}[b]{0.31\textwidth}
\includegraphics[width=1\textwidth]{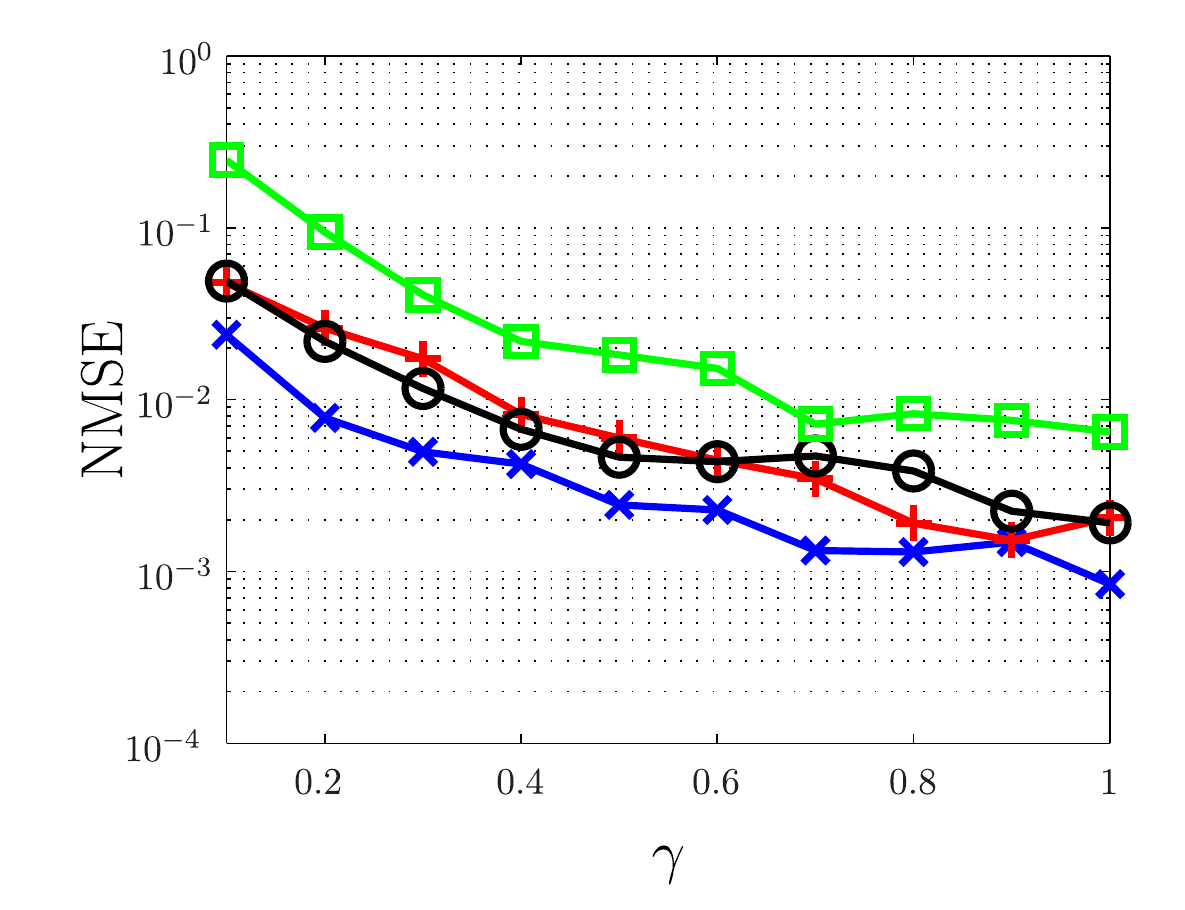}
\subcaption{\small\sl 1625 Symbols: $n=6$, $q=3$}\label{fig:compare_nmse_c6_3_3_1625}
\end{minipage}\\
\begin{minipage}[b]{0.31\textwidth}
\includegraphics[width=1\textwidth]{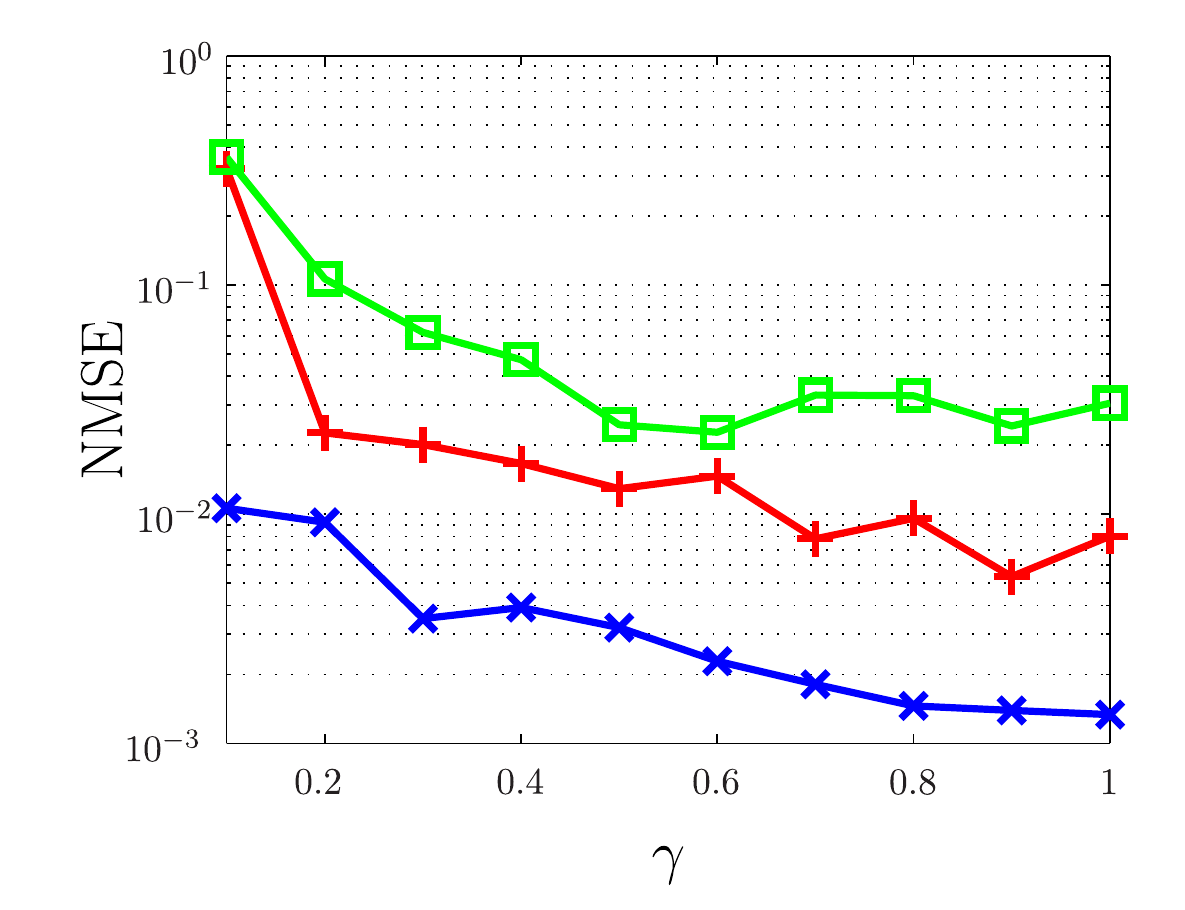}
\subcaption{\small\sl 800 Symbols: $n=4$, $q=0$}\label{fig:compare_nmse_c4_0_3_800}
\end{minipage}
\begin{minipage}[b]{0.31\textwidth}
\includegraphics[width=1\textwidth]{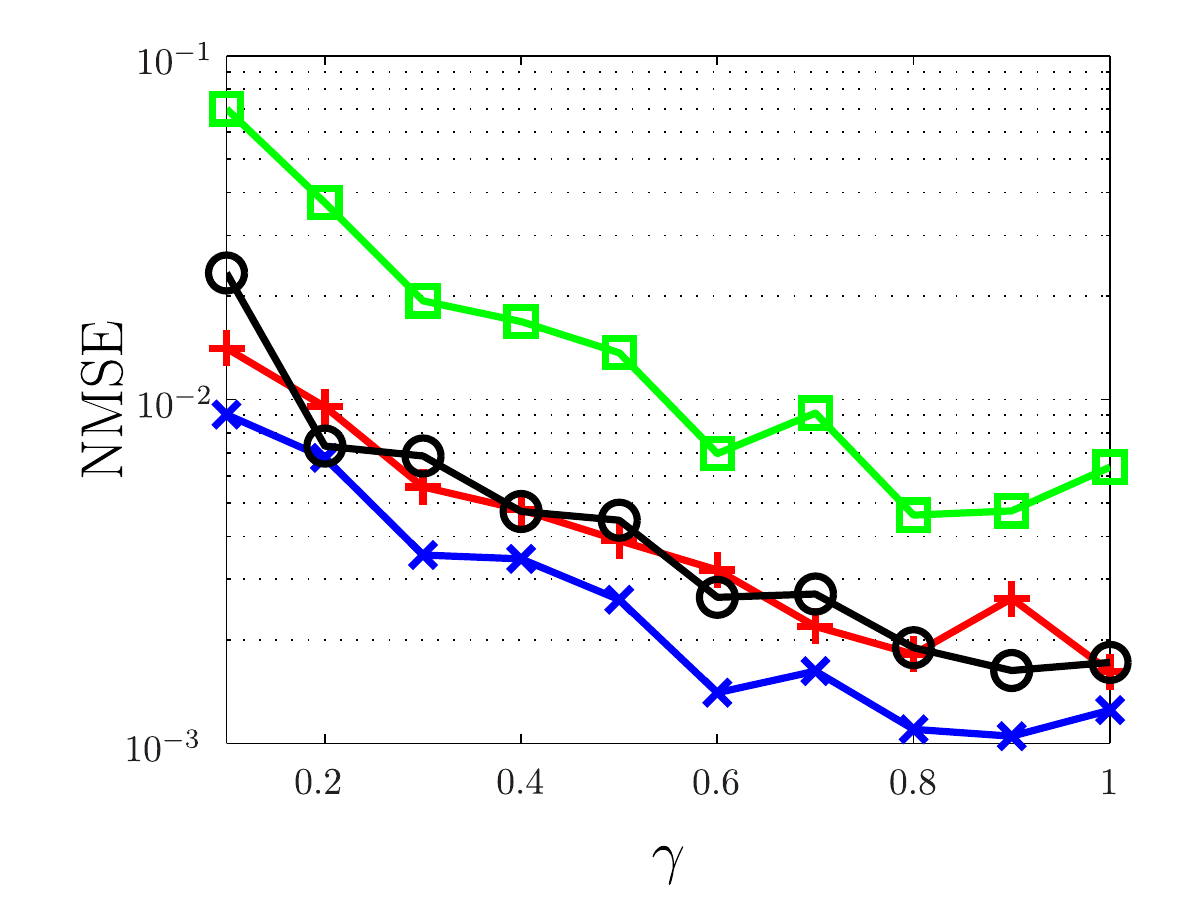}
\subcaption{\small\sl 800 Symbols: $n=4$, $q=2$}\label{fig:compare_nmse_c4_2_3_800}
\end{minipage}
\begin{minipage}[b]{0.31\textwidth}
\includegraphics[width=1\textwidth]{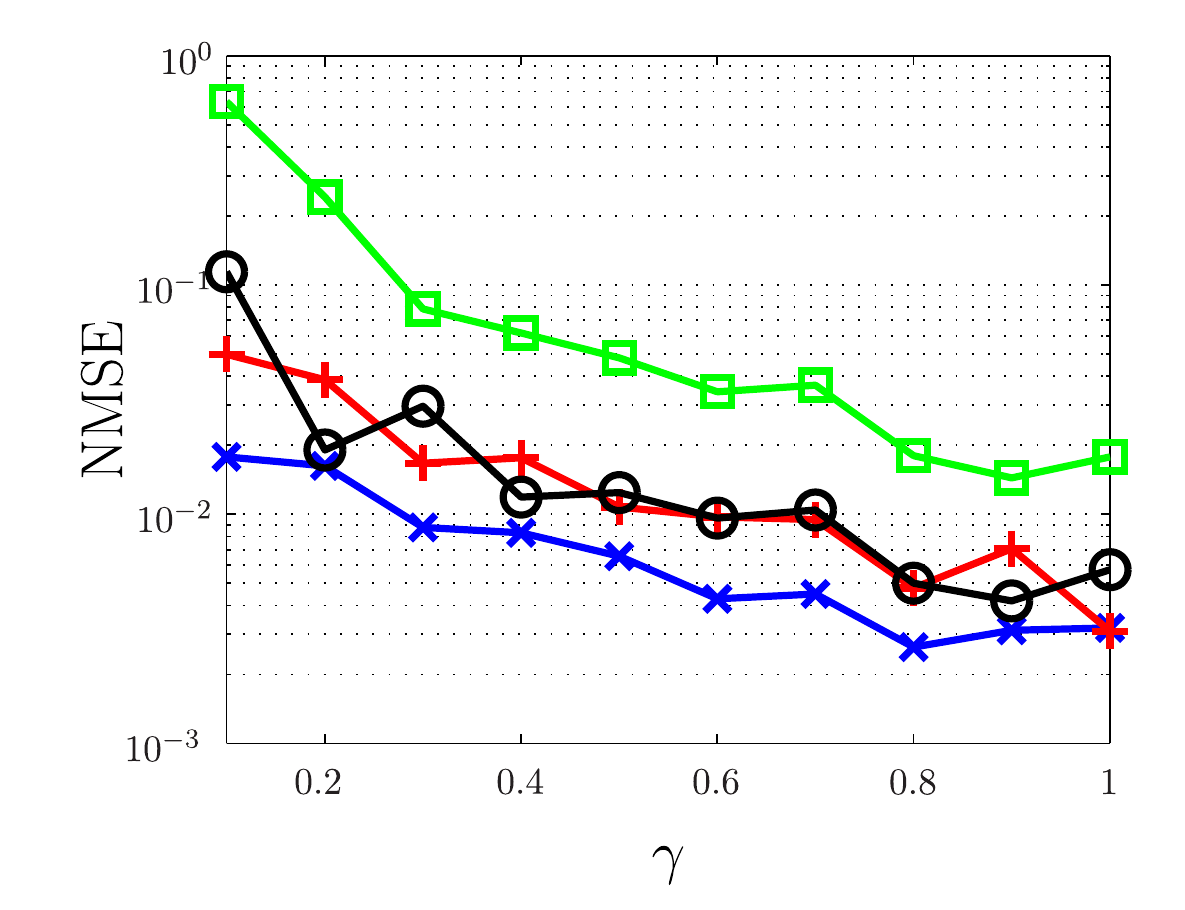}
\subcaption{\small\sl 800 Symbols: $n=6$, $q=3$}\label{fig:compare_nmse_c6_3_3_800}
\end{minipage}\\
\begin{minipage}[b]{0.31\textwidth}
\includegraphics[width=1\textwidth]{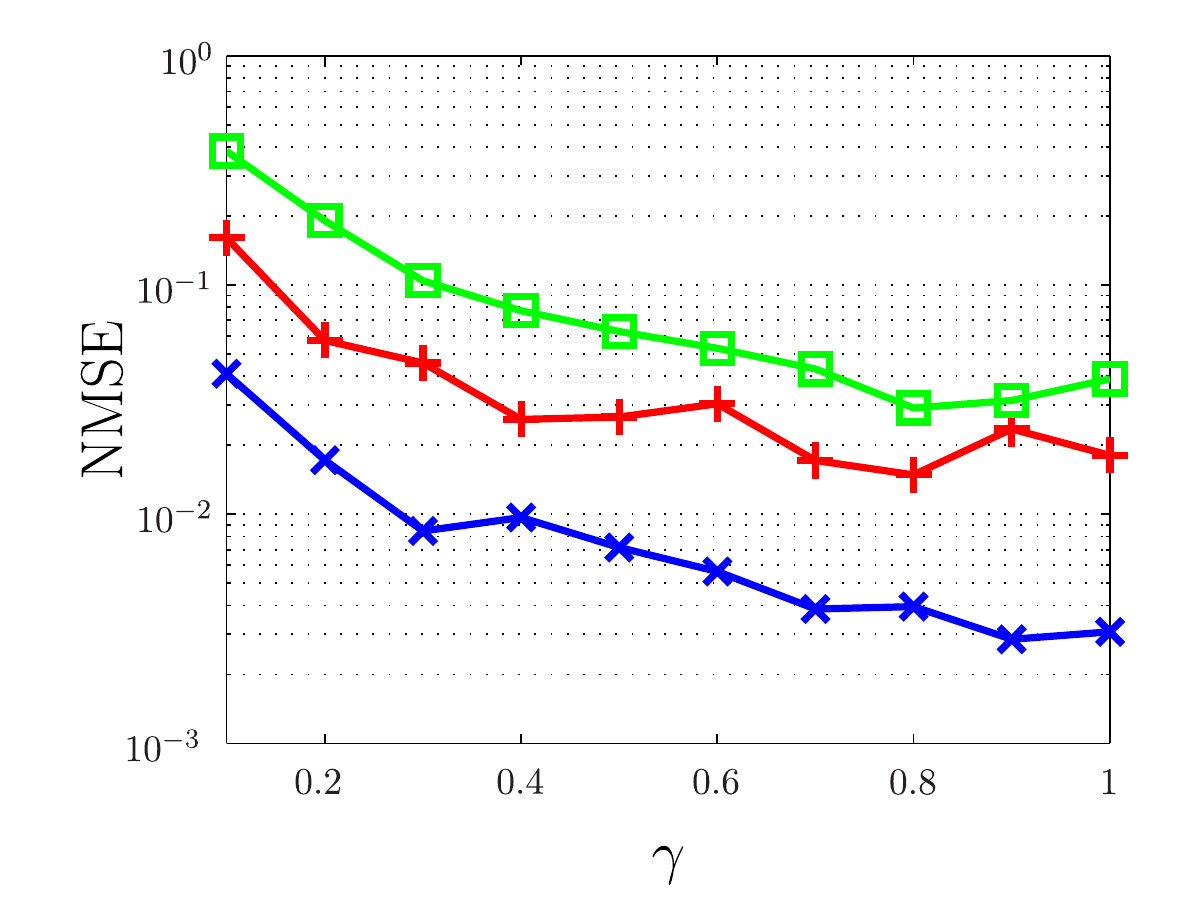}
\subcaption{\small\sl 400 Symbols: $n=4$, $q=0$}\label{fig:compare_nmse_c4_0_3_400}
\end{minipage}
\begin{minipage}[b]{0.31\textwidth}
\includegraphics[width=1\textwidth]{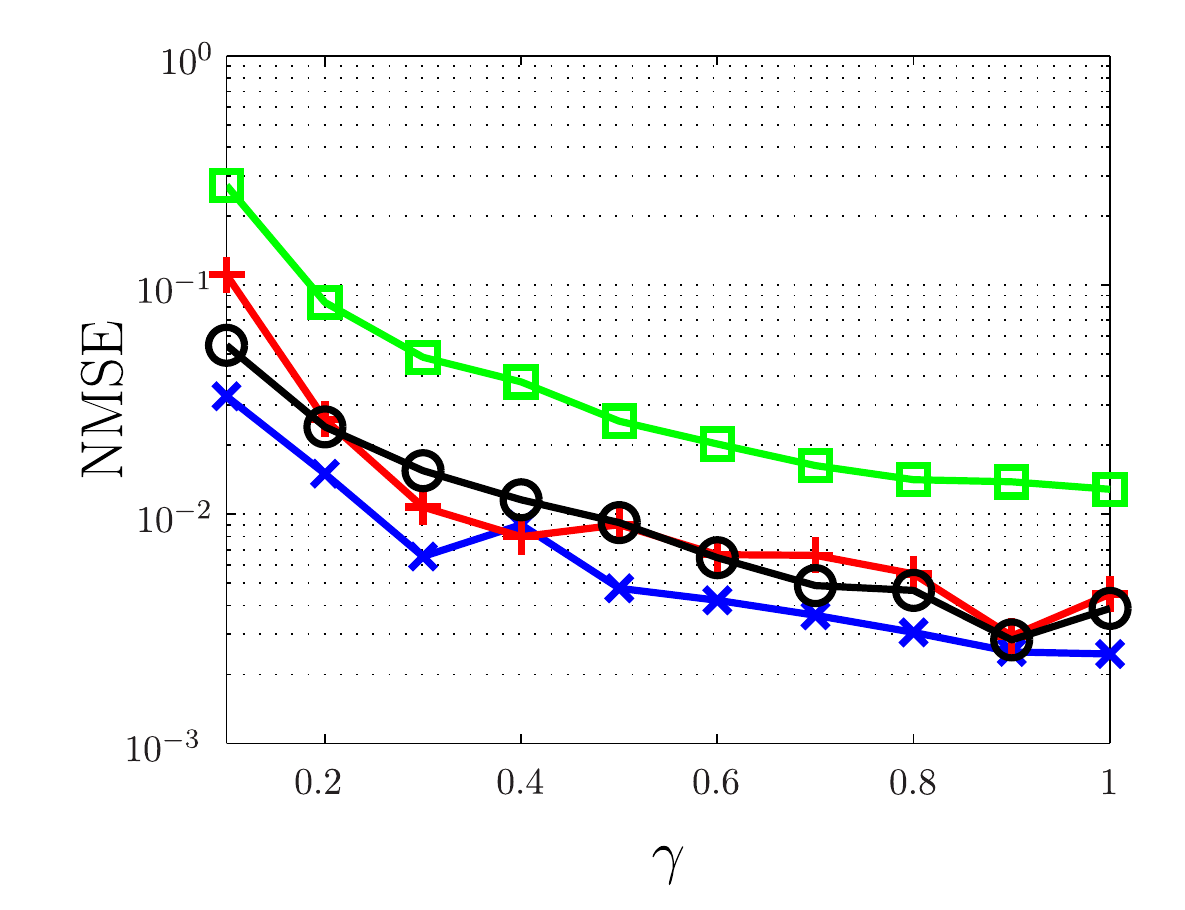}
\subcaption{\small\sl 400 Symbols: $n=4$, $q=2$}\label{fig:compare_nmse_c4_2_3_400}
\end{minipage}
\begin{minipage}[b]{0.31\textwidth}
\includegraphics[width=1\textwidth]{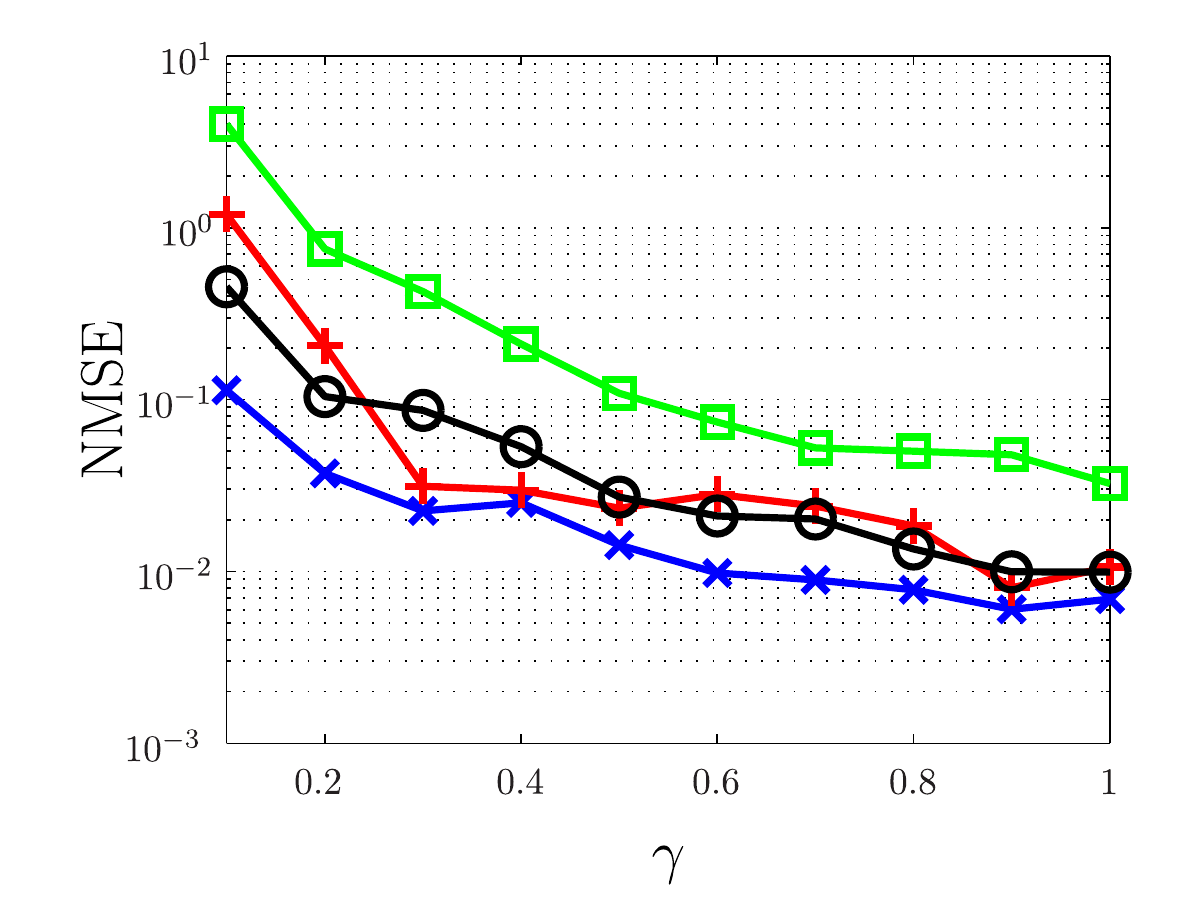}
\subcaption{\small\sl 400 Symbols: $n=6$, $q=3$}\label{fig:compare_nmse_c6_3_3_400}
\end{minipage}\\
\caption{\small\sl NMSE of compressive cyclic cumulants against cyclic cumulants. Across each row, the plots show NMSE of compressive cyclic cumulants versus cyclic cumulants for select values of $n$ and $q$ as a function of $\gamma$ for select processed data length when CNR=3dB.}\label{fig:compare_chocs_mse_3dB}
\end{figure*}

Before we proceed to establish the signal selectivity of compressive cyclic cumulants, in this section, we validate their inference quality. Specifically, we compare estimated compressive cyclic cumulants using various processing lengths (in terms of number of symbols) for select values of $n$, $q$ across a broad range of $\gamma$ against cyclic cumulants (which correspond to $\gamma=1$ in the preceding plots) using the normalized mean square error (NMSE)
\begin{equation}
\text{NMSE}=\frac{1}{N_{\text{tr}}}\sum_{i=1}^{N_{\text{tr}}}\frac{\left[\widehat{C}^{\beta,c}_{s_{\text{ncs}}}(\boldsymbol{0})_{n,q}-C^{\beta,c}_{s_{\text{ncs}}}(\boldsymbol{0})_{n,q}\right]^2}{\left[C^{\beta,c}_{s_{\text{ncs}}}(\boldsymbol{0})_{n,q}\right]^2},
\end{equation}
where $N_{\text{tr}}$ is the number of trials, $\beta=(n-2q)\Delta f_c$ and $C^{\beta,c}_{s_{\text{ncs}}}(\boldsymbol{0})_{n,q}$ is computed using \eqref{eqn:tvcc_2}. We note that since there is only one unique value of $\tau_i=0$, the proposed multi-channel sampling scheme of Figure~\ref{fig:multi_ch_nus_sampling} collapses into a single channel and subsequent compression rates obtained represent the best compression rates achievable. The plots in Figures~\ref{fig:compare_chocs_mse_noiseless},~\ref{fig:compare_chocs_mse_9dB},~\ref{fig:compare_chocs_mse_6dB}, and~\ref{fig:compare_chocs_mse_3dB} show the NMSE for the situations when no noise is added to the signal and also when noise is added such that resulting carrier-to-noise ratio (CNR) is 9dB, 6dB and 3dB respectively. The NMSE for each 8PSK signal has been omitted from the plots in the first column in Figures~\ref{fig:compare_chocs_mse_noiseless}--\ref{fig:compare_chocs_mse_3dB} since an 8PSK cyclic cumulant for $n=4$ and $q=0$ is zero.

We make the following observations with regards to the inference quality of compressive cyclic cumulants. First, as $n$ increases, the NMSE of compressive cyclic cumulant estimates increases across the entire range of $\gamma$ values considered. This also occurs in the case of cyclic cumulant estimates. Second, the NMSE of compressive cyclic cumulant estimates increases as the number of processing symbols decreases. This also occurs in the case of cyclic cumulant estimates. Third, as the compression factor increases (i.e., as $\gamma$ decreases), the NMSE of the compressive cyclic cumulant estimates increases  at a minimal rate of $\frac{1}{\gamma}$ as explained in the sequel. Equation~\eqref{eqn:ccm_var} gives the time-averaged variance of the compressive cyclic \emph{moment} estimator which can be used to predict the rate of increase in the NMSE of compressive cyclic \emph{cumulant} estimates (Figures~\ref{fig:compare_chocs_mse_noiseless}--\ref{fig:compare_chocs_mse_3dB}), for 4PSK and 16QAM signals (when $n=4$ and $q=0$) since for these signals, $C^{\alpha,c}_{s_{\textnormal{ncs}}}(\btau)_{4,0}=R^{\alpha,c}_{s_{\textnormal{ncs}}}(\btau)_{4,0}$. As the number of processing symbols decreases (i.e., as $\widehat{T}$ decreases), for a fixed $\gamma$, the increase in NMSE can be explained by the increase in time-averaged variance of the compressive cyclic cumulant estimates due to the $\frac{1}{\widehat{T}^2}$ scaling in both terms of \eqref{eqn:ccm_var_1} and the increase in bias due to \eqref{eq:meanBias}. On the other hand, for a fixed number of processing symbols (i.e., fixed $\widehat{T}$) the NMSE increases at rate $\frac{1}{\gamma}$ due to the $\frac{1}{\gamma}$ scaling in \eqref{eqn:ccm_var}. For the other compressive cyclic cumulants such as $C_{s_{\textnormal{ncs}}}^{\beta,c}(\btau)_{4,2}$ and $C_{s_{\textnormal{ncs}}}^{\beta,c}(\btau)_{6,3}$, their NMSE increases at a rate greater than $\frac{1}{\gamma}$ which is expected since $C_{s_{\textnormal{ncs}}}^{\beta,c}(\btau)_{4,2}$ and $C_{s_{\textnormal{ncs}}}^{\beta,c}(\btau)_{6,3}$ are both functions of higher powers of lower order compressive cyclic moments.

As we shall discuss in the next section, in some regimes, compressive cyclic cumulants can give reliable approximations to their cyclic cumulant counterparts.

\subsubsection{Compressive Cyclic Cumulant Signal Selectivity Quality}
\label{results_discussion}

In this section, we establish the signal selectivity quality of $\widehat{F}_s$, an estimate of the compressive cyclic cumulant feature ${F}_s$ (cf. \eqref{eqn:c_4_0_feature}). Figures~\ref{tab:confuse_500}--\ref{tab:confuse_3} show the confusion matrix when the minimum distance classification scheme was used to classify a four class scenario in the noiseless setting and with CNR of 9dB, 6dB and 3dB respectively. In each of the confusion matrices, $\times$, +, $\ocircle$ and \Square~represent 2PSK, 4PSK, 8PSK and 16QAM respectively. Each row in these figures shows the confusion matrices for $\gamma=0.1$, 0.5 and 1.

We make the following observations with regards to the classification performance using compressive cyclic cumulant estimates. First, as the compression factor increases (i.e., as $\gamma$ decreases), the classification performance degrades for a given CNR and number of processed symbols. The degradation of classification performance due to the decreasing $\gamma$ value is expected since the strengths of the spectral peaks (compressive cyclic moments) decrease as predicted by Lemma~\ref{lm:comp_moments}, and this results in increased estimation errors of $\widehat{F}_s$. Second, the classification performance degrades with decreasing CNR as well as decreasing number of processed symbols. This is due to the increased estimation errors associated with increased noise power and a reduced observation interval. This trend is also observed in the classification performance of cyclic cumulant based classifiers.

\begin{figure*}[tp]
\vspace{-5ex}
\begin{minipage}[b]{0.31\textwidth}
\includegraphics[width=1\textwidth]{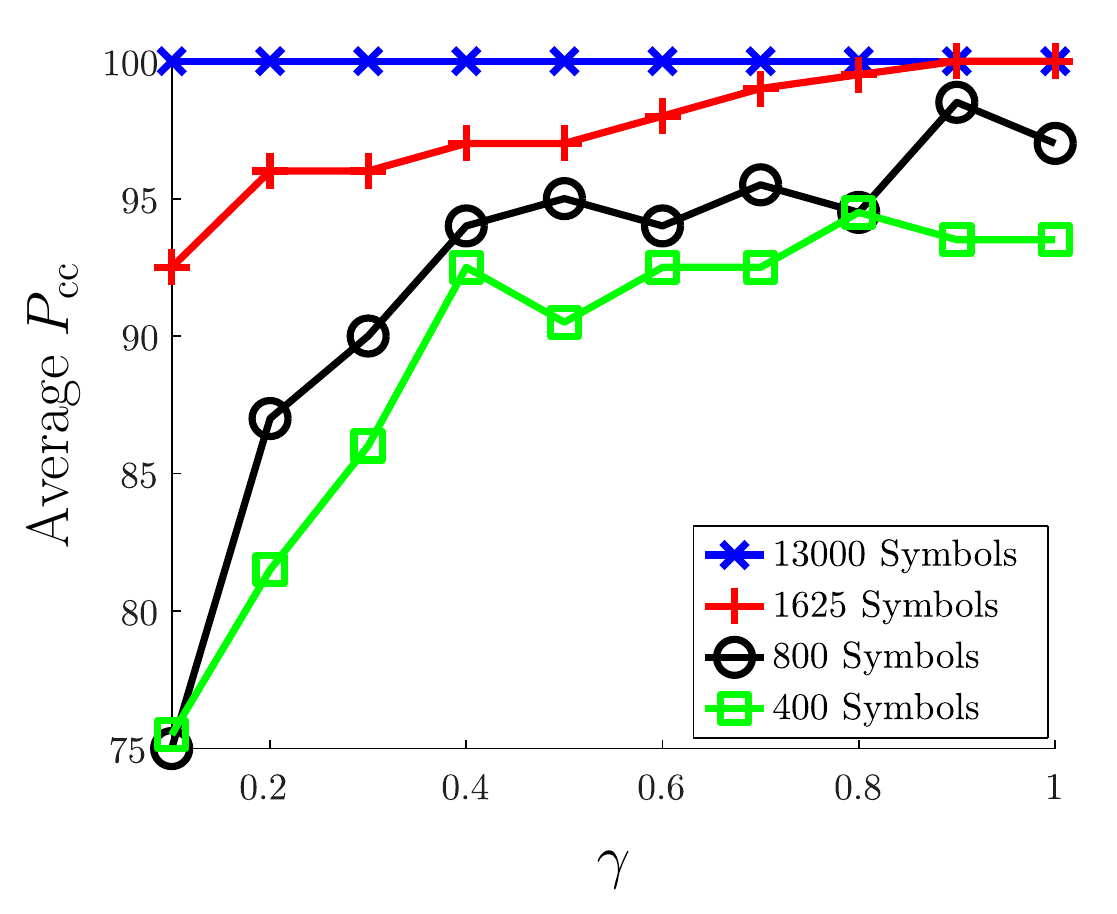}
\subcaption{\small\sl CNR:3dB}\label{fig:classify_c4_0_3}
\end{minipage}
\begin{minipage}[b]{0.31\textwidth}
\includegraphics[width=1\textwidth]{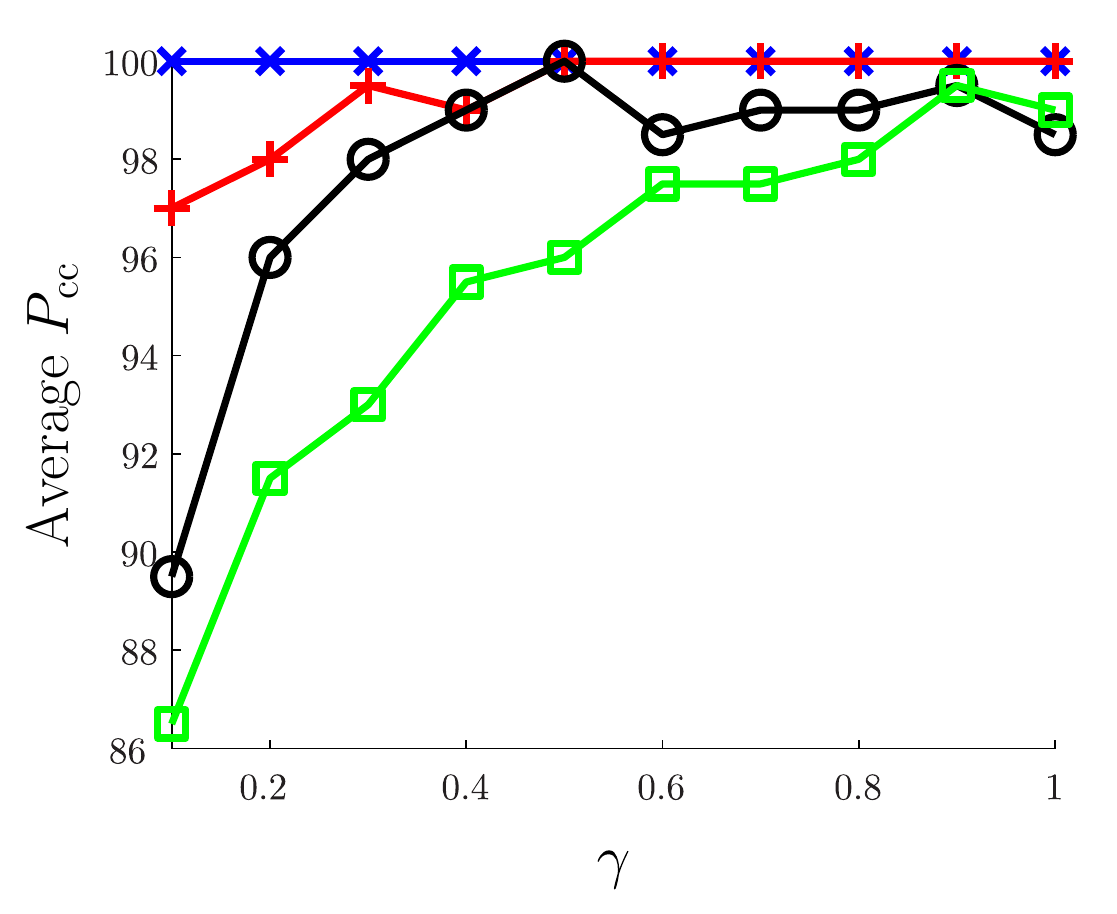}
\subcaption{\small\sl CNR:6dB}\label{fig:classify_c4_0_6}
\end{minipage}
\begin{minipage}[b]{0.31\textwidth}
\includegraphics[width=1\textwidth]{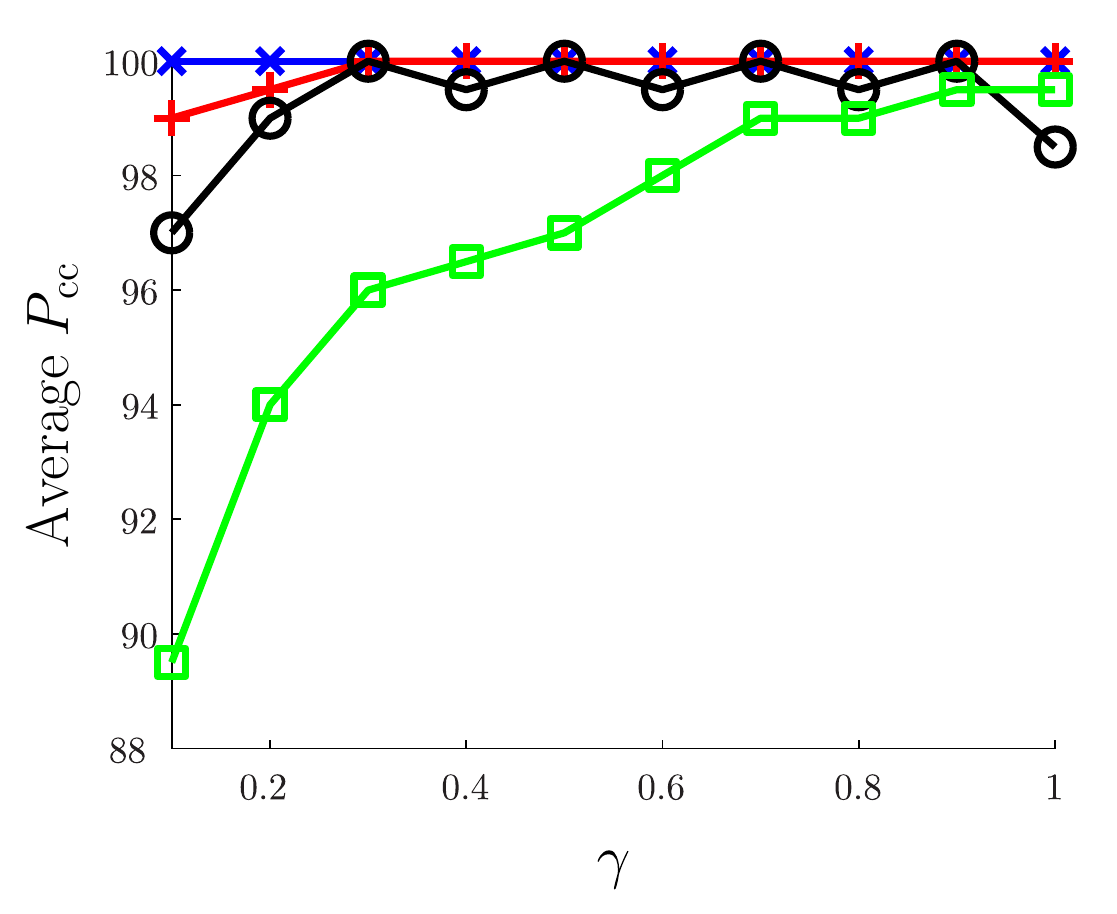}
\subcaption{\small\sl CNR:9dB}\label{fig:classify_c4_0_9}
\end{minipage}\\
\caption{\small\sl Comparison of average $P_{\text{cc}}$ for select values of processed data length when CNR=3dB, 6dB and 9dB.}\label{fig:classify_c4_0}
\end{figure*}

The observed trends are evident in Figure~\ref{fig:classify_c4_0}, which shows the average correct classification probability $P_{\text{cc}}$ for signals having CNRs of 3dB, 6dB and 9dB.

Table~\ref{tab:classify_chocs_hocs_c4_0} shows a comparison of classification performance between compressive cyclic cumulant based classifier ($\gamma=0.1$) and cyclic cumulant based classifier ($\gamma=1$) for various number of processed symbols. We note that when $\gamma=0.1$, processing 3250 symbols corresponds to using 3277 low-rate nonuniform samples to estimate compressive cyclic cumulants. On the other hand, when $\gamma=1$, processing 400 symbols corresponds to using 4096 uniform samples to estimate cyclic cumulants. Evidently, the classification performance of compressive cyclic cumulant based classifiers are similar to cyclic cumulant based classifiers when making pairwise comparisons, e.g., 3250 symbols versus 400 symbols, 6500 symbols versus 800 symbols, 13000 symbols versus 1625 symbols. Let us remind the reader that when $\gamma=0.1$, the average sampling rate commensurates with the information rate of the signal. Due to the compressibility of the lag product of the signal, compressive cyclic cumulant estimates can give reliable approximations of their cyclic cumulant counterparts at a fraction of the original sampling rate.

The results of Table~\ref{tab:classify_chocs_hocs_c4_0} also provide the following insights regarding the use of low-rate nonuniform sampling (in the CTHOCS case) versus the use of high rate uniform sampling (in the THOCS case). For a fixed observation time, CTHOCS-based classification performance degrades with decreasing $\gamma$. On the other hand, for an equivalent number of samples, nonuniform sampling and uniform sampling have comparable classification performance. While uniform sampling would allow a shorter observation time, a higher rate sampler is required which could be expensive or have limited resolution. Therefore, nonuniform sampling can viewed as a means to safely sample slower than Nyquist rate without risking aliasing (which is not supported by uniform sampling). If the degradation in CTHOCS-based classification performance is not acceptable (due to decreasing $\gamma$), an increase in observation interval would ultimately improve CTHOCS-based classification performance such that it is comparable to THOCS-based classification performance for the same number of measurements. We note that when acquiring signals of high symbol rates where a low-rate nonuniform sampling is truly useful, the effect of increasing the observation interval is mitigated since many symbols can still be observed in a relatively short time span for such signals. 

\begin{figure*}[tp]
\centering
\begin{minipage}[b]{0.31\textwidth}
\centering
\begin{tabular}{ccccc}
  &$\times$&+&$\ocircle$&\Square\\
  \hline		
  $\times$ 	& 50 &  0 &  0 &  0 \\
  +			&  0 & 50 &  0 &  0\\
  $\ocircle$ 	&  0 &  0 & 50 &  0\\
  \Square 	&  0 &  0 &  0 & 50\\
\end{tabular}
\subcaption{\small\sl 13000 Symbols: $\gamma=0.1$}
\label{tab:confuse_13000_500_0p1}
\end{minipage}
\begin{minipage}[b]{0.31\textwidth}
\centering
\begin{tabular}{ccccc}
  &$\times$&+&$\ocircle$&\Square\\
  \hline		
  $\times$ 	& 50 &  0 &  0 &  0 \\
  +			&  0 & 50 &  0 &  0\\
  $\ocircle$ 	&  0 &  0 & 50 &  0\\
  \Square 	&  0 &  0 &  0 & 50\\
\end{tabular}
\subcaption{ \small\sl 13000 Symbols: $\gamma=0.5$}
\label{tab:confuse_13000_500_0p5}
\end{minipage}
\begin{minipage}[b]{0.31\textwidth}
\centering
\begin{tabular}{ccccc}
  &$\times$&+&$\ocircle$&\Square\\
  \hline		
  $\times$ 	& 50 &  0 &  0 &  0 \\
  +			&  0 & 50 &  0 &  0\\
  $\ocircle$ 	&  0 &  0 & 50 &  0\\
  \Square 	&  0 &  0 &  0 & 50\\
\end{tabular}
\subcaption{ \small\sl 13000 Symbols: $\gamma=1$}
\label{tab:confuse_13000_500_1}
\end{minipage}
\\[2mm]
\begin{minipage}[b]{0.31\textwidth}
\centering
\begin{tabular}{ccccc}
  &$\times$&+&$\ocircle$&\Square\\
  \hline		
  $\times$ 	& 50 &  0 &  0 &  0 \\
  +			&  0 & 50 &  0 &  0\\
  $\ocircle$ 	&  0 &  0 & 50 &  0\\
  \Square 	&  0 &  0 &  0 & 50\\
\end{tabular}
\subcaption{ \small\sl 1625 Symbols: $\gamma=0.1$}
\label{tab:confuse_1625_500_0p1}
\end{minipage}
\begin{minipage}[b]{0.31\textwidth}
\centering
\begin{tabular}{ccccc}
  &$\times$&+&$\ocircle$&\Square\\
  \hline		
  $\times$ 	& 50 &  0 &  0 &  0 \\
  +			&  0 & 50 &  0 &  0\\
  $\ocircle$ 	&  0 &  0 & 50 &  0\\
  \Square 	&  0 &  0 &  0 & 50\\
\end{tabular}
\subcaption{ \small\sl 1625 Symbols: $\gamma=0.5$}
\label{tab:confuse_1625_500_0p5}
\end{minipage}
\begin{minipage}[b]{0.31\textwidth}
\centering
\begin{tabular}{ccccc}
  &$\times$&+&$\ocircle$&\Square\\
  \hline		
  $\times$ 	& 50 &  0 &  0 &  0 \\
  +			&  0 & 50 &  0 &  0\\
  $\ocircle$ 	&  0 &  0 & 50 &  0\\
  \Square 	&  0 &  0 &  0 & 50\\
\end{tabular}
\subcaption{ \small\sl 1625 Symbols: $\gamma=1$}
\label{tab:confuse_1625_500_1}
\end{minipage}
\\[2mm]
\begin{minipage}[b]{0.31\textwidth}
\centering
\begin{tabular}{ccccc}
  &$\times$&+&$\ocircle$&\Square\\
  \hline		
  $\times$ 	& 50 &  0 &  0 &  0 \\
  +			&  0 & 50 &  0 &  0\\
  $\ocircle$ 	&  0 &  0 & 50 &  0\\
  \Square 	&  0 &  1 &  0 & 49\\
\end{tabular}
\subcaption{ \small\sl 800 Symbols: $\gamma=0.1$}
\label{tab:confuse_800_500_0p1}
\end{minipage}
\begin{minipage}[b]{0.31\textwidth}
\centering
\begin{tabular}{ccccc}
  &$\times$&+&$\ocircle$&\Square\\
  \hline		
  $\times$ 	& 50 &  0 &  0 &  0 \\
  +			&  0 & 50 &  0 &  0\\
  $\ocircle$ 	&  0 &  0 & 50 &  0\\
  \Square 	&  0 &  0 &  0 & 50\\
\end{tabular}
\subcaption{ \small\sl 800 Symbols: $\gamma=0.5$}
\label{tab:confuse_800_500_0p5}
\end{minipage}
\begin{minipage}[b]{0.31\textwidth}
\centering
\begin{tabular}{ccccc}
  &$\times$&+&$\ocircle$&\Square\\
  \hline		
  $\times$ 	& 50 &  0 &  0 &  0 \\
  +			&  0 & 50 &  0 &  0\\
  $\ocircle$ 	&  0 &  0 & 50 &  0\\
  \Square 	&  0 &  1 &  0 & 49\\
\end{tabular}
\subcaption{ \small\sl 800 Symbols: $\gamma=1$}
\label{tab:confuse_800_500_1}
\end{minipage}
\\[2mm]
\begin{minipage}[b]{0.31\textwidth}
\centering
\begin{tabular}{ccccc}
  &$\times$&+&$\ocircle$&\Square\\
  \hline		
  $\times$ 	& 50 &  0 &  0 &  0 \\
  +			&  0 & 49 &  0 &  1\\
  $\ocircle$ 	&  0 &  0 & 50 &  0\\
  \Square 	&  0 &  2 &  1 & 47\\
\end{tabular}
\subcaption{ \small\sl 400 Symbols: $\gamma=0.1$}
\label{tab:confuse_400_500_0p1}
\end{minipage}
\begin{minipage}[b]{0.31\textwidth}
\centering
\begin{tabular}{ccccc}
  &$\times$&+&$\ocircle$&\Square\\
  \hline		
  $\times$ 	& 50 &  0 &  0 &  0 \\
  +			&  0 & 50 &  0 &  0\\
  $\ocircle$ 	&  0 &  0 & 50 &  0\\
  \Square 	&  0 &  3 &  0 &47\\
\end{tabular}
\subcaption{ \small\sl 400 Symbols: $\gamma=0.5$}
\label{tab:confuse_400_500_0p5}
\end{minipage}
\begin{minipage}[b]{0.31\textwidth}
\centering
\begin{tabular}{ccccc}
  &$\times$&+&$\ocircle$&\Square\\
  \hline		
  $\times$ 	& 50 &  0 &  0 &  0 \\
  +			&  0 & 50 &  0 &  0\\
  $\ocircle$ 	&  0 &  0 & 50 &  0\\
  \Square 	&  0 &  0 &  0 & 50\\
\end{tabular}
\subcaption{ \small\sl 400 Symbols: $\gamma=1$}
\label{tab:confuse_400_500_1}
\end{minipage}
\\
  \caption{\small\sl Confusion matrix using minimum distance classifier based on $\widehat{F}_s$ for signals under noiseless setting where $\times$, +, $\ocircle$ and \Square~represent 2PSK, 4PSK, 8PSK and 16QAM respectively.} \label{tab:confuse_500}
\end{figure*}

\begin{figure*}[tp]
\centering
\begin{minipage}[b]{0.31\textwidth}
\centering
\begin{tabular}{ccccc}
  &$\times$&+&$\ocircle$&\Square\\
  \hline		
  $\times$ 	& 50 &  0 &  0 &  0 \\
  +			&  0 & 50 &  0 &  0\\
  $\ocircle$ 	&  0 &  0 & 50 &  0\\
  \Square 	&  0 &  0 &  0 & 50\\
\end{tabular}
\subcaption{\small\sl 13000 Symbols: $\gamma=0.1$}
\label{tab:confuse_13000_9_0p1}
\end{minipage}
\begin{minipage}[b]{0.31\textwidth}
\centering
\begin{tabular}{ccccc}
  &$\times$&+&$\ocircle$&\Square\\
  \hline		
  $\times$ 	& 50 &  0 &  0 &  0 \\
  +			&  0 & 50 &  0 &  0\\
  $\ocircle$ 	&  0 &  0 & 50 &  0\\
  \Square 	&  0 &  0 &  0 & 50\\
\end{tabular}
\subcaption{ \small\sl 13000 Symbols: $\gamma=0.5$}
\label{tab:confuse_13000_9_0p5}
\end{minipage}
\begin{minipage}[b]{0.31\textwidth}
\centering
\begin{tabular}{ccccc}
  &$\times$&+&$\ocircle$&\Square\\
  \hline		
  $\times$ 	& 50 &  0 &  0 &  0 \\
  +			&  0 & 50 &  0 &  0\\
  $\ocircle$ 	&  0 &  0 & 50 &  0\\
  \Square 	&  0 &  0 &  0 & 50\\
\end{tabular}
\subcaption{ \small\sl 13000 Symbols: $\gamma=1$}
\label{tab:confuse_13000_9_1}
\end{minipage}
\\[2mm]
\begin{minipage}[b]{0.31\textwidth}
\centering
\begin{tabular}{ccccc}
  &$\times$&+&$\ocircle$&\Square\\
  \hline		
  $\times$ 	& 50 &  0 &  0 &  0 \\
  +			&  0 & 49 &  0 &  1\\
  $\ocircle$ 	&  0 &  0 & 50 &  0\\
  \Square 	&  0 &  1 &  0 & 49\\
\end{tabular}
\subcaption{ \small\sl 1625 Symbols: $\gamma=0.1$}
\label{tab:confuse_1625_9_0p1}
\end{minipage}
\begin{minipage}[b]{0.31\textwidth}
\centering
\begin{tabular}{ccccc}
  &$\times$&+&$\ocircle$&\Square\\
  \hline		
  $\times$ 	& 50 &  0 &  0 &  0 \\
  +			&  0 & 50 &  0 &  0\\
  $\ocircle$ 	&  0 &  0 & 50 &  0\\
  \Square 	&  0 &  0 &  0 & 50\\
\end{tabular}
\subcaption{ \small\sl 1625 Symbols: $\gamma=0.5$}
\label{tab:confuse_1625_9_0p5}
\end{minipage}
\begin{minipage}[b]{0.31\textwidth}
\centering
\begin{tabular}{ccccc}
  &$\times$&+&$\ocircle$&\Square\\
  \hline		
  $\times$ 	& 50 &  0 &  0 &  0 \\
  +			&  0 & 50 &  0 &  0\\
  $\ocircle$ 	&  0 &  0 & 50 &  0\\
  \Square 	&  0 &  0 &  0 & 50\\
\end{tabular}
\subcaption{ \small\sl 1625 Symbols: $\gamma=1$}
\label{tab:confuse_1625_9_1}
\end{minipage}
\\[2mm]
\begin{minipage}[b]{0.31\textwidth}
\centering
\begin{tabular}{ccccc}
  &$\times$&+&$\ocircle$&\Square\\
  \hline		
  $\times$ 	& 50 &  0 &  0 &  0 \\
  +			&  0 & 50 &  0 &  0\\
  $\ocircle$ 	&  0 &  0 & 50 &  0\\
  \Square 	&  0 &  6 &  0 & 44\\
\end{tabular}
\subcaption{ \small\sl 800 Symbols: $\gamma=0.1$}
\label{tab:confuse_800_9_0p1}
\end{minipage}
\begin{minipage}[b]{0.31\textwidth}
\centering
\begin{tabular}{ccccc}
  &$\times$&+&$\ocircle$&\Square\\
  \hline		
  $\times$ 	& 50 &  0 &  0 &  0 \\
  +			&  0 & 50 &  0 &  0\\
  $\ocircle$ 	&  0 &  0 & 50 &  0\\
  \Square 	&  0 &  0 &  0 & 50\\
\end{tabular}
\subcaption{ \small\sl 800 Symbols: $\gamma=0.5$}
\label{tab:confuse_800_9_0p5}
\end{minipage}
\begin{minipage}[b]{0.31\textwidth}
\centering
\begin{tabular}{ccccc}
  &$\times$&+&$\ocircle$&\Square\\
  \hline		
  $\times$ 	& 50 &  0 &  0 &  0 \\
  +			&  0 & 50 &  0 &  0\\
  $\ocircle$ 	&  0 &  0 & 50 &  0\\
  \Square 	&  0 &  3 &  0 & 47\\
\end{tabular}
\subcaption{ \small\sl 800 Symbols: $\gamma=1$}
\label{tab:confuse_800_9_1}
\end{minipage}
\\[2mm]
\begin{minipage}[b]{0.31\textwidth}
\centering
\begin{tabular}{ccccc}
  &$\times$&+&$\ocircle$&\Square\\
  \hline		
  $\times$ 	& 49 &  0 &  0 &  1 \\
  +			&  0 & 42 &  0 &  8\\
  $\ocircle$ 	&  0 &  0 & 50 &  0\\
  \Square 	&  0 &  12 &  1 & 38\\
\end{tabular}
\subcaption{ \small\sl 400 Symbols: $\gamma=0.1$}
\label{tab:confuse_400_9_0p1}
\end{minipage}
\begin{minipage}[b]{0.31\textwidth}
\centering
\begin{tabular}{ccccc}
  &$\times$&+&$\ocircle$&\Square\\
  \hline		
  $\times$ 	& 50 &  0 &  0 &  0 \\
  +			&  0 & 50 &  0 &  0\\
  $\ocircle$ 	&  0 &  0 & 50 &  0\\
  \Square 	&  0 &  6 &  0 &44\\
\end{tabular}
\subcaption{ \small\sl 400 Symbols: $\gamma=0.5$}
\label{tab:confuse_400_9_0p5}
\end{minipage}
\begin{minipage}[b]{0.31\textwidth}
\centering
\begin{tabular}{ccccc}
  &$\times$&+&$\ocircle$&\Square\\
  \hline		
  $\times$ 	& 50 &  0 &  0 &  0 \\
  +			&  0 & 50 &  0 &  0\\
  $\ocircle$ 	&  0 &  0 & 50 &  0\\
  \Square 	&  0 &  1 &  0 & 49\\
\end{tabular}
\subcaption{ \small\sl 400 Symbols: $\gamma=1$}
\label{tab:confuse_400_9_1}
\end{minipage}
\\
  \caption{\small\sl Confusion matrix using minimum distance classifier based on $\widehat{F}_s$ for signals having CNR of 9dB.} \label{tab:confuse_9}
\end{figure*}

\begin{figure*}[tp]
\centering
\begin{minipage}[b]{0.31\textwidth}
\centering
\begin{tabular}{ccccc}
  &$\times$&+&$\ocircle$&\Square\\
  \hline		
  $\times$ 	& 50 &  0 &  0 &  0 \\
  +			&  0 & 50 &  0 &  0\\
  $\ocircle$ 	&  0 &  0 & 50 &  0\\
  \Square 	&  0 &  0 &  0 & 50\\
\end{tabular}
\subcaption{\small\sl 13000 Symbols: $\gamma=0.1$}
\label{tab:confuse_13000_6_0p1}
\end{minipage}
\begin{minipage}[b]{0.31\textwidth}
\centering
\begin{tabular}{ccccc}
  &$\times$&+&$\ocircle$&\Square\\
  \hline		
  $\times$ 	& 50 &  0 &  0 &  0 \\
  +			&  0 & 50 &  0 &  0\\
  $\ocircle$ 	&  0 &  0 & 50 &  0\\
  \Square 	&  0 &  0 &  0 & 50\\
\end{tabular}
\subcaption{ \small\sl 13000 Symbols: $\gamma=0.5$}
\label{tab:confuse_13000_6_0p5}
\end{minipage}
\begin{minipage}[b]{0.31\textwidth}
\centering
\begin{tabular}{ccccc}
  &$\times$&+&$\ocircle$&\Square\\
  \hline		
  $\times$ 	& 50 &  0 &  0 &  0 \\
  +			&  0 & 50 &  0 &  0\\
  $\ocircle$ 	&  0 &  0 & 50 &  0\\
  \Square 	&  0 &  0 &  0 & 50\\
\end{tabular}
\subcaption{ \small\sl 13000 Symbols: $\gamma=1$}
\label{tab:confuse_13000_6_1}
\end{minipage}
\\[2mm]
\begin{minipage}[b]{0.31\textwidth}
\centering
\begin{tabular}{ccccc}
  &$\times$&+&$\ocircle$&\Square\\
  \hline		
  $\times$ 	& 50 &  0 &  0 &  0 \\
  +			&  0 & 47 &  0 &  3\\
  $\ocircle$ 	&  0 &  0 & 50 &  0\\
  \Square 	&  0 &  3 &  0 & 47\\
\end{tabular}
\subcaption{ \small\sl 1625 Symbols: $\gamma=0.1$}
\label{tab:confuse_1625_6_0p1}
\end{minipage}
\begin{minipage}[b]{0.31\textwidth}
\centering
\begin{tabular}{ccccc}
  &$\times$&+&$\ocircle$&\Square\\
  \hline		
  $\times$ 	& 50 &  0 &  0 &  0 \\
  +			&  0 & 50 &  0 &  0\\
  $\ocircle$ 	&  0 &  0 & 50 &  0\\
  \Square 	&  0 &  0 &  0 & 50\\
\end{tabular}
\subcaption{ \small\sl 1625 Symbols: $\gamma=0.5$}
\label{tab:confuse_1625_6_0p5}
\end{minipage}
\begin{minipage}[b]{0.31\textwidth}
\centering
\begin{tabular}{ccccc}
  &$\times$&+&$\ocircle$&\Square\\
  \hline		
  $\times$ 	& 50 &  0 &  0 &  0 \\
  +			&  0 & 50 &  0 &  0\\
  $\ocircle$ 	&  0 &  0 & 50 &  0\\
  \Square 	&  0 &  0 &  0 & 50\\
\end{tabular}
\subcaption{ \small\sl 1625 Symbols: $\gamma=1$}
\label{tab:confuse_1625_6_1}
\end{minipage}
\\[2mm]
\begin{minipage}[b]{0.31\textwidth}
\centering
\begin{tabular}{ccccc}
  &$\times$&+&$\ocircle$&\Square\\
  \hline		
  $\times$ 	& 50 &  0 &  0 &  0 \\
  +			&  0 & 42 &  0 &  8\\
  $\ocircle$ 	&  0 &  0 & 50 &  0\\
  \Square 	&  0 &  9 &  4 & 37\\
\end{tabular}
\subcaption{ \small\sl 800 Symbols: $\gamma=0.1$}
\label{tab:confuse_800_6_0p1}
\end{minipage}
\begin{minipage}[b]{0.31\textwidth}
\centering
\begin{tabular}{ccccc}
  &$\times$&+&$\ocircle$&\Square\\
  \hline		
  $\times$ 	& 50 &  0 &  0 &  0 \\
  +			&  0 & 50 &  0 &  0\\
  $\ocircle$ 	&  0 &  0 & 50 &  0\\
  \Square 	&  0 &  0 &  0 & 50\\
\end{tabular}
\subcaption{ \small\sl 800 Symbols: $\gamma=0.5$}
\label{tab:confuse_800_6_0p5}
\end{minipage}
\begin{minipage}[b]{0.31\textwidth}
\centering
\begin{tabular}{ccccc}
  &$\times$&+&$\ocircle$&\Square\\
  \hline		
  $\times$ 	& 50 &  0 &  0 &  0 \\
  +			&  0 & 50 &  0 &  0\\
  $\ocircle$ 	&  0 &  0 & 50 &  0\\
  \Square 	&  0 &  3 &  0 & 47\\
\end{tabular}
\subcaption{ \small\sl 800 Symbols: $\gamma=1$}
\label{tab:confuse_800_6_1}
\end{minipage}
\\[2mm]
\begin{minipage}[b]{0.31\textwidth}
\centering
\begin{tabular}{ccccc}
  &$\times$&+&$\ocircle$&\Square\\
  \hline		
  $\times$ 	& 49 &  1 &  0 &  0 \\
  +			&  0 & 39 &  0 &  11\\
  $\ocircle$ 	&  0 &  0 & 50 &  0\\
  \Square 	&  0 &  14 &  1 & 35\\
\end{tabular}
\subcaption{ \small\sl 400 Symbols: $\gamma=0.1$}
\label{tab:confuse_400_6_0p1}
\end{minipage}
\begin{minipage}[b]{0.31\textwidth}
\centering
\begin{tabular}{ccccc}
  &$\times$&+&$\ocircle$&\Square\\
  \hline		
  $\times$ 	& 50 &  0 &  0 &  0 \\
  +			&  0 & 47 &  0 &  3\\
  $\ocircle$ 	&  0 &  0 & 50 &  0\\
  \Square 	&  0 &  5 &  0 &45\\
\end{tabular}
\subcaption{ \small\sl 400 Symbols: $\gamma=0.5$}
\label{tab:confuse_400_6_0p5}
\end{minipage}
\begin{minipage}[b]{0.31\textwidth}
\centering
\begin{tabular}{ccccc}
  &$\times$&+&$\ocircle$&\Square\\
  \hline		
  $\times$ 	& 50 &  0 &  0 &  0 \\
  +			&  0 & 50 &  0 &  0\\
  $\ocircle$ 	&  0 &  0 & 50 &  0\\
  \Square 	&  0 &  2 &  0 & 48\\
\end{tabular}
\subcaption{ \small\sl 400 Symbols: $\gamma=1$}
\label{tab:confuse_400_6_1}
\end{minipage}
\\
  \caption{\small\sl Confusion matrix using minimum distance classifier based on $\widehat{F}_s$ for signals having CNR of 6dB.} \label{tab:confuse_6}
\end{figure*}

\begin{figure*}[tp]
\centering
\begin{minipage}[b]{0.31\textwidth}
\centering
\begin{tabular}{ccccc}
  &$\times$&+&$\ocircle$&\Square\\
  \hline		
  $\times$ 	& 50 &  0 &  0 &  0 \\
  +			&  0 & 50 &  0 &  0\\
  $\ocircle$ 	&  0 &  0 & 50 &  0\\
  \Square 	&  0 &  0 &  0 & 50\\
\end{tabular}
\subcaption{\small\sl 13000 Symbols: $\gamma=0.1$}
\label{tab:confuse_13000_3_0p1}
\end{minipage}
\begin{minipage}[b]{0.31\textwidth}
\centering
\begin{tabular}{ccccc}
  &$\times$&+&$\ocircle$&\Square\\
  \hline		
  $\times$ 	& 50 &  0 &  0 &  0 \\
  +			&  0 & 50 &  0 &  0\\
  $\ocircle$ 	&  0 &  0 & 50 &  0\\
  \Square 	&  0 &  0 &  0 & 50\\
\end{tabular}
\subcaption{ \small\sl 13000 Symbols: $\gamma=0.5$}
\label{tab:confuse_13000_3_0p5}
\end{minipage}
\begin{minipage}[b]{0.31\textwidth}
\centering
\begin{tabular}{ccccc}
  &$\times$&+&$\ocircle$&\Square\\
  \hline		
  $\times$ 	& 50 &  0 &  0 &  0 \\
  +			&  0 & 50 &  0 &  0\\
  $\ocircle$ 	&  0 &  0 & 50 &  0\\
  \Square 	&  0 &  0 &  0 & 50\\
\end{tabular}
\subcaption{ \small\sl 13000 Symbols: $\gamma=1$}
\label{tab:confuse_13000_3_1}
\end{minipage}
\\[2mm]
\begin{minipage}[b]{0.31\textwidth}
\centering
\begin{tabular}{ccccc}
  &$\times$&+&$\ocircle$&\Square\\
  \hline		
  $\times$ 	& 50 &  0 &  0 &  0 \\
  +			&  0 & 42 &  0 &  8\\
  $\ocircle$ 	&  0 &  0 & 50 &  0\\
  \Square 	&  0 &  6 &  1 & 43\\
\end{tabular}
\subcaption{ \small\sl 1625 Symbols: $\gamma=0.1$}
\label{tab:confuse_1625_3_0p1}
\end{minipage}
\begin{minipage}[b]{0.31\textwidth}
\centering
\begin{tabular}{ccccc}
  &$\times$&+&$\ocircle$&\Square\\
  \hline		
  $\times$ 	& 50 &  0 &  0 &  0 \\
  +			&  0 & 47 &  0 &  3\\
  $\ocircle$ 	&  0 &  0 & 50 &  0\\
  \Square 	&  0 &  3 &  0 & 47\\
\end{tabular}
\subcaption{ \small\sl 1625 Symbols: $\gamma=0.5$}
\label{tab:confuse_1625_3_0p5}
\end{minipage}
\begin{minipage}[b]{0.31\textwidth}
\centering
\begin{tabular}{ccccc}
  &$\times$&+&$\ocircle$&\Square\\
  \hline		
  $\times$ 	& 50 &  0 &  0 &  0 \\
  +			&  0 & 50 &  0 &  0\\
  $\ocircle$ 	&  0 &  0 & 50 &  0\\
  \Square 	&  0 &  0 &  0 & 50\\
\end{tabular}
\subcaption{ \small\sl 1625 Symbols: $\gamma=1$}
\label{tab:confuse_1625_3_1}
\end{minipage}
\\[2mm]
\begin{minipage}[b]{0.31\textwidth}
\centering
\begin{tabular}{ccccc}
  &$\times$&+&$\ocircle$&\Square\\
  \hline		
  $\times$ 	& 50 &  0 &  0 &  0 \\
  +			&  1 & 31 &  15 &  3\\
  $\ocircle$ 	&  0 &  0 & 50 &  0\\
  \Square 	&  1 &  21 &  9 & 19\\
\end{tabular}
\subcaption{ \small\sl 800 Symbols: $\gamma=0.1$}
\label{tab:confuse_800_3_0p1}
\end{minipage}
\begin{minipage}[b]{0.31\textwidth}
\centering
\begin{tabular}{ccccc}
  &$\times$&+&$\ocircle$&\Square\\
  \hline		
  $\times$ 	& 50 &  0 &  0 &  0 \\
  +			&  0 & 45 &  0 &  5\\
  $\ocircle$ 	&  0 &  0 & 50 &  0\\
  \Square 	&  0 &  5 &  0 & 45\\
\end{tabular}
\subcaption{ \small\sl 800 Symbols: $\gamma=0.5$}
\label{tab:confuse_800_3_0p5}
\end{minipage}
\begin{minipage}[b]{0.31\textwidth}
\centering
\begin{tabular}{ccccc}
  &$\times$&+&$\ocircle$&\Square\\
  \hline		
  $\times$ 	& 50 &  0 &  0 &  0 \\
  +			&  0 & 50 &  0 &  0\\
  $\ocircle$ 	&  0 &  0 & 50 &  0\\
  \Square 	&  0 &  6 &  0 & 44\\
\end{tabular}
\subcaption{ \small\sl 800 Symbols: $\gamma=1$}
\label{tab:confuse_800_3_1}
\end{minipage}
\\[2mm]
\begin{minipage}[b]{0.31\textwidth}
\centering
\begin{tabular}{ccccc}
  &$\times$&+&$\ocircle$&\Square\\
  \hline		
  $\times$ 	& 44 &  6 &  0 &  0 \\
  +			&  6 & 33 &  3 &  8\\
  $\ocircle$ 	&  0 &  0 & 50 &  0\\
  \Square 	&  2 &  15 &  9 & 24\\
\end{tabular}
\subcaption{ \small\sl 400 Symbols: $\gamma=0.1$}
\label{tab:confuse_400_3_0p1}
\end{minipage}
\begin{minipage}[b]{0.31\textwidth}
\centering
\begin{tabular}{ccccc}
  &$\times$&+&$\ocircle$&\Square\\
  \hline		
  $\times$ 	& 50 &  0 &  0 &  0 \\
  +			&  0 & 43 &  0 &  7\\
  $\ocircle$ 	&  0 &  0 & 50 &  0\\
  \Square 	&  0 &  12 &  0 &38\\
\end{tabular}
\subcaption{ \small\sl 400 Symbols: $\gamma=0.5$}
\label{tab:confuse_400_3_0p5}
\end{minipage}
\begin{minipage}[b]{0.31\textwidth}
\centering
\begin{tabular}{ccccc}
  &$\times$&+&$\ocircle$&\Square\\
  \hline		
  $\times$ 	& 50 &  0 &  0 &  0 \\
  +			&  0 & 44 &  0 &  6\\
  $\ocircle$ 	&  0 &  0 & 50 &  0\\
  \Square 	&  0 &  7 &  0 & 43\\
\end{tabular}
\subcaption{ \small\sl 400 Symbols: $\gamma=1$}
\label{tab:confuse_400_3_1}
\end{minipage}
\\
  \caption{\small\sl Confusion matrix using minimum distance classifier based on $\widehat{F}_s$ for signals having CNR of 3dB.} \label{tab:confuse_3}
\end{figure*}

\begin{table*}[tp]
\centering

\begin{tabular}{cc|c|c|c|c|c|c|l}
\cline{3-8}
& & 3250 & 400 & 6500 & 800& 13000 & 1625& Symbols\\ \cline{3-8}
& & 0.1 & 1 & 0.1 & 1& 0.1 & 1& $\gamma$\\ \cline{3-8}
& & 3277 & 4096 & 6554 & 8192& 13107 & 16384& Samples\\ \cline{2-8}
\multicolumn{1}{ c }{\multirow{3}{*}{CNR} } &
\multicolumn{1}{ |c| }{3dB} & 96\% & 93.5\% & 96.5 \% & 97\% & 99.5\% & 100\% &   \\ \cline{2-8}
\multicolumn{1}{ c  }{}                        &
\multicolumn{1}{ |c| }{6dB} & 100\% & 99\% & 99.5\% & 98.5\% & 100\% & 100\% &    \\ \cline{2-8}
\multicolumn{1}{ c  }{}                        &
\multicolumn{1}{ |c| }{9dB} & 100\% & 99.5\% & 100\% & 98.5\% & 100\% & 100\% &    \\ \cline{2-8}
\end{tabular}
\caption{Comparison of average $P_{\text{cc}}$ for select values of processed symbol length between $\gamma=0.1$ and $\gamma=1$ for signals having CNRs of 3dB, 6dB and 9dB.}
\label{tab:classify_chocs_hocs_c4_0}
\end{table*}

\section{Conclusion}
\label{sec:conclusions}

In extending the theory of THOCS to CTHOCS, we have proposed the use of a low-rate nonuniform sampling protocol and a low complexity technique for estimating compressive cyclic cumulants from these nonuniform samples. We also considered specifically the signal selectivity quality of compressive cyclic cumulants and saw promising classification results for a four class scenario even at sampling rates which commensurate with the information rate of the signal. Our work could potentially enable modulation classification for much higher bandwidth signals than is currently possible due to sampling hardware limitations.

There are open questions not yet addressed by our paper. In our signal model, we have not accounted for potential oscillator drifts common in most communication radios and sensing systems. Hence, the impact of carrier frequency drifts and timing offset drifts on CTHOCS remains to be seen. In our simulations, we have also not accounted for drifts in the sampling time instants of the nonuniform sampler, although in other contexts CS tests on actual nonuniform sampling hardware have been promising~\cite{Wakin12}. Further theoretical analysis of CTHOCS could also allow a better characterization of how small one can actually choose $\gamma$ in practice.

In this paper, we also did not thoroughly discuss the detection aspect of spectral peaks (compressive cyclic moments) which is fundamental for the accurate estimation of compressive cyclic cumulants or cyclic cumulants. Though the potential locations of the spectral peaks were assumed to be known in our simulations $\big(\alpha = (n-2q)\Delta f_c +\frac{k}{T}\big)$, a peak detection scheme was employed to detect the peaks since the number of peaks ($k$) varies for each signal configuration. That is, in our simulations we knew {\em where} to look for peaks but we did not assume the presence or absence of any peak was known a priori. In practice, it would be necessary to compute the power of these spectral peaks (given various signal parameters) and accordingly, design robust peak detection algorithms when noise is present in the incoming signal for a given level of detection probability, false alarm rate and  compression rate ($\gamma$). We have used the standard Cell-averaging Constant False Alarm Rate (CA-CFAR) algorithm to detect the spectral peaks in our simulations. 

\appendix
\section{Proof of Lemma~\ref{lm:estimate_error}}
\label{pf:estimate_error}

Our argument uses the following result.

\begin{lemma}[\cite{5550495}]
Let $u$, $v$ $\in \mathbb{R}^L$ be given, and suppose that a matrix $A$ satisfies the RIP of order $\text{max}(\|u+v\|_0,\|u-v\|_0)$ with isometry constant $\delta$. Then
\begin{equation}
\big|\langle Au,Av\rangle-\langle u,v\rangle\big|\le \delta\|u\|_2\|v\|_2.
\label{eqn:omp_rip_lemma}
\end{equation}
\label{lem:omp_rip_lemma}
\end{lemma}

Let $e_k \in \mathbb{R}^L$ denote the $k$th cardinal basis vector, equal to 1 in its $k$th entry and 0 elsewhere. Letting $v = e_k$ in Lemma~\ref{lem:omp_rip_lemma}, we have that
\begin{equation}
\big|\langle Au,Ae_k\rangle-\langle u,e_k\rangle\big|\le \delta\|u\|_2,
\label{eqn:omp_rip_ei_lemma}
\end{equation}
as long as $A$ satisfies the RIP of order $\|u\|_0+1$ with isometry constant $\delta$. We first derive an upper bound on the estimation error $\widehat{\eta}_{\text{lag}}[k]-\eta_{lag}[k]$:
\allowdisplaybreaks{
\begin{align*}
\widehat{\eta}_{\text{lag}}[k]-\eta_{\text{lag}}[k]&=e_k^H\widehat{\eta}_{\text{lag}}-e_k^H\eta_{\text{lag}}\\
&=e_k^HA^Hw_{\text{lag}}-e_k^H\eta_{\text{lag}}\\
&=\langle w_{\text{lag}}, Ae_k\rangle-\langle \eta_{\text{lag}}, e_k\rangle\\
&=\langle A\eta_{\text{lag}}, Ae_k\rangle-\langle \eta_{\text{lag}}, e_k\rangle\\
&=\langle A\eta_{\text{lag},s}, Ae_k\rangle-\langle \eta_{\text{lag},s}, e_k\rangle+\langle A\eta_{\text{lag},t}, Ae_k\rangle-\langle \eta_{\text{lag},t}, e_k\rangle\\
&\le \delta \|\eta_{\text{lag},s}\|_2+\langle A\eta_{\text{lag},t}, Ae_k\rangle-\langle \eta_{\text{lag},t}, e_k\rangle\\
&= \delta \|\eta_{\text{lag},s}\|_2+\sum_{j=1}^{m-1}\langle A\eta_{\text{lag},t,j}, Ae_k\rangle-\langle \eta_{\text{lag},t,j}, e_k\rangle\\
&\le \delta \|\eta_{\text{lag},s}\|_2+\delta \|\eta_{\text{lag},t,1}\|_2+\delta \|\eta_{\text{lag},t,2}\|_2+\cdots+\delta \|\eta_{\text{lag},t,m-1}\|_2\\
&\le \delta \|\eta_{\text{lag},s}\|_2+\delta \sqrt{s} \|\eta_{\text{lag},t}\|_\infty+\delta \sqrt{s} \|\eta_{\text{lag},t}\|_\infty+\cdots+\delta \sqrt{s} \|\eta_{\text{lag},t}\|_\infty\\
&= \delta \|\eta_{\text{lag},s}\|_2+\delta (m-1) \sqrt{s} \|\eta_{\text{lag},t}\|_\infty.\\
\end{align*}
}
\noindent Here, the fifth line follows from the linearity of the inner product, the sixth line follows from direct application of \eqref{eqn:omp_rip_ei_lemma}, the seventh line follows from breaking $\eta_{\text{lag},t}$ into $m-1$ components $\eta_{\text{lag},t,1}, \eta_{\text{lag},t,2}, \dots, \eta_{\text{lag},t,m-1}$, each of sparsity $s$, the eighth line follows from direct application of \eqref{eqn:omp_rip_ei_lemma}, and the ninth line follows from a standard bound relating the $\ell_2$ norm of an $s$-sparse vector to its $\ell_\infty$ norm.

Using similar arguments, one can derive a lower bound for the error estimate as $\widehat{\eta}_{\text{lag}}[k]-\eta_{\text{lag}}[k]\ge -\delta \|\eta_{\text{lag},s}\|_2-\delta (m-1) \sqrt{s} \|\eta_{\text{lag},t}\|_\infty$. This completes the proof of Lemma~\ref{lm:estimate_error}.

\section{Proof of Lemma~\ref{lm:comp_moments}}
\label{pf:comp_moments}

The result follows from the following equalities:\footnote{Using techniques found in the FOT literature such as \cite{Gardner_2_94,Napolitano03}, Lemma \ref{lm:comp_moments} also holds using the usual Dirac impulse train as the sampling operator and one would arrive at the same conclusion.} 
\allowdisplaybreaks{
\begin{align*}
{R}_{x_{\text{ncs}}}(t,\btau)_{n,q} &=\ealpha\left[x_{\text{lag ncs}}(t,\btau)_{n,q}\right]\\
&=\ealpha\left[\prod_{i=1}^n x^{(*)_i}_{i\:\text{ncs}}(t,\tau_i)\right]\\
&=\ealpha\left[\left(\prod_{i=1}^n x^{(*)_i}(t-\tau_i)\right)\left(\prod_{i=1}^n\sum^{\infty}_{m=-\infty} a_mp_{\textnormal{s/h}} (t-mT_s)\right)\right]\\
&=\ealpha\left[\prod_{i=1}^n x^{(*)_i}(t-\tau_i)\right]\ealpha\left[\prod_{i=1}^n\sum^{\infty}_{m=-\infty} a_mp_{\textnormal{s/h}}(t-mT_s)\right]\\
&=\ealpha\left[\prod_{i=1}^n x^{(*)_i}(t-\tau_i)\right]\ealpha\left[\sum^{\infty}_{m=-\infty}\prod_{i=1}^n a_mp_{\textnormal{s/h}}(t-mT_s)\right]\\
&=\ealpha\left[\prod_{i=1}^n x^{(*)_i}(t-\tau_i)\right] \cdot \\
& ~~~~ \ealpha\left[\sum^{\infty}_{m=-\infty}\left(\prod_{i=1}^n a_m\right)\left(\prod_{i=1}^np_{\textnormal{s/h}}(t-mT_s)\right)\right]\\
&=\ealpha\left[\prod_{i=1}^n x^{(*)_i}(t-\tau_i)\right]\cdot \\
& ~~~~ \left[\sum^{\infty}_{m=-\infty}\mathbb{E}\left(\prod_{i=1}^n  a_m\right)\left(\prod_{i=1}^np_{\textnormal{s/h}}(t-mT_s)\right)\right]\\
&=\ealpha\left[\prod_{i=1}^n x^{(*)_i}(t-\tau_i)\right]\left[\sum^{\infty}_{m=-\infty}\gamma \left(\prod_{i=1}^np_{\textnormal{s/h}}(t-mT_s)\right)\right]\\
&=\gamma\ealpha\left[\prod_{i=1}^n x^{(*)_i}(t-\tau_i)\right]\left[\prod_{i=1}^n\sum^{\infty}_{m=-\infty}p_{\textnormal{s/h}}(t-mT_s)\right]\\
&=\gamma\ealpha\left[\prod_{i=1}^n x^{(*)_i}(t-\tau_i)\sum^{\infty}_{m=-\infty}p_{\textnormal{s/h}}(t-mT_s)\right]\\
&=\gamma\ealpha\left[\prod_{i=1}^n x^{(*)_i}_{i\:\textnormal{s/h}}(t,\tau_i)\right]\\
&=\gamma\ealpha\left[x_{\text{lag s/h}}(t,\btau)\right]\\
&=\gamma R_{x_{\textnormal{s/h}}}(t,\btau)_{n,q},
\end{align*}
}
where the fourth equality follows due to the independence of the time series, the fifth equality follows since the s/h pulses do not overlap, the seventh equality follows since the probabilistic temporal moment function is equal to the FOT temporal moment function, the eighth equality follows since $\mathbb{E}(a^n_m)=\gamma$, and the ninth equality follows since the s/h pulses do not overlap. 

\section{Proof of Lemma~\ref{lm:ccm_msav}}
\label{pf:ccm_msav}

The result follows from the following equalities:
\allowdisplaybreaks{
\begin{align}
&\left\langle\left| \widehat{R}_{x_{\textnormal{ncs}}}^{\alpha,c}(t,\btau)_{n,q}\right|^2\right\rangle_t\nonumber\\
&=\left\langle\frac{1}{\gamma^2 \widehat{T}^2} \int_{t-\frac{\widehat{T}}{2}+\frac{\tau_0^*}{2}}^{t+\frac{\widehat{T}}{2}-\frac{\tau_0^*}{2}}\int_{t-\frac{\widehat{T}}{2}+\frac{\tau_0^*}{2}}^{t+\frac{\widehat{T}}{2}-\frac{\tau_0^*}{2}} x_{\textnormal{lag ncs}}(r,\btau)_{n,q}\Big(x_{\textnormal{lag ncs}}(s,\btau)_{n,q}\Big)^* e^{-j 2 \pi \alpha (r-s)}drds\right\rangle_t\nonumber\\
&=\left\langle\frac{1}{\gamma^2 \widehat{T}^2} \int_{-\frac{\widehat{T}}{2}+\frac{\tau_0^*}{2}}^{\frac{\widehat{T}}{2}-\frac{\tau_0^*}{2}} \int_{-\frac{\widehat{T}}{2}+\frac{\tau_0^*}{2}}^{\frac{\widehat{T}}{2}-\frac{\tau_0^*}{2}}\prod_{i=1}^{n}x^{(*)_i}_{i\:\textnormal{ncs}}(r+t,\tau_i)\left(x^{(*)_i}_{i\:\textnormal{ncs}}(s+t,\tau_i)\right)^* e^{-j 2 \pi \alpha (r-s)} drds\right\rangle_t\nonumber\\
&=\frac{1}{\gamma^2 \widehat{T}^2} \int_{-\frac{\widehat{T}}{2}+\frac{\tau_0^*}{2}}^{\frac{\widehat{T}}{2}-\frac{\tau_0^*}{2}} \int_{-\frac{\widehat{T}}{2}+\frac{\tau_0^*}{2}}^{\frac{\widehat{T}}{2}-\frac{\tau_0^*}{2}}\left\langle\prod_{i=1}^{n}x^{(*)_i}_{i\:\textnormal{ncs}}(r+t,\tau_i)\left(x^{(*)_i}_{i\:\textnormal{ncs}}(s+t,\tau_i)\right)^* e^{-j 2 \pi \alpha (r-s)}\right\rangle_t drds\nonumber\\
&=\frac{1}{\gamma^2 \widehat{T}^2} \int_{-\frac{\widehat{T}}{2}+\frac{\tau_0^*}{2}}^{\frac{\widehat{T}}{2}-\frac{\tau_0^*}{2}} \int_{-\frac{\widehat{T}}{2}+\frac{\tau_0^*}{2}}^{\frac{\widehat{T}}{2}-\frac{\tau_0^*}{2}}\left\langle\prod_{i=1}^{n}x^{(*)_i}_{i\:\textnormal{ncs}}(r+t,\tau_i)\left(x^{(*)_i}_{i\:\textnormal{ncs}}(s+t,\tau_i)\right)^*\right\rangle_t e^{-j 2 \pi \alpha (r-s)} drds\nonumber\\
&=\frac{1}{\gamma^2 \widehat{T}^2} \int_{-\frac{\widehat{T}}{2}+\frac{\tau_0^*}{2}}^{\frac{\widehat{T}}{2}-\frac{\tau_0^*}{2}}\int_{-\frac{\widehat{T}}{2}+\frac{\tau_0^*}{2}}^{\frac{\widehat{T}}{2}-\frac{\tau_0^*}{2}} \Bigg\langle\prod_{i=1}^{n}x^{(*)_i}(r+t-\tau_i)\sum_{m=-\infty}^{\infty}a_mp_{\textnormal{s/h}}(r+t-mT_s)\nonumber\\
&\times\left(x^{(*)_i}(s+t-\tau_i)\sum_{m=-\infty}^{\infty}a_mp_{\textnormal{s/h}}(s+t-mT_s)\right)^*\Bigg\rangle_t e^{-j 2 \pi \alpha (r-s)}drds\nonumber\\
&=\frac{1}{\gamma^2 \widehat{T}^2} \int_{-\frac{\widehat{T}}{2}+\frac{\tau_0^*}{2}}^{\frac{\widehat{T}}{2}-\frac{\tau_0^*}{2}}\int_{-\frac{\widehat{T}}{2}+\frac{\tau_0^*}{2}}^{\frac{\widehat{T}}{2}-\frac{\tau_0^*}{2}} \left\langle\prod_{i=1}^{n}x^{(*)_i}(r+t-\tau_i)\left(x^{(*)_i}(s+t-\tau_i)\right)^*\right\rangle_t\nonumber\\
&\times\left\langle\prod_{i=1}^{n}\left(\sum_{m=-\infty}^{\infty}a_mp_{\textnormal{s/h}}(r+t-mT_s)\sum_{m=-\infty}^{\infty}a_mp_{\textnormal{s/h}}(s+t-mT_s)\right)\right\rangle_t e^{-j 2 \pi \alpha (r-s)}drds\nonumber\\
&=\frac{1}{\gamma^2 \widehat{T}^2} \int_{-\frac{\widehat{T}}{2}+\frac{\tau_0^*}{2}}^{\frac{\widehat{T}}{2}-\frac{\tau_0^*}{2}}\int_{-\frac{\widehat{T}}{2}+\frac{\tau_0^*}{2}}^{\frac{\widehat{T}}{2}-\frac{\tau_0^*}{2}} \left\langle\prod_{i=1}^{n}x^{(*)_i}(r+t-\tau_i)\left(x^{(*)_i}(s+t-\tau_i)\right)^*\right\rangle_t\nonumber\\
&\times\left\langle\left[\sum_{m=-\infty}^{\infty}a^n_m p^n_{\textnormal{s/h}}(r+t-mT_s)\right]\left[\sum_{m=-\infty}^{\infty}a^n_m p^n_{\textnormal{s/h}}(s+t-mT_s)\right]\right\rangle_t e^{-j 2 \pi \alpha (r-s)}drds\nonumber\\
&=\frac{1}{\gamma^2 \widehat{T}^2} \int_{-\frac{\widehat{T}}{2}+\frac{\tau_0^*}{2}}^{\frac{\widehat{T}}{2}-\frac{\tau_0^*}{2}}\int_{-\frac{\widehat{T}}{2}+\frac{\tau_0^*}{2}}^{\frac{\widehat{T}}{2}-\frac{\tau_0^*}{2}} \left\langle\prod_{i=1}^{n}x^{(*)_i}(r+t-\tau_i)\left(x^{(*)_i}(s+t-\tau_i)\right)^*\right\rangle_t\nonumber\\
&\times\left\langle\sum_{m=-\infty}^{\infty}a^{2n}_m p^{n}_{\textnormal{s/h}}(r+t-mT_s)p^{n}_{\textnormal{s/h}}(s+t-mT_s)+\sum_{u\ne v}a^n_u a^n_v p^n_{\textnormal{s/h}}(r+t-uT_s)p^n_{\textnormal{s/h}}(s+t-vT_s)\right\rangle_t e^{-j 2 \pi \alpha (r-s)}drds\nonumber\\
&=\frac{1}{\gamma^2 \widehat{T}^2} \int_{-\frac{\widehat{T}}{2}+\frac{\tau_0^*}{2}}^{\frac{\widehat{T}}{2}-\frac{\tau_0^*}{2}}\int_{-\frac{\widehat{T}}{2}+\frac{\tau_0^*}{2}}^{\frac{\widehat{T}}{2}-\frac{\tau_0^*}{2}} \left\langle\prod_{i=1}^{n}x^{(*)_i}(r+t-\tau_i)\left(x^{(*)_i}(s+t-\tau_i)\right)^*\right\rangle_t\nonumber\\
&\times\left[\left\langle\sum_{m=-\infty}^{\infty}a^{2n}_m p^{n}_{\textnormal{s/h}}(r+t-mT_s)p^{n}_{\textnormal{s/h}}(s+t-mT_s)\right\rangle_t+\left\langle\sum_{u\ne v}a^n_u a^n_v p^n_{\textnormal{s/h}}(r+t-uT_s)p^n_{\textnormal{s/h}}(s+t-vT_s)\right\rangle_t\right]\nonumber\\
&\times e^{-j 2 \pi \alpha (r-s)}drds\nonumber\\
&=\frac{1}{\gamma^2 \widehat{T}^2} \int_{-\frac{\widehat{T}}{2}+\frac{\tau_0^*}{2}}^{\frac{\widehat{T}}{2}-\frac{\tau_0^*}{2}}\int_{-\frac{\widehat{T}}{2}+\frac{\tau_0^*}{2}}^{\frac{\widehat{T}}{2}-\frac{\tau_0^*}{2}} \left\langle\prod_{i=1}^{n}x^{(*)_i}(r+t-\tau_i)\left(x^{(*)_i}(s+t-\tau_i)\right)^*\right\rangle_t\nonumber\\
&\times\left[\left\langle\sum_{m=-\infty}^{\infty}\mathbb{E}\left(a^{2n}_m\right) p^{n}_{\textnormal{s/h}}(r+t-mT_s)p^{n}_{\textnormal{s/h}}(s+t-mT_s)\right\rangle_t+\left\langle\sum_{u\ne v}\mathbb{E}\left(a^n_u a^n_v\right) p^n_{\textnormal{s/h}}(r+t-uT_s)p^n_{\textnormal{s/h}}(s+t-vT_s)\right\rangle_t\right]\nonumber\\
&\times e^{-j 2 \pi \alpha (r-s)}drds\nonumber\\
&=\frac{1}{\gamma^2 \widehat{T}^2} \int_{-\frac{\widehat{T}}{2}+\frac{\tau_0^*}{2}}^{\frac{\widehat{T}}{2}-\frac{\tau_0^*}{2}}\int_{-\frac{\widehat{T}}{2}+\frac{\tau_0^*}{2}}^{\frac{\widehat{T}}{2}-\frac{\tau_0^*}{2}} \left\langle\prod_{i=1}^{n}x^{(*)_i}(r+t-\tau_i)\left(x^{(*)_i}(s+t-\tau_i)\right)^*\right\rangle_t\nonumber\\
&\times\left[\left\langle\sum_{m=-\infty}^{\infty}\gamma p^{n}_{\textnormal{s/h}}(r+t-mT_s)p^{n}_{\textnormal{s/h}}(s+t-mT_s)\right\rangle_t+\left\langle\sum_{u\ne v}\gamma^2 p^n_{\textnormal{s/h}}(r+t-uT_s)p^n_{\textnormal{s/h}}(s+t-vT_s)\right\rangle_t\right]\nonumber\\
&\times e^{-j 2 \pi \alpha (r-s)}drds\nonumber\\
&=\frac{1}{\widehat{T}^2} \int_{-\frac{\widehat{T}}{2}+\frac{\tau_0^*}{2}}^{\frac{\widehat{T}}{2}-\frac{\tau_0^*}{2}}\int_{-\frac{\widehat{T}}{2}+\frac{\tau_0^*}{2}}^{\frac{\widehat{T}}{2}-\frac{\tau_0^*}{2}} \left\langle\prod_{i=1}^{n}x^{(*)_i}(r+t-\tau_i)\left(x^{(*)_i}(s+t-\tau_i)\right)^*\right\rangle_t\nonumber\\
&\times\left[\frac{1}{\gamma}\left\langle\sum_{m=-\infty}^{\infty} p^{n}_{\textnormal{s/h}}(r+t-mT_s)p^{n}_{\textnormal{s/h}}(s+t-mT_s)\right\rangle_t+\left\langle\sum_{u\ne v} p^n_{\textnormal{s/h}}(r+t-uT_s)p^n_{\textnormal{s/h}}(s+t-vT_s)\right\rangle_t\right]\nonumber\\
&\times e^{-j 2 \pi \alpha (r-s)}drds,
\end{align}}
where the second equality follows from changes of variables, the third equality follows from using the Dominated Convergence Theorem assuming the time series considered here are uniformly bounded from above, the fourth equality follows since the complex exponential term is independent of $t$, the sixth equality follows from the independence of the time series and that both $a_m$ and $p_{\textnormal{s/h}}(t)$ are real, the seventh equality follows since the s/h pulses do not overlap, the ninth equality follows from linearity of the time-averaging operator, the tenth equality follows since the probabilistic temporal moment function is equal to the FOT temporal moment function for both terms within the square brackets, and the eleventh equality follows since $\mathbb{E}\left(a^{2n}_m\right)=\gamma$ and $\mathbb{E}\left(a^{n}_ua^{n}_v\right)=\gamma^2\;\;\forall u\ne v$.

\bibliographystyle{plain}
\bibliography{References}

\end{document}